\documentclass[fleqn,usenatbib]{mnras}

\usepackage[T1]{fontenc}

\DeclareRobustCommand{\VAN}[3]{#2}
\let\VANthebibliography\thebibliography
\def\thebibliography{\DeclareRobustCommand{\VAN}[3]{##3}\VANthebibliography}

\usepackage{graphicx}	% Including figure files
\usepackage{amsmath}	% Advanced maths commands
\usepackage{amssymb}	% Extra maths symbols
\usepackage{appendix}
\usepackage{lscape}

\def\mum{\,$\mu$m}
\def\deg{^{\circ}}
\def\sun{$_{\odot}$}
\def\Hyp{\textit{Hyper}}
\def\Cut{\textit{Cutex}}

\def\Msun{\,M$_{\odot}$}

\def\LM{\,L$_\odot$/M$_\odot$}

\def\12co{$^{12}$CO}
\def\co13{$^{13}$CO}
\def\NH2{N$_{\mathrm{H}_{2}}$}
\def\n2h{N$_{2}$H$^{+}$}
\def\d2n{N$_{2}$D$^{+}$}
\def\nh3{NH$_{3}$}
\def\hco{HCO$^{+}$}
\def\h13co{H$^{13}$CO}
\def\h20{H$_{2}$O}

\def\rms{\textit{r.m.s.}}

\def\avir{$\alpha_{vir}$}
\def\0avir{$\alpha_{0}$}

\def\a_eff{$\tilde{\alpha}_{eff}$}

\def\rms{\textit{r.m.s.}}

\title[The SQUALO project]{The SQUALO project (Star formation in QUiescent And Luminous Objects) I: clump-fed accretion mechanism in high-mass star-forming objects}

\author[A. Traficante]{
A. Traficante,$^{1}$\thanks{E-mail: alessio.traficante@inaf.it},
B. M. Jones$^{2,3}$, A. Avison$^{2,4,5}$, G. A. Fuller$^{2,3,6}$, M. Benedettini$^{1}$, D. Elia$^{1}$, S. Molinari$^{1}$, 
\newauthor N. Peretto$^{7}$, S. Pezzuto$^{1}$, T. Pillai$^{8}$, K. L. J. Rygl$^{9}$ E. Schisano$^{1}$ and R. J. Smith$^{2}$
\\
\\
$^{1}$IAPS-INAF, Via Fosso del Cavaliere, 100, I-00133, Rome, Italy\\
$^{2}$Jodrell Bank Centre for Astrophysics, Department of Physics and Astronomy, The University of Manchester, Manchester, M13 9PL, UK\\
$^{3}$ I. Physikalisches Institut, Universit\"at zu K\"oln, Z\"ulpicher Str.77, D-50937 K\"oln, Germany\\
$^{4}$UK ALMA Regional Centre Node, M13 9PL, UK\\
$^{5}$SKA Observatory, Jodrell Bank, Lower Withington, Macclesfield, SK11 9FT\\
$^6$Intituto de Astrof\'isica de Andalucia (CSIC), Glorieta de al Astronomia s/n E-18008, Granada, Spain\\
$^{7}$School of Physics and Astronomy, Cardiff University, Queen’s Buildings, The Parade, Cardiff, CF243AA, UK\\
$^{8}$Institute for Astrophysical Research, Boston University, 725 Commonwealth Avenue, Boston MA 02215, USA\\
$^{9}$INAF Institute of Radio Astronomy, Bologna, Italy
}
\date{Accepted XXX. Received YYY; in original form ZZZ}

\pubyear{2023}

\begin{document}
\label{firstpage}
\pagerange{\pageref{firstpage}--\pageref{lastpage}}
\maketitle

% Abstract of the paper
\begin{abstract}
\noindent
The formation mechanism of the most massive stars is far from completely understood. It is still unclear if the formation is core-fed or clump-fed, i.e. if the process is an extension of what happens in low-mass stars, or if the process is more dynamical
such as a continuous, multi-scale accretion from the gas at parsec (or even larger) scales. In this context we introduce the SQUALO project, an ALMA 1.3 mm and 3 mm survey designed to investigate the properties of 13 massive clumps selected at various evolutionary stages, with the common feature that they all show evidence for accretion at the clump scale. In this work we present the results obtained from the 1.3 mm continuum data. Our observations identify 55 objects with masses in the range $0.4\leq M\leq309$ \Msun, with evidence that the youngest clumps already present some degree of fragmentation. The data show that physical properties such as mass and surface density of the fragments and their parent clumps are tightly correlated. The minimum distance between fragments decreases with evolution, suggesting a dynamical scenario in which massive clumps first fragment under the influence of non-thermal motions driven by the competition between turbulence and gravity. With time gravitational collapse takes over and the fragments organize themselves into more thermally supported objects while continuing to accrete from their parent clump. Finally, one source does not fragment, suggesting that the support of other mechanisms (such as magnetic fields) is crucial only in specific star-forming regions.

%. All our evidences support the \textit{clump-fed} scenario, where the physics of the fragment is tightly correlated with that of the parent clump.

\end{abstract}

% Select between one and six entries from the list of approved keywords.
% Don't make up new ones.

%%%%%%%%%%%%%%%%%%%%%%%%%%%%%%%%%%%%%%%%%%%%%%%%%%

%%%%%%%%%%%%%%%%% BODY OF PAPER %%%%%%%%%%%%%%%%%%

\section{Introduction}
The formation of the most massive stars ($M>8$ \Msun) remains still a highly debated question. A critical aspect to understand is the mechanism that feeds the massive stars from their surroundings. The so-called \textit{core-fed} scenario suggests that the stars gather their mass from their parent core \citep{McKee03,Tan14}, relatively pressure-confined from its natal clump, in a similar fashion to the low-mass star formation mechanism \citep{Shu77}. In this scenario the  pre-stellar cores must be quite massive: accounting for a star-formation efficiency of $\simeq40\%$ from cores to stars \citep{Konyves15,Konyves20}, to form a $\simeq10$ \Msun\ star we must observe at least a $\simeq 25$ \Msun\ pre-stellar core. The hunt for such high-mass pre-stellar cores has been carried out for years now \citep{Motte18}, but the results are inconclusive. Studies based on the analysis of far-infrared (FIR)/sub-mm Galactic Plane surveys suggest that massive pre-stellar cores are extremely rare, if ever observed \citep{Ginsburg12,Traficante18_PI}. Moreover, studies of clump fragmentation at sub-parsec scales, which are significantly increasing in the ALMA era, show that massive, pre-stellar objects with no hint of ongoing star-formation activity do not seem to exist at all \citep{Svoboda19, Sanhueza19}, except for a handful of very isolated cases \citep{Nony18}. This massive pre-stellar phase must be extremely fast, and therefore very rare to identify \citep{Ginsburg12}. 

Alternatively, massive pre-stellar cores may not exist at all: the formation of massive stars may be a much more dynamical process, small seeds compete to accrete and form massive protostellar objects \citep{Bonnell06} and they are themselves fed by gas in the parent clump. This is a hierarchical process which forms dynamically connected structures at various scales from the parsec, clump scales (or even larger) down to the formation of the single cores in a global collapse, \textit{clump-fed} scenario \citep{Smith09b,Wang10,Peretto13,Vazquez-Semadeni19,Peretto20}. This interpretation would explain why in some sub-parsec resolutions surveys of massive clumps we found resolved fragments at different scales, from ~0.06 pc \citep{Csengeri17} down to 1000-2000 AU scales \citep{Beuther18,Sanhueza19,Svoboda19}. And this model is corroborated by a large number of single-dish observations of clumps at various evolutionary stages that show high accretion rates of the order of $\dot{M}\simeq 10^{-3}-10^{-4}$ \Msun\ yr$^{-1}$ \citep{Fuller05,Rygl10,He15,Wyrowski16,Traficante18_PI}. 

% accreting from parsec (or even larger) scales, in a global collapse, \textit{clump-fed} scenario \citep{Smith09b,Wang10,Peretto13,Vazquez-Semadeni19,Peretto20}. \textbf{The global collapse mechanism is a hierarchical process which forms dynamically connected structures at various scales from the parsec, clump scales down to the formation of the single cores \citep{Vazquez-Semadeni19}. 

A direct link from accretion at clump scales and the formation of fragments at sub-parsec scales within each star-forming clumps, however, has still not yet been identified. There are examples in the literature of regions that show accretion at core scales \citep{Yuan18,Cortes19,Neupane20}, which sometimes have simultaneous accretion at the clump scales, with further evidence of global collapse of the parent filaments \citep{Peretto13,Yuan18}. However, this large-scale accretion may not necessarily be responsible for the observed small-scale accretion, which could remain relatively distinct from the parsec-scales dynamics \citep{Henshaw14}.  

In this paper we discuss the first results from the SQUALO (Star formation in QUiescent And Luminous Objects) survey. SQUALO is an ALMA project approved in Cycle 6 (project ID 2018.1.00443.S) whose main goal is to investigate the connection between clump scales and core scales properties at various evolutionary phases for massive objects that show clear signatures of gas accretion at the parsec scale. We present in particular the ALMA sample of the 13 clumps selected in the project and the results obtained from the ALMA observations in the continuum at 1.3mm. These new ALMA data are used to investigate the fragmentation properties of the clumps, and to correlate them with the properties of the parent clumps, presenting a first attempt to connect parsec and sub-parsec scales on a sample of objects that are undergoing clump-scales accretion at various evolutionary stages. A detailed study of the accretion properties using ALMA Band 3 data will be the subject of a forthcoming SQUALO paper (Paper II, Traficante et al. $in\ prep.$).

The work is divided as follows: in Section \ref{sec:data_observations} we present the sample selection, based on the catalogue of Hi-GAL clumps analyzed in \citet{Traficante17} and \citet{Traficante18_PII}, and the new SQUALO observations; in Section \ref{sec:clumps_cores_analysis} we describe how we have analyzed the clumps data and we detail how we have extracted the main properties of the fragments in each clump; in Section \ref{sec:fragment_properties} we present the results derived from the analysis of the ALMA data alone, while in Section \ref{sec:fragment_clump_props} we correlate the clumps properties with those of their inner fragments. Here we show how the fragment and the clump properties are intimately connected. Finally, in Section \ref{sec:conclusions} we summarize our results and draw our conclusions.

% Massive star formation and how to gather material to form clusters. Multi-scale collapse down to 0.2 pc scale. All evolutionary phases, citing Molinari and Merello works.

%%%%%%%%%%%%%%%%%%%%%%%%%%%%%%%%%%%%%%%%%%%%%%%%%%%%%%%%%%%%%%%%%%%%%%%%%%%%%%%%%%%%%%%
\section{Sources selection and observations}\label{sec:data_observations}
%The SQUALO project aims to investigate the evolution of the gas dynamics in massive star-forming regions from parsec to sub-parsec scales. 

%To reach the goal of the SQUALO project we observed with ALMA in band 3 and band 6 thirteen clumps selected to cover a wide range of evolutionary phases and to show clear signatures of collapse at the clump scales.

We started the sample selection from the massive clumps analyzed in \citet{Traficante18_PII}, who combined the Hi-GAL clumps identified in the inner Galaxy by \citet{Elia17} with the sample of 3 mm molecular line transitions observed with the MALT90 survey \citep{Jackson13}. The \citet{Elia17} catalogue contains more than 100000 clumps identified in the Galactic plane with the Hi-GAL survey \citep{Molinari10_PASP} in the longitude range $-71\deg\leq l\leq67\deg$ with well defined spectral energy distributions (SEDs), i.e. with at least three consecutive photometric points among the 5 available within the Hi-GAL survey (70\mum, 160\mum, 250\mum, 350\mum\ and 500\mum). The MALT90 survey mapped 16 molecular lines in the 3mm band with the MOPRA telescope in a $3.4\arcmin\times3.4\arcmin$ region for 2012 clumps in the longitude range $3\deg\leq l\leq20\deg$ in the first quadrant and $-60\deg\leq l\leq-3\deg$ in the fourth quadrant, selected from the ATLASGAL survey \citep{Schuller09}. The spatial resolution is 38\arcsec, the spectral resolution of 0.11 km s$^{-1}$ and the typical system temperature is $180\leq T_{sys}\leq300$K, for a typical \rms\ noise of 250 mK per channel \citep{Jackson13}.

\citet{Traficante18_PII} combined these two catalogues and identified 213 clumps with well defined dust properties and gas kinematics, determined from their \n2h\ $(1-0)$ spectra, among which 21 showed blue-asymmetric spectra in their \hco\ $(1-0)$ line profiles interpreted as evidence of parsec-scale infall motions \citep{Fuller05}. 

For the SQUALO project we extracted a sub-sample of 10 massive clumps with these infall profiles and with the following properties: mass $M\geq170$ M\sun, surface density $\Sigma\geq1$ g cm$^{-2}$ \citep[a good indicator that clumps may form high-mass stars, e.g.][]{Tan14}, at a distance $d\leq5.5$ kpc and relatively isolated from other clumps after visual inspection of the Hi-GAL column density maps. 

These 10 clumps cover a luminosity to mass ratio $(L/M)$ range of $4\leq L/M \leq 107$ \LM. Assuming $L/M$ as evolutionary indicator \citep{Molinari08,Molinari16_l_m}, this initial sample missed the youngest objects \citep[$L/M$<1, e.g.][]{Molinari16_l_m}, often observed in absorption or with very weak emission in the Herschel 70\mum\ maps and therefore defined as 70\mum-quiet clumps. We then added three sources from the survey of massive 70\mum-quiet clumps of \citet{Traficante17} following the same selection criteria of the more evolved sources, including that these 3 clumps also show blue-asymmetric \hco\ $(1-0)$ line profiles. The kinematics of the gas in these objects have been observed with the IRAM 30m telescope at a spatial resolution of 27\arcsec\ and a variable spectral resolution of 0.06 km s$^{-1}$ for the \n2h\ $(1-0)$ and of $\simeq0.2$ km s$^{-1}$ for the \hco\ $(1-0)$ with a typical \rms\ noise per channel of $130-320$ mK. 

The final SQUALO sample contains 13 sources with a $L/M$ range $0.1\leq L/M \leq 107$ \LM, with sources classified from 70\mum-quiet up to HII regions, all of them that will likely form high-mass stars and that exhibit parsec-scales infall motions with accretion rates $\dot{M}$ in the range  $0.7\leq\dot{M}\leq27.3 \times\ 10^{-3}$ \Msun\ yr$^{-1}$ at the clump scales.

%%%%%%%%%%%%%%%%%%%%%%%%%%%%%%%%%%%%%%%%%%%%%%%%%%%%%%%%%%%%%%%%%%%%%%%%%%%%%%%%%%%%%%%
\subsection{ALMA observations}
%Adam wrote loads of scripts, Beth ran the scripts then transcribed the sacred casa hacks to words. Alessio supervised.

\subsection{Data calibration}
The single-pointings observations for ALMA project 2018.1.00443.S were conducted from 05 October 2018 to 11 January 2019 with two tunings, in bands 3 and 6 respectively. The band 6 science goal was observed as 3 observing blocks on the 12m array, 16 observing blocks on the 7m array and 28 observations with the ALMA Total Power (hereafter TP) dishes. The band 3 data were taken as 7 blocks on the 12m array and 20 blocks on the 7m array. For continuum imaging, only the interferometric data is used.

The band 3 and band 6 tunings were set to have 4 spectral windows (SPWs) that have been further split to optimize the analysis of the continuum and to trace specific lines with high spectral resolution. The central frequency, channel width and total bandwidth of the observed SPWs is given in Table \ref{tab:obs_props}. The name of each SPW denotes the key purpose, although other notable lines such as the $J=1-0$ transitions of $^{12}$CO and $^{13}$CO, are also included in the wide bandwidth band 6 `continuum' windows. 

% The band 3 tuning was set to have 4 spectral windows (SPW), and the band 6 tuning consisted of 7 SPWs. The central frequency, channel width and total bandwidth of the observed SPWs is given in Table \ref{tab:obs_props}. The name of each SPW denotes the key purpose, although other notable lines such as the $J=1-0$ transitions of $^{12}$CO and $^{13}$CO, are also included in the wide bandwidth band 6 `continuum' windows. 

\begin{table*}
\begin{tabular}{l l c c c}
\hline
Band & Name & Central freq. & Bandwidth & Channel width \\
& & (GHz) & (MHz) & (MHz) \\ \hline\hline
3   & Continuum         & 87.800 & 1875.0 & 1.129 \\
    & HCO$^+$ 1-0       & 89.189 & 117.19 & 0.071 \\
    & H$^{13}$CO$^+$ 1-0 & 86.754 & 58.59 & 0.071 \\
    & SiO \textit{v}=1 2-1       & 86.243 & 58.59 & 0.071 \\
    & HCN 1-0 & 88.631 & 117.19 & 0.071 \\
    \hline
6   & Continuum 1       & 231.100 & 1875.0 & 1.129 \\
    & Continuum 2       & 234.000 & 1875.0 & 1.129 \\
    & CH$_3$CN 12$_0$-11$_0$, F=12-11 & 220.5300 & 468.75 & 0.282 \\
    & OCS 18-17         & 218.903 & 58.59 & 0.141 \\
    & H$_2$CO 3$_{21}$-2$_{20}$ & 218.760 & 58.59 & 0.141 \\
    & H$_2$CO 3$_{22}$-2$_{21}$ & 218.476 & 58.59 & 0.141 \\
    & H$_2$CO 3$_{03}$-2$_{02}$ & 218.222 & 58.59 & 0.141 \\
    \hline
\end{tabular}
\caption{Spectral tuning for the SQUALO observations made with ALMA. Unless otherwise specified, all targeted transitions are the v=0 states.}
\label{tab:obs_props}
\end{table*}

After the observations were complete, the data were downloaded from the ALMA Archive and calibrated using the ALMA pipeline scripts provided by the archive using CASA version 5.4 \citep{McMullin07}. This resulted in fully calibrated measurement sets for the 12m and 7m arrays. Beyond the ALMA pipeline calibration, further processing and imaging was undertaken in CASA version 5.6 and is described below.

%%%%%%%%%%%%%%%%%%%%%%%%%%%%%%%%%%%%%%%%%%%%%%%%%%%%%%%%%%%%%%%%%%%%%%%%%%%%%%%%%%%%%%%
\subsection{Interferometric data preparation}
In addition to slight differences in tuning between the observation dates, the 7\,m and 12\,m correlators output a different bandwidth, with excess in the 7\,m observations. The spectral windows were first matched into groups corresponding to the same target frequency over all dates, before separating the data set to contain only matching windows and a single array using CASA's task \texttt{split}.
Regridding is performed with \texttt{mstransform} on each to transform the channels to equally spaced in LSRK (Kinematic Local Standard of Rest) velocity, with the reference frequency set equal to the rest frequency of the target spectral line, or the centre of the window for continuum spectral windows.

The overlap between the 7\,m and 12\,m windows is calculated by plotting all data with \texttt{plotms} and outputting the spectral axis for each array to a text file.
The maximum lower limit and the minimum upper limit of the spectral axes were used to define the common channels in all observations of the source.
Each of the data sets were then trimmed to the common spectral range using a further \texttt{split} and recombined into a single data set taken with a heterogeneous array for each spectral window.

%%%%%%%%%%%%%%%%%%%%%%%%%%%%%%%%%%%%%%%%%%%%%%%%%%%%%%%%%%%%%%%%%%%%%%%%%%%%%%%%%%%%%%%
\subsection{Identification of line-free channels}
To successfully create continuum images, the channels without lines must first be identified for each spectral window. 
For the majority of sources in the sample, line identification can be done visually using \texttt{plotms} to inspect the amplitude against velocity averaged over all observations.
For some sources, human inspection is neither an efficient nor accurate method for identifying continuum channels.
The automated line identification algorithm LumberJack is instead used to identify the continuum channels\footnote{LumberJack available at \url{https://github.com/adam-avison/LumberJack}}. LumberJack utilises a sigma clip and `gradient' analysis (and cross matches the results) of spectral channel flux density values extracted at the location of identified sources to determine regions spectral line free channel regions within the data on a spectral window by spectral window basis.  
% {\color{red}Adam, would you like to add a few words about LJ here?}
A buffer of 5 channels is also added on either side of detected spectral line regions to ensure that all line emission is removed. See Avison et al. (MNRAS, $submitted$) for further details.

% (check which sources had LJ)
% {\color{red}TEMPO Paper} and 

%%%%%%%%%%%%%%%%%%%%%%%%%%%%%%%%%%%%%%%%%%%%%%%%%%%%%%%%%%%%%%%%%%%%%%%%%%%%%%%%%%%%%%%
\subsection{Joint deconvolution imaging}
The image size and pixel size are determined by the combined properties of the interferometric configurations.
All imaging is completed in 'mosaic' mode to ensure the variable primary beam with baseline is accounted for, necessary for heterogeneous arrays.
The field of view is given by the combined primary beam weighted by number of baselines, and the field is imaged to the 20\% power level.
The image size is set to twice the field of view as mosaic mode requires excess padding to ensure the edges are treated correctly.
The pixel size used is 0.2\arcsec\ and samples the major axis of the beam with $\simeq6$ pixels for the band 6 data. 

Briggs' weighting with a robust parameter of 0.5 is used, as recommended by ALMA for combined array imaging. 
Multiscale \texttt{tclean} was used for all imaging, with the scales to image set to 0 (point source), the size of the synthesised beam and three times the beam size.
Cleaning and masking was performed interactively with the task \texttt{tclean}. 
All line-free channels from all spectral windows are combined for maximum sensitivity to create the aggregate bandwidth continuum images used in this paper. In the following, we will consider for the scientific analysis the primary beam (PB) corrected images produced with the described pipeline.

The synthesized beam, linear resolution and achieved \rms\ of each source are given in Table \ref{tab:alma_setup}. Our data allow us to trace sub-pc fragments of sizes down to $\simeq0.01$ pc ($\simeq2130$ AU) in the closest objects, and down to $\simeq0.04$ pc ($\simeq7700$ AU) in the furthest clumps, smaller than the expected core size \citep[0.1 pc, e.g.][]{DiFrancesco01}. The \rms\ varies significantly across the sample (from $\simeq0.8$ mJy/beam to $\simeq15.2$ mJy/beam), and it increases in sources with very bright fragments, where due to the limited dynamical range of the maps the local \rms\ is higher than the nominal one.

\begin{table*}
\begin{tabular}{cccccc}
\hline
DESIGNATION & B$_{min}$ & B$_{maj}$ & B$_{PA}$ & Linear Res. & \rms\ \\
 & ($\mathrm{{}^{\prime\prime}}$) & ($\mathrm{{}^{\prime\prime}}$) & ($\mathrm{{}^{\circ}}$) & ($\mathrm{AU}$) & ($\mathrm{mJy\ beam^{-1}}$) \\
\hline
\hline
HIGALBM327.3918+0.1996 & 1.10 & 1.25 & $-$54.5 & 6050 & 2.68 \\
HIGALBM327.4022+0.4449 & 1.10 & 1.25 & $-$54.9 & 5423 & 8.67 \\
HIGALBM331.1314$-$0.2438 & 1.08 & 1.26 & $-$64.7 & 5752 & 11.53 \\
HIGALBM332.6045$-$0.1674 & 1.07 & 1.27 & $-$67.8 & 3579 & 1.15 \\
HIGALBM338.9260+0.6340 & 1.03 & 1.12 & 77.6 & 4475 & 3.26 \\
HIGALBM341.2149$-$0.2359 & 1.03 & 1.11 & 80.3 & 3683 & 1.87 \\
HIGALBM343.5212$-$0.5172 & 1.02 & 1.11 & 85.8 & 3223 & 1.07 \\
HIGALBM343.7560$-$0.1629 & 1.02 & 1.11 & 86.3 & 2126 & 15.18 \\
HIGALBM344.1032$-$0.6609 & 1.02 & 1.11 & 86.7 & 2128 & 1.64 \\
HIGALBM344.2210$-$0.5932 & 1.02 & 1.11 & 87.3 & 2130 & 3.81 \\
HIGALBM24.0116+0.4897 & 1.23 & 1.46 & $-$88.9 & 6988 & 1.42 \\
HIGALBM28.1957$-$0.0724 & 1.30 & 1.49 & $-$87.7 & 7460 & 0.78 \\
HIGALBM31.9462+0.0759 & 1.32 & 1.49 & 89.9 & 7715 & 0.83 \\
\hline
\end{tabular}
\caption{\textit{Col. 1}: Source designation ID from the \citet{Elia21} catalogue; \textit{Cols. 2-4}: Synthesized beam properties (beam minor axis, beam major axis and position angle, respectively) for each of the 13 SQUALO sources. \textit{Col 5}: linear resolution of the average synthesized beam assuming the source distance showed in Table \ref{tab:tab_prop}; \textit{Col 6}: achieved \rms\ in each map.}
\label{tab:alma_setup}
\end{table*}

%%%%%%%%%%%%%%%%%%%%%%%%%%%%%%%%%%%%%%%%%%%%%%%%%%%%%%%%%%%%%%%%%%%%%%%%%%%%%%%%%%%%%%%
\section{Clumps and fragments analysis}\label{sec:clumps_cores_analysis}

In this Section we describe how we extract the physical and kinematic properties of the clumps and the inner fragments from the ancillary catalogues and the new ALMA data. In the hierarchical process of star-formation, the structures we detect could represent some intermediate step between clumps and the innermost cores (the mass concentrations currently forming protostars), and this is the reason why we refer to them with the generic term "fragments".

%%%%%%%%%%%%%%%%%%%%%%%%%%%%%%%%%%%%%%%%%%%%%%%%%%%%%%%%%%%%%%%%%%%%%%%%%%%%%%%%%%%%%%%
\subsection{Clump analysis}
The physical properties of the 13 clumps have been re-evaluated according to the most recent version of the Hi-GAL clumps catalogue \citep{Elia21} in order to produce a coherent sample of parameters. All these clumps have been identified in the Hi-GAL catalogue and all but one, HIGALBM31.9462+0.0759, have a distance assigned based on the \citet{Mege21} Hi-GAL distances catalogue. In the work of \citet{Elia21} the physical properties of all the sources with unknown distance have been evaluated assuming a reference value of 1 kpc \citep{Elia21}. In our case, HIGALBM31.9462+0.0759 is part of a well known infrared dark cloud complex located at a distance of 5.5 kpc \citep{Battersby14}. Therefore, for this source we re-scaled the \citet{Elia21} distance-dependent properties to this value. The kinematics of the gas have been taken from \citet{Traficante18_PII} and \citet{Traficante17} for the ten MALT90 sources and the three 70\mum-quiet clumps respectively. The main properties of the 13 clumps considered for this work are summarized in Table \ref{tab:tab_prop}.

% For the 10 objects with MALT90 counterparts the clump physical properties have been taken from \citet{Traficante18_PII}, which considers the source properties from the Hi-GAL catalogue of \citet{Elia17}. The distance dependent parameters have been re-scaled according to the latest version of the Hi-GAL distances described in \citet{Elia21} and \citet[][]{Mege21}. The physical properties of the three 70\mum-quiet clumps selected from \citet{Traficante17} have been also taken from the \citet{Elia17} catalogue in order to make a coherent sample of source properties. The distance dependent parameters have been in this case re-scaled according to the distances used in \citet{Traficante17}, that were refined for each single source. The kinematics of the gas have been taken from \citet{Traficante18_PII} and \citet{Traficante17} for the ten MALT90 sources and the three 70\mum-quiet clumps respectively. The main properties of the 13 clumps considered for this work are summarized in Table \ref{tab:tab_prop}.

\begin{table*}
\begin{tabular}{cccccccccccc}
\hline
DESIGNATION & $M_{cl}$ & $L_{cl}$ & $L_{cl}/M_{cl}$ & $\Sigma_{cl}$ & $\dot{M}_{cl}$ & $R_{cl}$ & $T_{cl}$ & $\alpha_{vir,cl}$ & $\alpha$ & $\delta$ & $d$ \\
 & ($\mathrm{M_{\odot}}$) & ($\mathrm{L_{\odot}}$) & ($\mathrm{L_{\odot}\,M_{\odot}^{-1}}$) & ($\mathrm{g\,cm^{-2}}$) & ($\mathrm{10^{-3}\,M_{\odot}\,yr^{-1}}$) & ($\mathrm{pc}$) & ($\mathrm{K}$) &  & ($\deg$) & ($\deg$)  & ($\mathrm{kpc}$) \\
 \hline
 \hline
HIGALBM327.3918+0.1996 & 1121 & 14146 & 12.6 & 3.82 & 4.67 & 0.14 & 22.9 & 0.3 & 15:50:18.646 & -53:57:03.254 & 5.16 \\
HIGALBM327.4022+0.4449 & 2157 & 73841 & 34.2 & 5.57 & 27.29 & 0.16 & 29.6 & 0.24 & 15:49:19.454 & -53:45:13.190 & 4.63 \\
HIGALBM331.1314-0.2438 & 2467 & 54845 & 22.2 & 8.99 & 11.78 & 0.14 & 31.3 & 0.26 & 16:10:59.686 & -51:50:23.104 & 4.95 \\
HIGALBM332.6045-0.1674 & 177 & 1405 & 7.9 & 2.42 & 1.63 & 0.07 & 19.6 & 0.73 & 16:17:29.486 & -50:46:09.646 & 3.07 \\
HIGALBM338.9260+0.6340 & 2689 & 10280 & 3.8 & 4.93 & 62.1 & 0.19 & 19.3 & 0.23 & 16:40:13.786 & -45:38:28.313 & 4.16 \\
HIGALBM341.2149-0.2359 & 700 & 7210 & 10.3 & 1.37 & 2.32 & 0.18 & 21.2 & 0.37 & 16:52:23.515 & -44:27:54.338 & 3.45 \\
HIGALBM343.5212-0.5172 & 347 & 7039 & 20.3 & 2.37 & 3.51 & 0.10 & 21.3 & 0.52 & 17:01:33.986 & -42:50:18.98 & 3.03 \\
HIGALBM343.7560-0.1629 & 461 & 6778 & 14.7 & 4.41 & 3.83 & 0.09 & 23.8 & 0.32 & 17:00:49.831 & -42:26:09.665 & 2.0 \\
HIGALBM344.1032-0.6609 & 230 & 3586 & 15.6 & 1.84 & 8.86 & 0.09 & 21.5 & 1.22 & 17:04:06.982 & -42:27:56.441 & 2.0 \\
HIGALBM344.2210-0.5932 & 202 & 21645 & 107.0 & 1.88 & 1.75 & 0.09 & 34.1 & 0.86 & 17:04:12.830 & -42:19:51.704 & 2.0 \\
HIGALBM24.0116+0.4897 & 1334 & 158 & 0.1 & 1.93 & 4.0 & 0.21 & 10.7 & 0.16 & 18:33:18.535 & -7:42:23.857 & 5.22 \\
HIGALBM28.1957-0.0724 & 1598 & 594 & 0.4 & 0.97 & 0.74 & 0.33 & 12.8 & 0.27 & 18:43:02.561 & -4:14:50.150 & 5.36 \\
HIGALBM31.9462+0.0759 & 1392 & 133 & 0.1 & 2.5 & 8.21 & 0.19 & 10.3 & 0.23 & 18:49:22.207 & -0:50:33.536 & 5.5 \\
\hline
\end{tabular}
\caption{Physical properties of the 13 Hi-GAL clumps selected in the SQUALO survey. These properties have been extracted from the latest release of the Hi-GAL clump catalogue \citet{Elia21}. The physical properties of HIGALBM31.9462+0.0759 have been re-scaled from the Hi-GAL catalogue assuming a distance of 5.5 kpc \citep{Battersby14}. \textit{Col.1}: Source designation ID; \textit{Cols.2-3}: mass $M_{cl}$ and luminosity $L_{cl}$ of our sources; \textit{Col.4}: $L_{cl}/M_{cl}$ of the clump, used as evolutionary indicator; \textit{Col. 5}: surface density $\Sigma_{cl}$ of the clumps; \textit{Col. 6}: mass accretion rate derived from \hco\ $(1-0)$ or HNC $(1-0)$ blue-asymmetric spectra; \textit{Col. 7}: deconvolved radius of the clumps, evaluated from the 2d-Gaussian fit to each source done at 250\mum\ \citep{Elia21}; \textit{Col. 8}: temperature of the clumps; \textit{Col. 9}: virial parameter of the clumps; \textit{Cols. 10-11}: source peak identified in the 160\mum\ Hi-GAL map; \textit{Col.12}: adopted source distance.}

%\citet{Traficante17} and \citet{Traficante18_PII} and re-scaled for the source distances adopted in this work. \textit{Col.1}: Source ID; \textit{Cols.2-3}: mass $M_{cl}$ and luminosity $L_{cl}$ of our sources; \textit{Col.4}: $L_{cl}/M_{cl}$ of the clump, used as evolutionary indicator; \textit{Col. 5}: surface density $\Sigma_{cl}$ of the clumps; \textit{Col. 6}: mass accretion rate derived from \hco\ $(1-0)$ or HNC $(1-0)$ blue-asymmetric spectra; \textit{Col. 7}: deconvolved radius of the clumps, evaluated from the 2d-Gaussian fit to each source done at 250\mum \citep{Elia21}; \textit{Col. 8}: temperature of the clumps; \textit{Col. 9}: virial parameter of the clumps; \textit{Cols. 10-11}: source peak identified in the 160\mum\ Hi-GAL map; \textit{Col.12}: adopted sources distance.}
\label{tab:tab_prop}
\end{table*}

%%%%%%%%%%%%%%%%%%%%%%%%%%%%%%%%%%%%%%%%%%%%%%%%%%%%%%%%%%%%%%%%%%%%%%%%%%%%%%%%%%%%%%%
\subsection{Fragment analysis}
The first step to determine the physical properties of our objects is to detect the fragments in each ALMA dust continuum image and extract their photometry. There are many possible approaches to extract the photometry from continuum images, so we initially decided to proceed with an "agnostic" approach and perform the photometry using two different algorithms, Astrodendro and \Hyp. In the following, we will consider the results obtained with the \textit{Hyper} code \citep{Traficante15a}. We chose this code because it has been specifically designed to: 1) deal with highly variable backgrounds, which is often the case of some of these objects that are embedded in elongated structures, and 2) deblend coupled sources, a situation that we have encountered in our sample (see Figure~\ref{fig:alma_cores}). A detailed comparisons of the different outputs of the two codes are described in Appendix \ref{app:codes}. 

We have tested several thresholds for the source identification with \Hyp\ and, after visual inspection of all the different runs, we decided to identify the fragments considering as real objects all the compact sources with peaks above 3 times the \rms\ of each cleaned, PB corrected image, estimated with a sigma-clipping procedure. Here we consider compact sources all objects with maximum aspect ratio equal to 1.5 and with FWHMs not larger than 2 times the beam size. With these identification criteria we found a total of 60 fragments distributed in our 13 clumps. After further inspection we identified 5 fragments, the faintest of the sample, that were located at the edges of the primary beam corrected images (and/or significantly away from the Hi-GAL 250\mum\ clump footprint), for which the integrated fluxes could not be accurately recovered by \Hyp. We considered these objects as spurious identifications and removed them from our final selection. In Figures \ref{fig:alma_cores} and \ref{fig:alma_cores_continuum} we show the 13 ALMA images with the 55 fragments included in the final selection (in blue crosses) and the 5 fragments excluded from the analysis (in magenta crosses). In each image are also shown the Hi-GAL 160\mum\ and 250\mum\ footprints obtained by the \Cut\ 2d-Gaussian fitting (the green-dotted ellipses). Out of these 55 fragments, 44 are observed within the 160\mum\ footprint (the shortest Hi-GAL wavelength in common to all our sources). Of the remaining 11 fragments, 7 are within the 250\mum\ footprint and 4 lie just at the edges of the clump as observed at 250\mum. Although the majority of the fragment formation occurs within the inner regions of the clump traced by the 160\mum\ emission, $\simeq20\%$ of objects are associated with the cold outskirts of the clump, traced by the 250\mum\ (and longer wavelengths) emission.

\begin{figure*}
	\includegraphics[width=7.5cm]{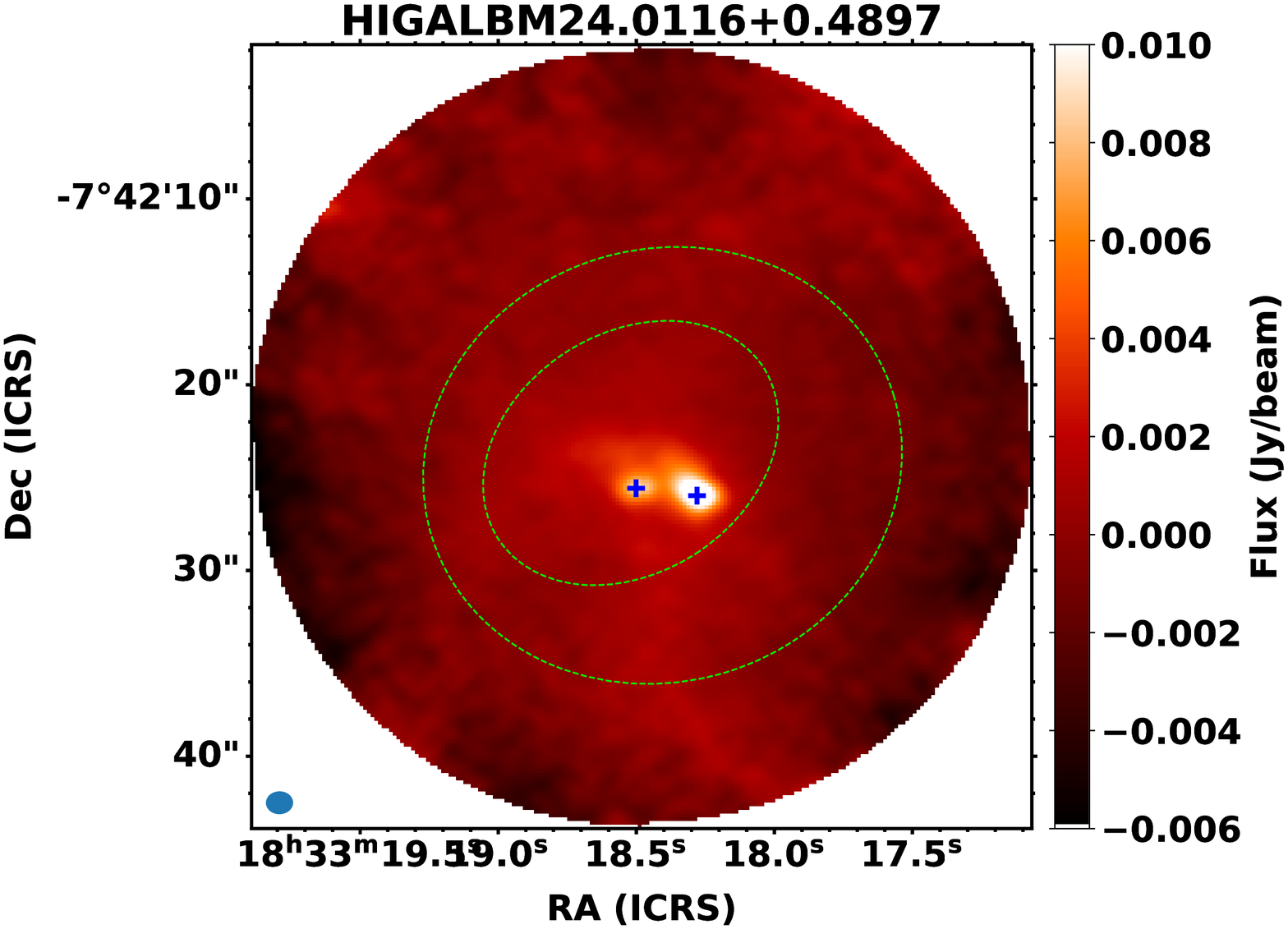} \qquad \qquad
	\includegraphics[width=7.5cm]{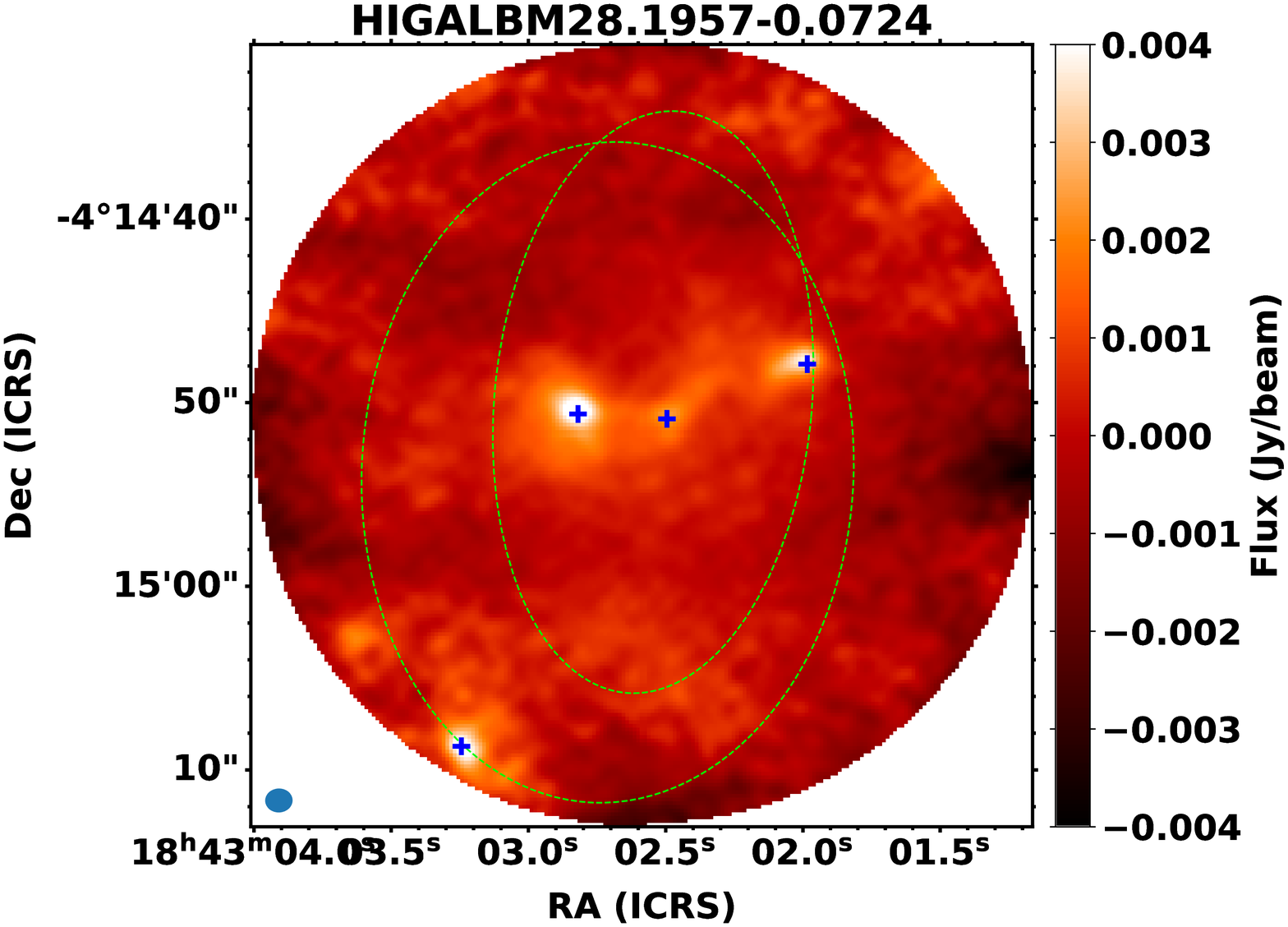} 
	\includegraphics[width=7.5cm]{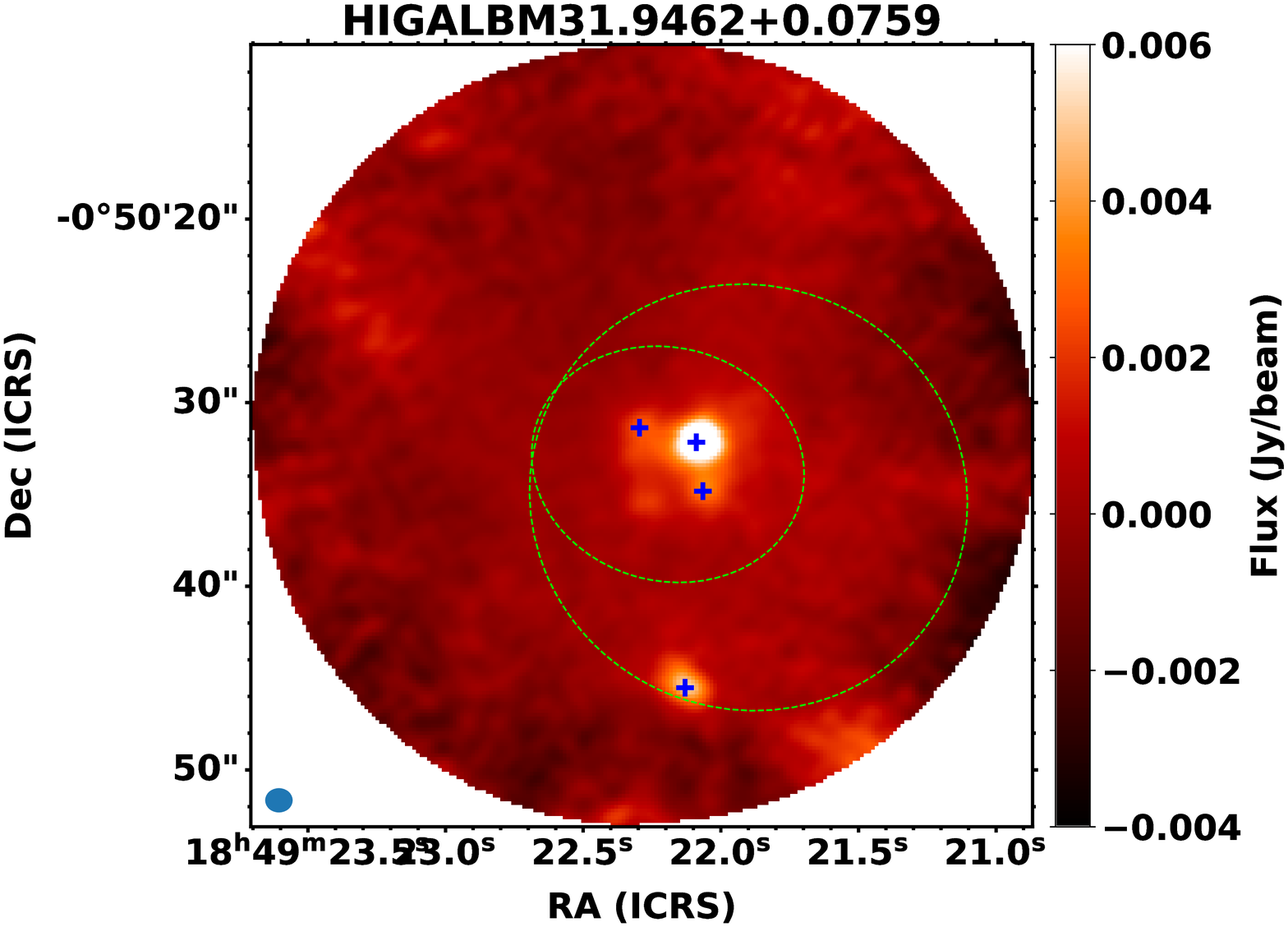}  \qquad  \qquad
	\includegraphics[width=7.5cm]{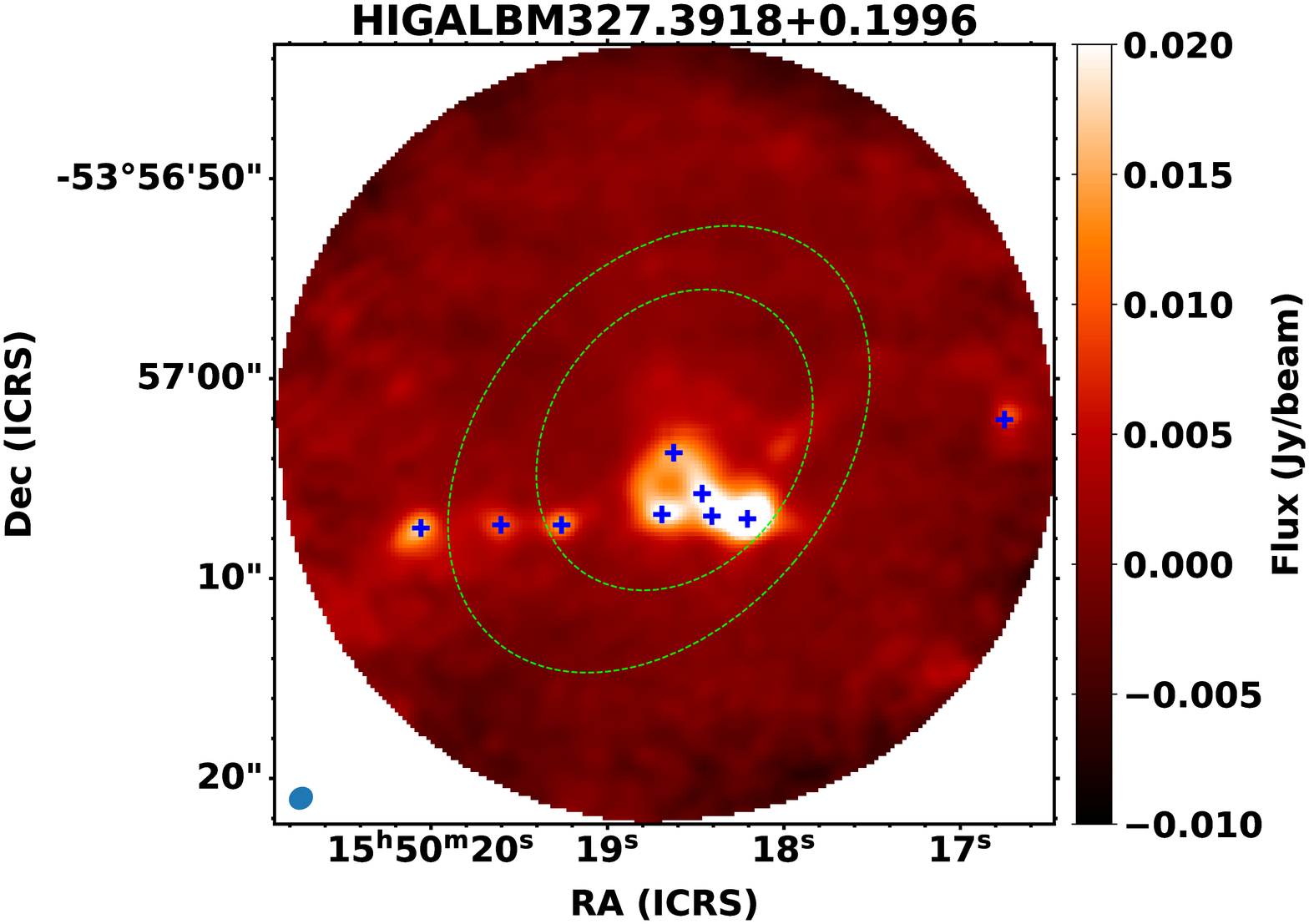} 
	\includegraphics[width=7.5cm]{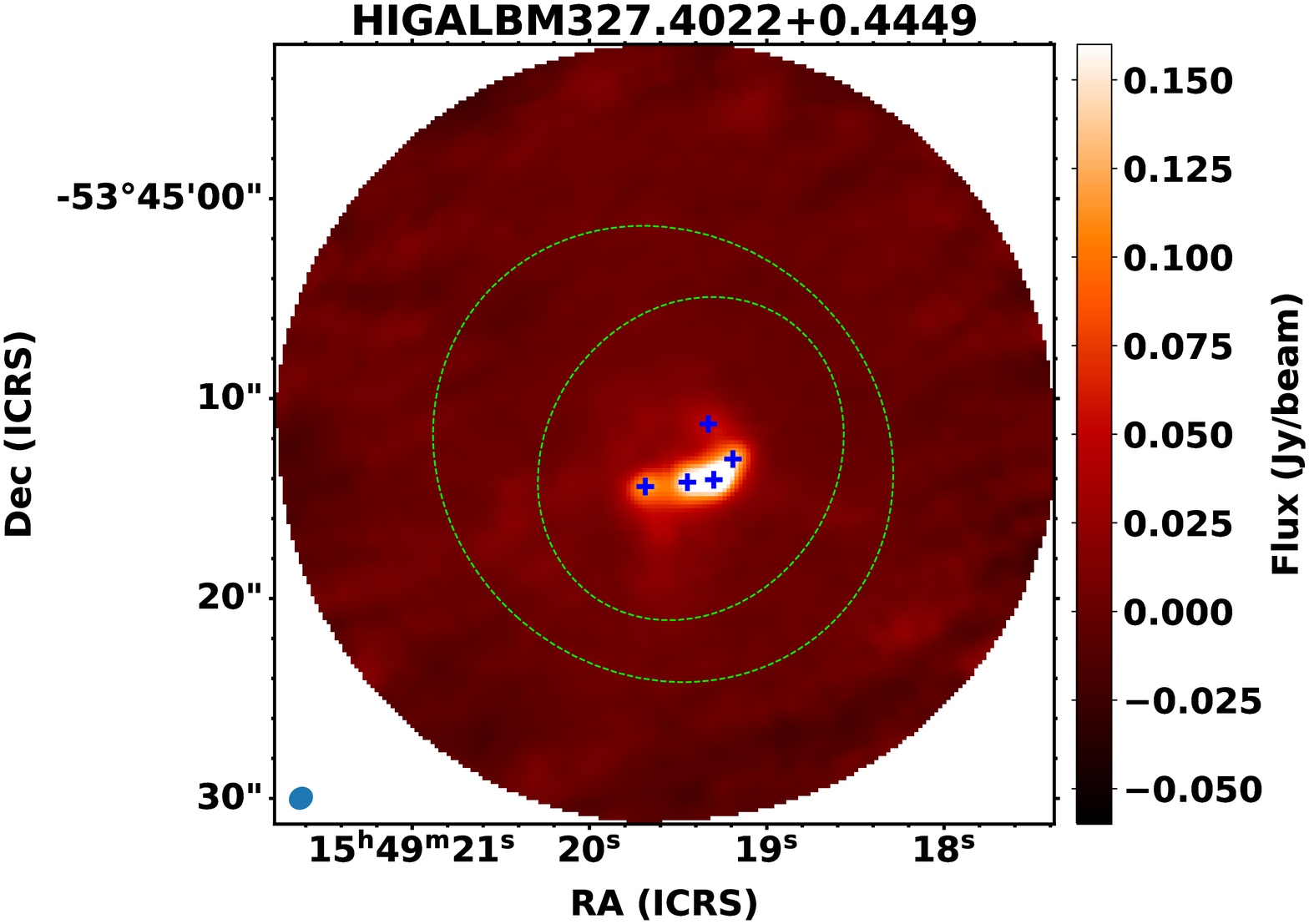} \qquad \qquad
	\includegraphics[width=7.5cm]{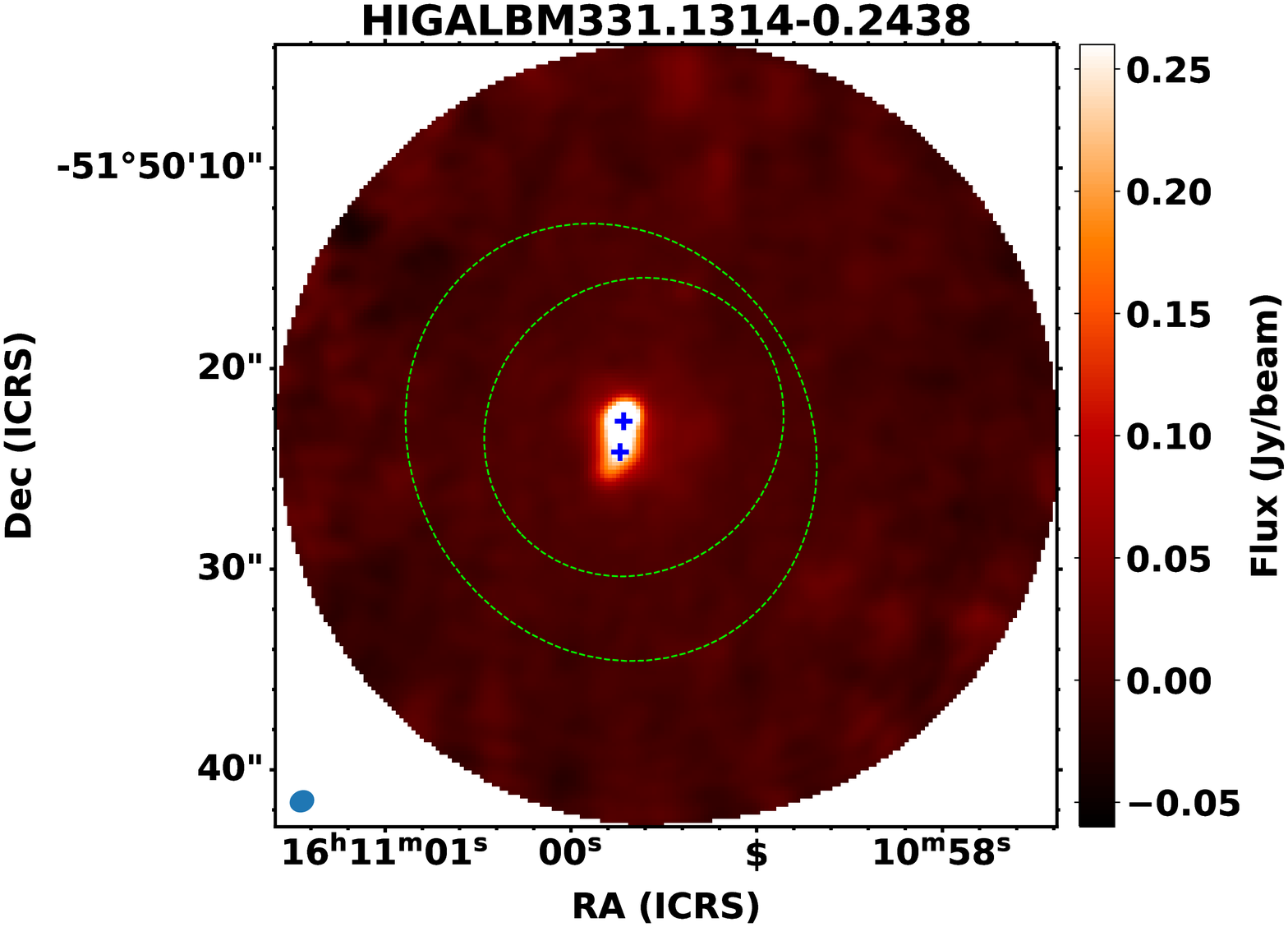} 
	\includegraphics[width=7.5cm]{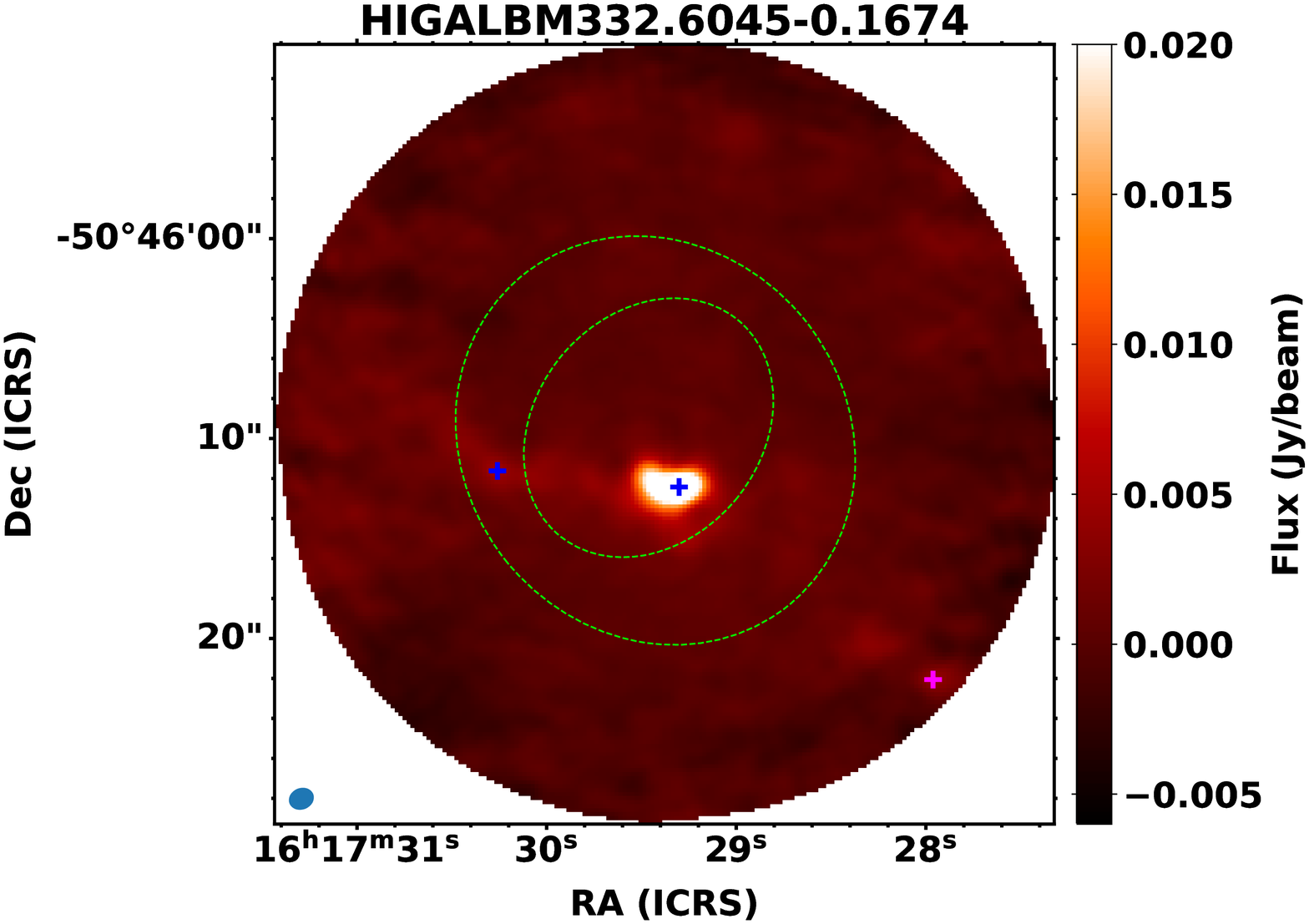} \qquad \qquad
	\includegraphics[width=7.5cm]{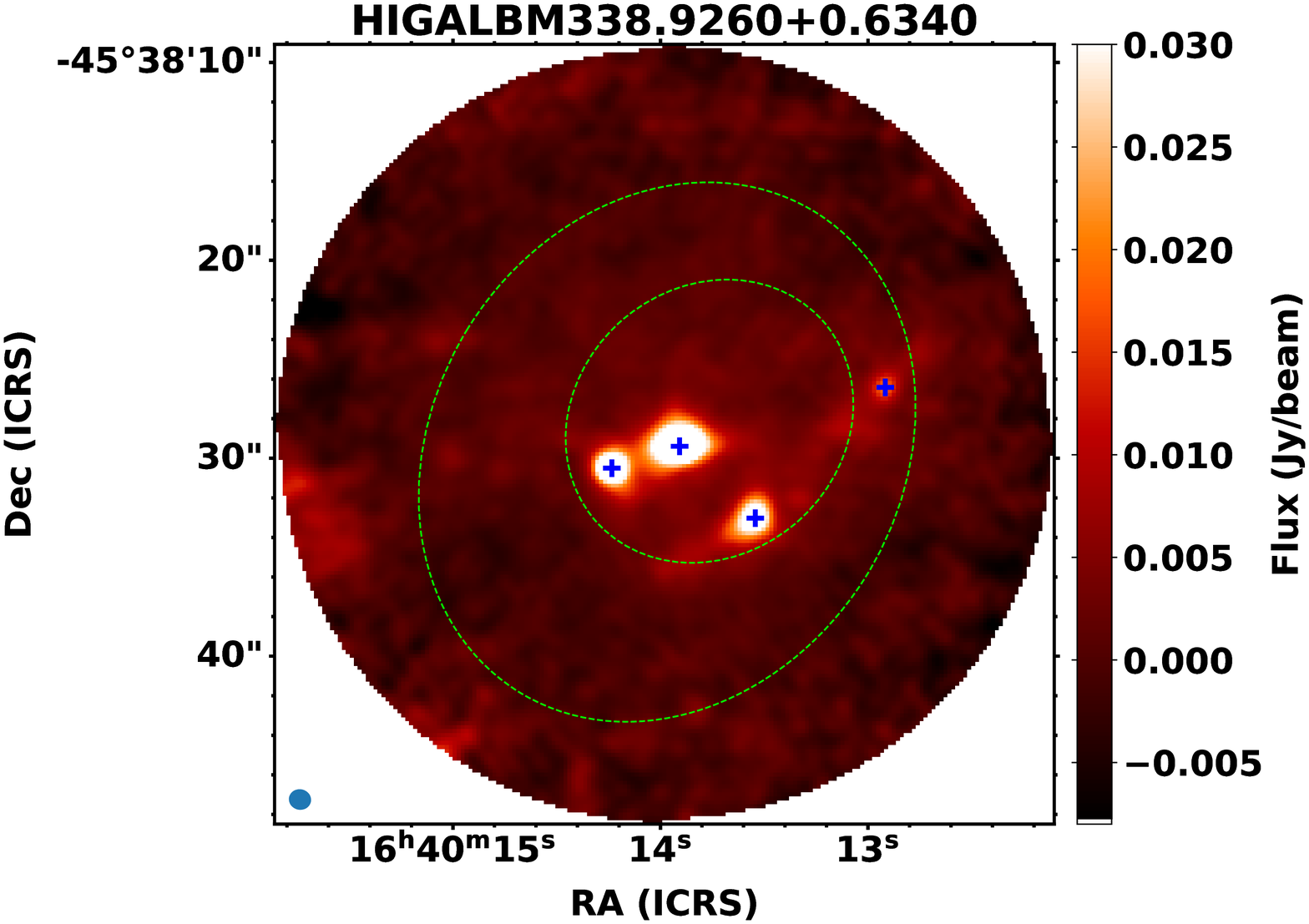} 
    \caption{ALMA 1.3 continuum images of the 13 clumps observed in the SQUALO survey. The green ellipses are the Hi-GAL 160\mum\ and 250\mum\ footprints. The blue crosses are the centroids of the 55 \Hyp\ fragments considered in the scientific analysis, and the magenta crosses are the 5 fragments excluded from the analysis. Sources HIGALBM344.1032-0.6609 and HIGALBM344.2210-0.5932 had a minor shift of the phase center with respect to the position of the Hi-GAL 160\mum\ clumps.}
    \label{fig:alma_cores}
\end{figure*}

\begin{figure*}
	\includegraphics[width=7.5cm]{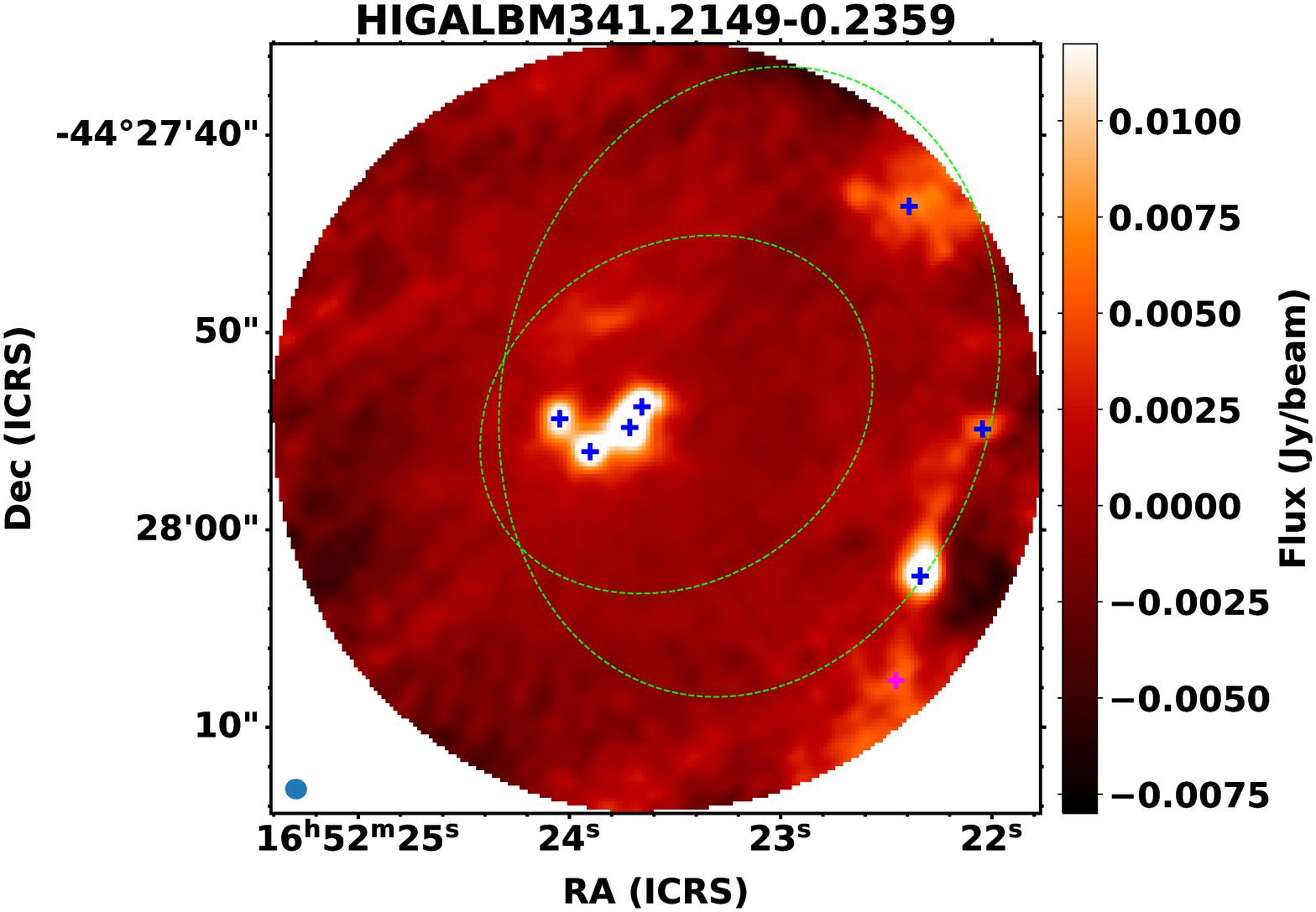}\qquad \qquad
	\includegraphics[width=7.5cm]{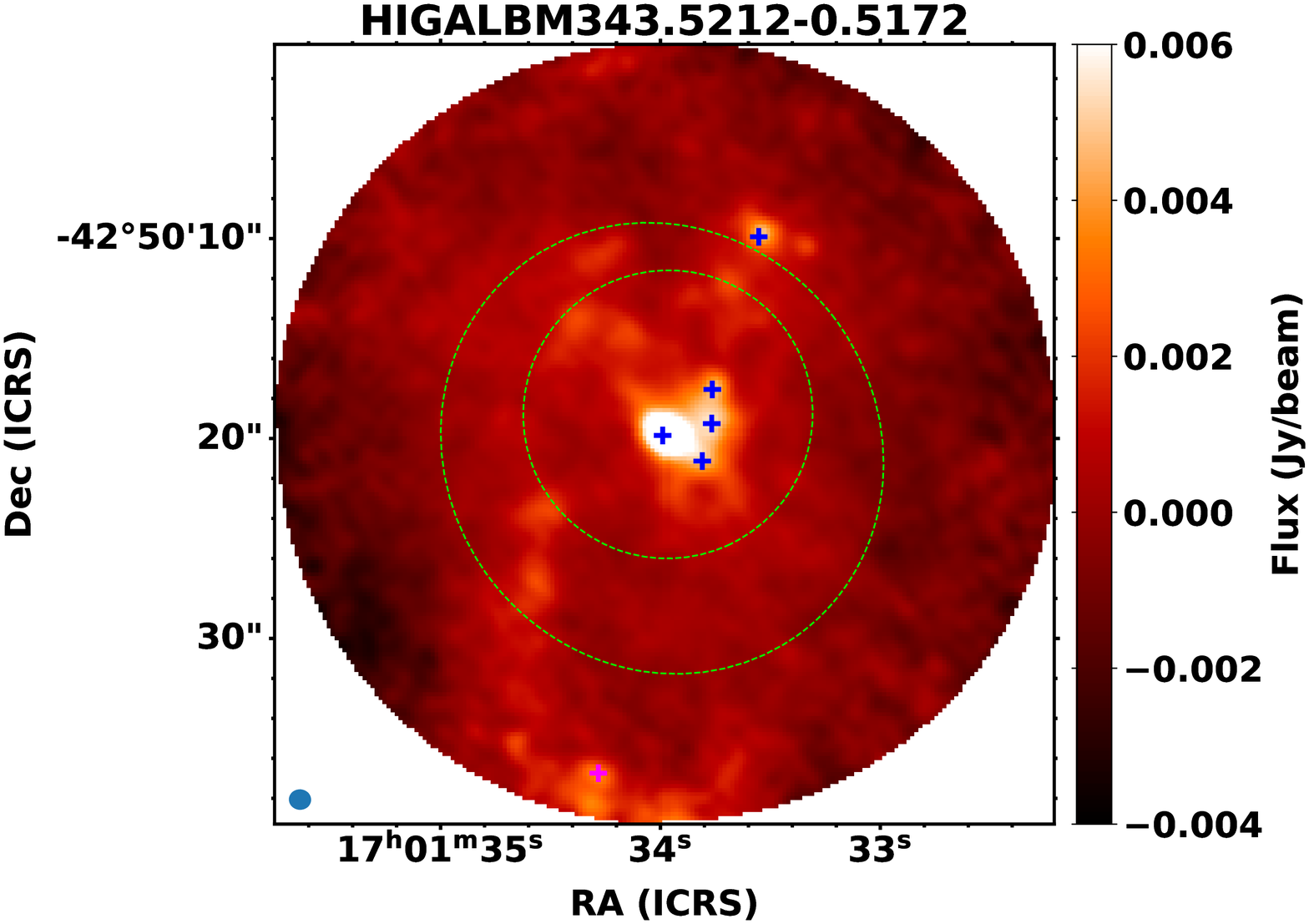} 
	\includegraphics[width=7.5cm]{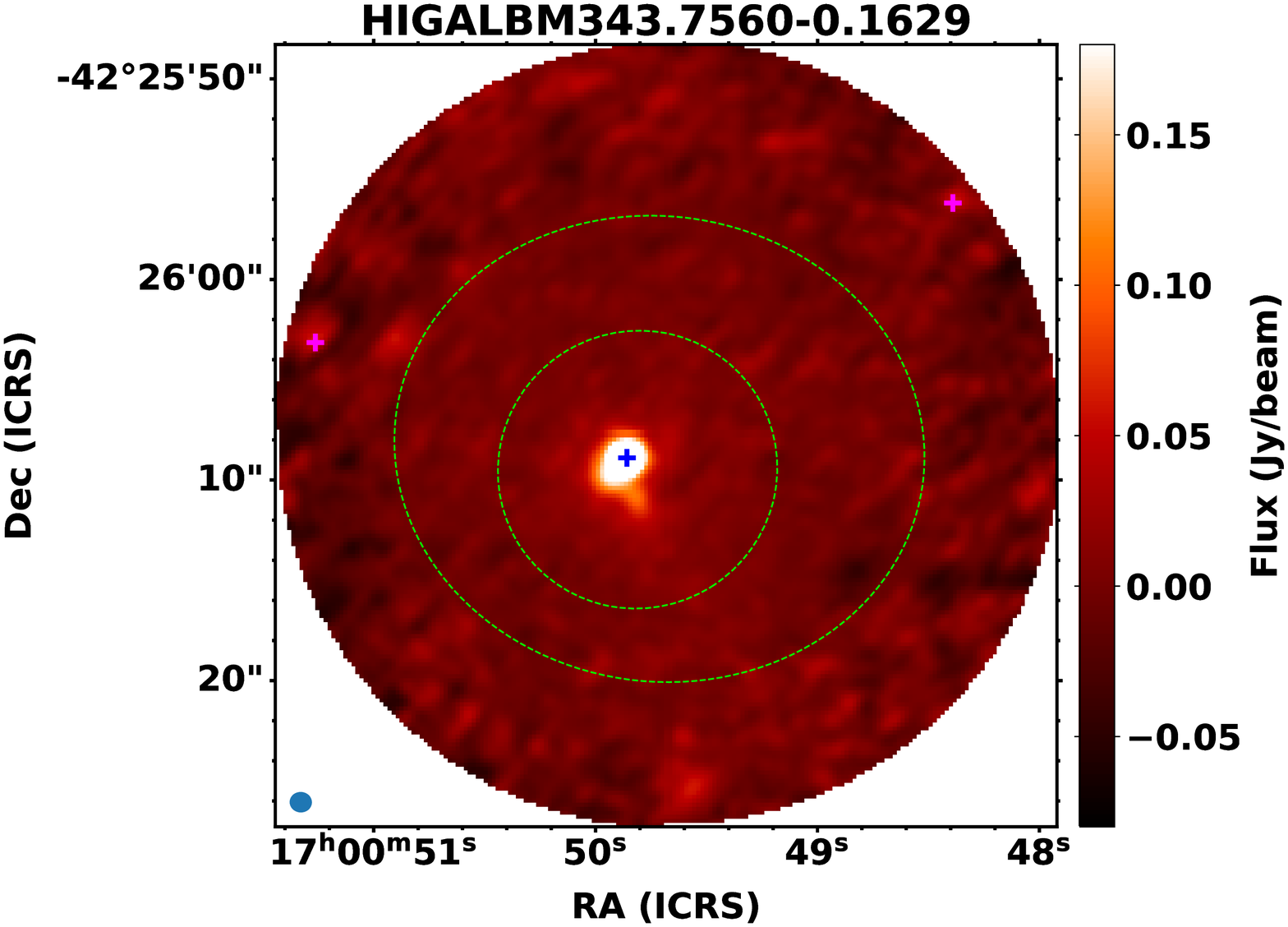} \qquad \qquad
	\includegraphics[width=7.5cm]{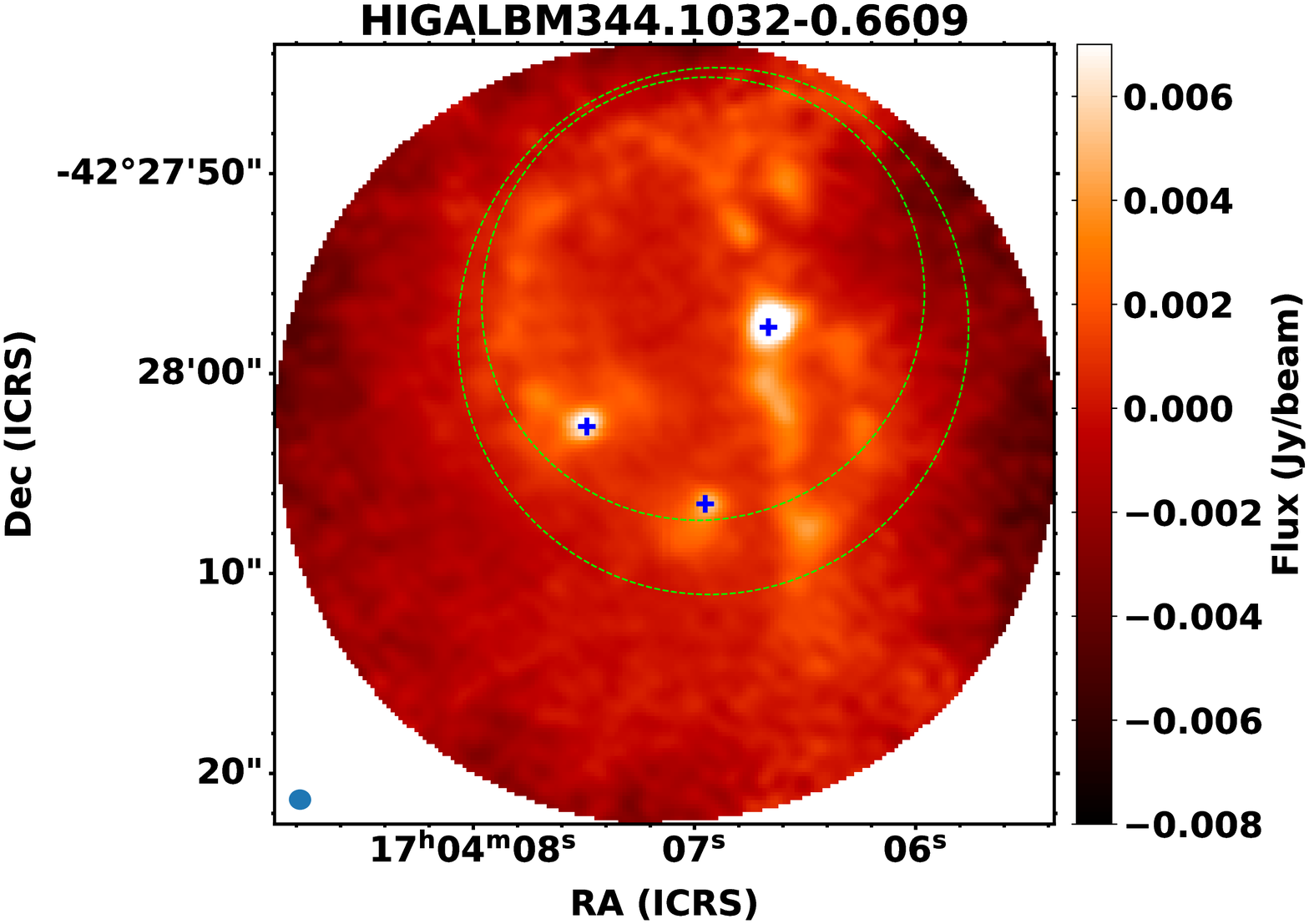} 
	\includegraphics[width=7.5cm]{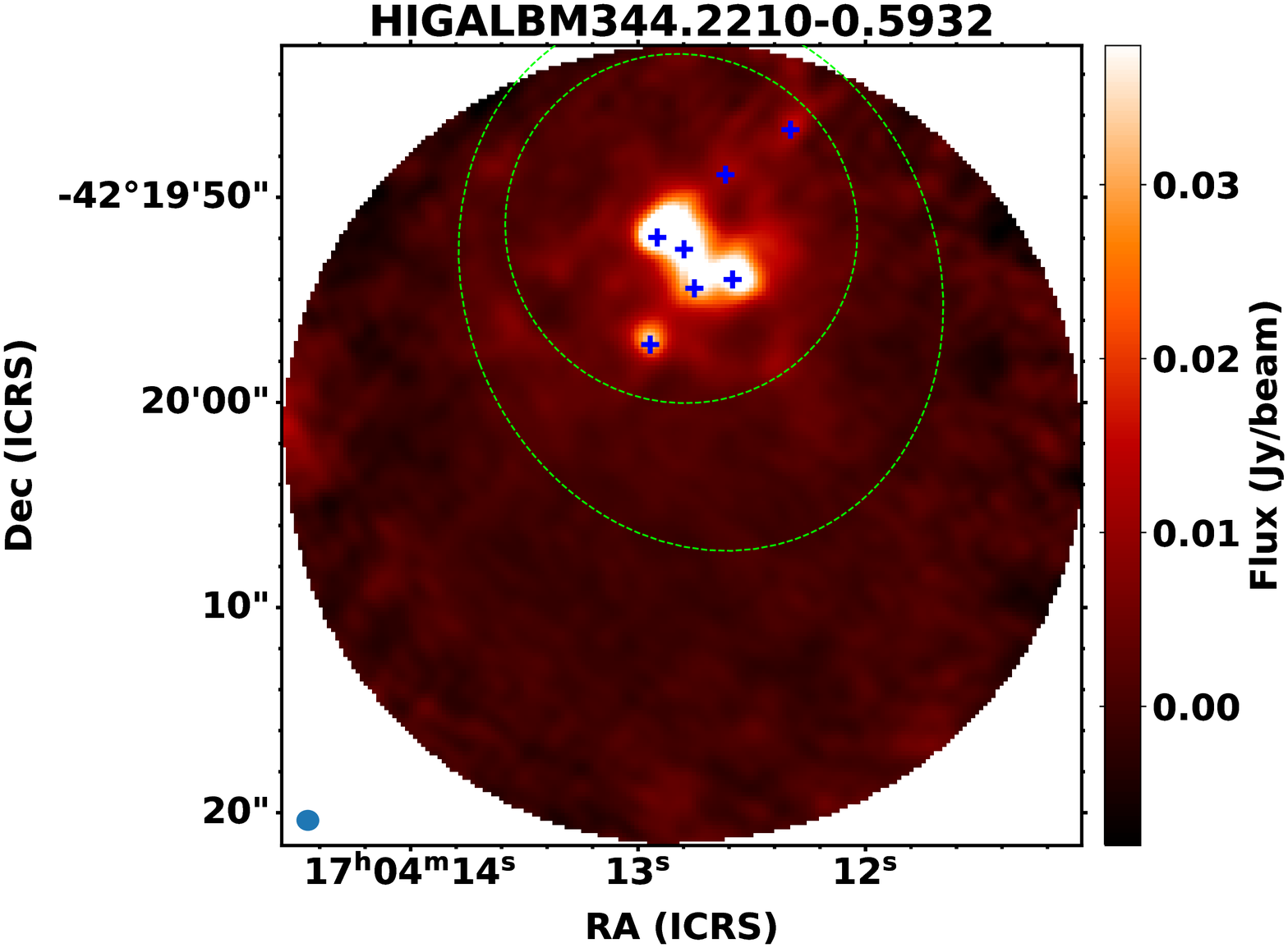} 
    \caption{Figure \ref{fig:alma_cores} continuum.}
    \label{fig:alma_cores_continuum}
\end{figure*}

%%%%%%%%%%%%%%%%%%%%%%%%%%%%%%%%%%%%%%%%%%%%%%%%%%%%%%%%%%%%%%%%%%%%%%%%%%%%%%%%%%%%%%%
\subsubsection{Radius and mass of the fragments}
The FWHMs estimated with the \Hyp\ 2D-Gaussian fitting are used to define the major and minor semi-axes of the ellipses across which we perform the aperture photometry of each fragment. The geometrical mean of the two FWHMs is used to estimate the radius of the fragments $R_{f}$, which spans the range $0.013\leq R_{f}\leq 0.049$ pc. The majority of our fragments are not point-like, but rather moderately resolved structures (Figure ~\ref{fig:alma_cores}) but it is possible that some of the objects that we have identified will further fragment into smaller pieces, as observed in higher resolution observations \citep[e.g.][]{Beuther18, Pouteau22}. %Our objects therefore could represent some intermediate step in the hierarchical process between clumps and actual cores, and this is the reason why we refer to them with the generic term "fragments". 

The integrated 1.3 mm flux of each fragment has been converted into mass following the formula \citep[e.g.][]{Svoboda19,Zhang20}:

\begin{equation}\label{eq:mass_estim}
    M_{f}=\frac{D^{2}S_{1.3}}{B_{1.3}(T_{f})\kappa_{1.3}}
\end{equation}

\noindent
where $D$ is the clump distance, $S_{1.3}$ the integrated flux at 1.3 mm, and $\kappa_{1.3}$ the dust absorption coefficient at 1.3 mm, fixed to $\kappa_{1.3}=0.005$  cm$^{2}$ g$^{-1}$ \citep[][]{Preibisch93} that also includes gas-to-dust mass ratio of 100, and $T_{f}$, the dust temperature of the dust envelopes of each fragment, which is a key parameter of this formula. Ideally, it should be measured from line emissions that are not strongly affected by outflows or feedback, such as NH$_{3}$ \citep[e.g.][]{Palau15}, which are not available for our sample. There are other lines in our spectral setup that may be used as temperature tracers, such as H$_{2}$CO or CH$_{3}$OH. All these lines, however, have been shown to also trace the hot gas around each object associated with outflows and shocks (\citet[][]{Zhang15}), not just the cold dust envelopes, and they could likely over-estimate the fragment temperatures. 

Instead, to infer the temperature of the dust envelopes of the fragments we started from the clump properties, similar to what has been done in recent ALMA works \citep[][]{Csengeri17,Svoboda19,Sanhueza19,Zhang20}. However, instead of using a single temperature for the whole sample we have assumed three different dust temperatures that account for the different evolution of our sources. We split the clumps in three groups based on their $L/M$, following the three evolutionary phases described in \citet{Molinari16_l_m}: $L/M<1$ \LM, when they are still in a pristine, pre-stellar phase, with temperatures of the innermost regions of the clumps T < 30 K; $1\leq L/M\leq10$ \LM, when they are forming the first protostars and the temperatures rises to\ T$\simeq30$ K; $L/M>10$ \LM, when a relatively evolved protostar is heating up the dust envelope and the inner cores could have already formed an ultra-compact HII region, with temperatures significantly above 30 K. In particular, some of our fragments may be even warmer than the 40 K used as the highest temperature in the clump analysis \citep{Elia21}. At the same time, the mm dust emission traces principally the (relatively) cold dust envelope, therefore we do not expect a significantly higher temperature of the 1.3 mm dust emission observed with ALMA. The temperatures at these scales are expected to vary in the range $15\leq T\leq 30$ K from pre-stellar to intermediate-mass protostellar fragments \citep{Motte22}. The mass-averaged dust temperatures in the cores identified at $\simeq2500$ AU scales in W43-MM2$\&$MM3 are in the range $20\leq\mathrm{T}\leq65$ K \citep{Pouteau22}. The fragments identified in the hub-filament systems of \citet{Anderson21}, which resolves spatial scales down to $\simeq0.03-0.07$ pc and are therefore more similar to our objects, span a temperature range $15\leq\mathrm{T}\leq46$ K. Without data available for lines sensitive to the temperature of the dust envelopes of our fragments, and based on the previous considerations, we fixed the fragment temperatures $T_{f}$ for the three groups to 20 K, 30 K and 40 K respectively. The physical parameters derived for each fragment with our assumptions are listed in Appendix \ref{app:frag_properties}.

%%%%%%%%%%%%%%%%%%%%%%%%%%%%%%%%%%%%%%%%%%%%%%%%%%%%%%%%%%%%%%%%%%%%%%%%%%%%%%%%%%%%%%%
\subsubsection{Temperature uncertainties and their implications}\label{sec:temperature_uncertainties}

The uncertainties on the temperature of the dust envelopes of the fragments are the largest source of uncertainties that can affect our results. In order to quantify these values we assumed an error in the temperature estimation up to 20 K, except for the lowest temperature considered for the group with $L/M<10$ \LM, for which we considered a lower limit of 10 K. The range of temperatures $T_{lim}$ explored for each group are in Table \ref{tab:temperature_limits}. 

% Therefore, for each group we explored the masses variation assuming these limits in temperatures: group $T_{d,f}= 20K$: $10\leq T_{lim}\leq40$ K; $T_{d,f}= 30K$: $10\leq T_{lim}\leq50$ K; $T_{d,f}= 40K$: $20\leq T_{lim}\leq60$ K (see Table \ref{tab:temperature_limits}). 

These assumptions allow us to explore a range of temperatures similar, and even larger than the values observed in other cores/fragments surveys \citep{Pouteau22,Anderson21}. At the same time, we do not expect a particularly warm fragment in extremely young star-forming regions (maximum $T_{lim}= 40$ K for fragments in clumps with $L/M<1$ \LM), neither a particularly cold fragment in very luminous and warmed-up clumps ((minimum $T_{lim}= 20$ K for fragments in clumps with $L/M>10$ \LM).

These temperature limits can be converted into mass limits $M_{lim}$ from Equation \ref{eq:mass_estim}. In particular, it follows that the mass ratio $M_{f/lim}$ evaluated assuming two different temperatures, $T_{f}$ and $T_{lim}$, is equivalent to the ratio of the blackbody part of Equation \ref{eq:mass_estim}:

\begin{equation}\label{eq:mass_limits_blackbody_ratio}
M_{f/lim} = \frac{M_{f}}{M_{lim}} = \frac{B_{1.3}(T_{lim})}{B_{1.3}(T_{f})}
\end{equation}

\begin{table}
\begin{center}
\begin{tabular}{ccc}
\hline
$T_{f}$ & $T_{lim}$ & $M_{f/lim}$  \\
(K) & (K) & \\
\hline
\hline
20 & 10 - 40 & 2.74 - 0.43 \\
% 20 & 40 & 0.43 \\
30 & 10 - 50 & 4.54 - 0.56 \\
% 30 & 50 & 0.56 \\
40 & 20 - 60 & 2.32 - 0.64 \\
% 40 & 60 & 0.64 \\
\hline
\end{tabular}
\caption{\textit{Col. 1}: reference temperature of the dust used to estimate the fragment mass for the scientific analysis; \textit{Col. 2}: range of temperatures used to estimate the mass uncertainties; \textit{Col. 3}: range of $M_{f/lim}$ for the various fragments estimated from Equation \ref{eq:mass_limits_blackbody_ratio}.}
\label{tab:temperature_limits}
\end{center}
\end{table}

The values of $M_{f/lim}$ for the various $T_{lim}$ are in Table \ref{tab:temperature_limits}. We obtain mass differences of $\simeq$ a factor of 2 and up to $M_{f/lim}\simeq4.5$ (when considering $T_{lim}= 10K$ instead of $T_{f}= 30K$). Overall, these temperature variations may significantly affect the mass estimates for each single fragment. These variations are not affecting the statistical studies that we will do in the following sections. As we show in details in Appendix \ref{app:Pearson_MC_statistics}, we have run a series of Monte Carlo simulations in which we have assigned to each fragment a random temperature within the ranges described before. For each new temperature we have re-evaluated the temperature-dependent parameters described in Section \ref{sec:fragments_temperature_dependent_clump_props} and compared the results with those obtained using T = $T_{f}$ . These tests shown that our results, based on the correlation between different quantities, are robust against these temperature variations. Nonetheless, in the analysis in Section \ref{sec:fragment_clump_props}, where we will compare the fragment properties with those of the parent clumps, we will split the results in two groups. The ones derived from the spatial distribution of the fragments, which are independent from the dust temperature, and the results that are sensitive to the temperature estimation such as the ones derived from the mass and surface density of the fragments.

% Nonetheless, in the analysis in Section \ref{sec:fragment_clump_props}, where we will compare the fragment properties with those of the parent clumps, we will split the results derived from the spatial distribution of the fragments, which are independent from the dust temperature, from the results that are sensitive to the temperature estimation such as the ones derived from the mass and surface density of the fragments.

% we have considered a range of temperatures for each core to re-estimate the masses of the fragments from Equation \ref{eq:mass_estim}. W

% Note that the uncertainties on the temperature of the dust envelopes can significantly affect these results. If we estimate the masses of the fragments from Equation \ref{eq:mass_estim} assuming first 20 K and then 40 K for the whole sample, we found differences of above a factor of 2 in mass. In the analysis in Section \ref{sec:fragment_clump_props}, where we will compare the fragment properties with those of the parent clumps, we will split the results derived from the spatial distribution of the fragments, which are independent from the dust temperature, from the results that are sensitive to the temperature estimation such as the ones derived from the mass and surface density of the fragments. 

%%%%%%%%%%%%%%%%%%%%%%%%%%%%%%%%%%%%%%%%%%%%%%%%%%%%%%%%%%%%%%%%%%%%%%%%%%%%%%%%%%%%%%%
\subsubsection{Mass completeness limits}\label{sec:mass_completeness}
A crucial part of the analysis of the properties of these fragments is to understand how good is our recovery of the faintest objects in each clump. The mass completeness limit $M_{com}$ for each clump depends on both the \rms\ of the map and the source distance, which in our sample vary in the ranges $\simeq0.8$ mJy/beam and $\simeq15.2$ mJy/beam  (Table \ref{tab:alma_setup}) and $2.0\leq d\leq 5.5$ kpc (Table \ref{tab:obs_props}) respectively. 

Note that due to the background-subtracted flux estimation the intensity of the faintest peak that we are able to recover can be significantly lower than the $3\sigma$ value of the \rms\ of each map used to identify the peaks in each image. Several tests done with \Hyp\ or similar extraction algorithms such as \Cut\ \citep[][]{Molinari11} showed that we can assume a conservative estimate of $1\sigma$ value of the local \rms\ as the lower limit to the peak emission of a fragment that we can recover in each map \citep[e.g.][]{Molinari16_cat}. These flux values can be converted into mass limits assuming that they correspond to the emission of a point-like source. Using Equation \ref{eq:mass_estim} and assuming for each clump the dust temperature defined in the previous section, we found $M_{com}$ in the range $0.1\leq M_{com}\leq4.7$ M\sun. These limits are reported in Table \ref{tab:completeness_limits}, together with the minimum recovered mass of the fragments for each clump ($M_{min}$), estimated as discussed in the previous section, the ratio between these two values ($M_{ratio}$), and the total number of robust fragments identified in each source $\#_{f}$. The mass of the least massive fragment is at least 2.2 times higher than the mass limits (and up to $\simeq72$ times greater for the single massive fragment identified in HIGALBM343.7560-0.1629), suggesting that if more fragments are present in each map, they should be on average several times less massive than the observed ones.

% Note that we consider as real fragments all objects with a peak emission above $3\sigma$ level of the \rms\ of each map, which however includes the flux of the source and of the local background. The background-subtracted flux of the source can be significantly lower than the $3\sigma$ level of the \rms. Several tests done with \Hyp\ or similar extraction algorithms such as \Cut\ \citep[][]{Molinari11} showed that we can assume a conservative estimate of $1\sigma$ value of the local \rms\ as the lower limit to the peak emission of a fragment that we can recover in each map \citep[e.g.][]{Molinari16_cat}. These flux values can be converted into mass limits assuming that they corresponds to the emission of a point-like source with a dust temperature equal to the value used to estimated the mass of the fragments in each clump. With these assumptions we obtain $M_{com}$ in the range $0.2\leq M_{com}\leq4.8$ M\sun. These limits are reported in Table \ref{tab:completeness_limits}, together with the minimum recovered mass of the fragments for each clump ($M_{min}$), estimated as discussed in the previous section, the ratio between these two values ($M_{ratio}$), and the total number of robust fragments identified in each source $\#_{f}$. The mass of the least massive fragment is at least 2.2 times higher than the mass limits (and up to $\simeq78$ times greater for the single massive fragment identified in HIGALBM343.7560-0.1629), suggesting that if more fragments are present in each map, they should be on average several times less massive than the observed ones.

\begin{table}
\begin{center}
\begin{tabular}{cccccc}
\hline
DESIGNATION & $M_{com}$ & $M_{min}$ & $M_{ratio}$ & $\#_{f}$ \\
 & ($\mathrm{M_{\odot}}$) & ($\mathrm{M_{\odot}}$) & ($\mathrm{}$) &  \\
 \hline
 \hline
%G327.393+00.199 & 1.19 & 9.75 & 8.2 & 9 \\
%G327.403+00.444 & 2.25 & 45.25 & 20.2 & 5 \\
%G331.132-00.245 & 3.30 & 98.25 & 29.7 & 2 \\
%G332.604-00.168 & 0.18 & 2.59 & 14.4 & 2 \\
%G338.927+00.632 & 1.17 & 9.66 & 8.3 & 4 \\
%G341.215-00.236 & 0.30 & 0.74 & 2.5 & 7 \\
%G343.520-00.519 & 0.20 & 0.99 & 5.0 & 5 \\
%G343.756-00.163 & 1.30 & 105.62 & 81.2 & 1 \\
%G344.101-00.661 & 0.12 & 0.51 & 4.3 & 3 \\
%G344.221-00.594 & 0.38 & 2.15 & 5.6 & 7 \\
%24.013+0.488 & 0.84 & 8.88 & 10.6 & 2 \\
%28.178-0.091 & 0.44 & 2.51 & 5.7 & 4 \\
%31.946+0.076 & 0.50 & 2.61 & 5.3 & 4 \\

HIGALBM327.3918+0.1996 & 1.20 & 7.13 & 6.0 & 9 \\
HIGALBM327.4022+0.4449 & 3.12 & 45.84 & 14.7 & 5 \\
HIGALBM331.1314-0.2438 & 4.74 & 104.49 & 22.0 & 2 \\
HIGALBM332.6045-0.1674 & 0.25 & 2.71 & 10.6 & 2 \\
HIGALBM338.9260+0.6340 & 1.33 & 9.48 & 7.1 & 4 \\
HIGALBM341.2149-0.2359 & 0.37 & 0.81 & 2.2 & 7 \\
HIGALBM343.5212-0.5172 & 0.16 & 0.71 & 4.3 & 5 \\
HIGALBM343.7560-0.1629 & 1.02 & 73.35 & 71.9 & 1 \\
HIGALBM344.1032-0.6609 & 0.11 & 0.42 & 3.8 & 3 \\
HIGALBM344.2210-0.5932 & 0.26 & 1.27 & 5.0 & 7 \\
HIGALBM24.0116+0.4897 & 1.51 & 8.95 & 5.9 & 2 \\
HIGALBM28.1957-0.0724 & 0.87 & 2.57 & 2.9 & 4 \\
HIGALBM31.9462+0.0759 & 0.98 & 2.61 & 2.7 & 4 \\

\hline
\end{tabular}
\caption{Mass completeness values for each source. \textit{Col 1}: Clump designation; \textit{Col2}: Mass completeness limit $M_{com}$ derived for each clump and based on the \rms\ of each map as described in Section \ref{sec:mass_completeness}; \textit{Col3:} Minimum mass $M_{min}$ identified in each clump in our ALMA Band6 maps; \textit{Col4}: $M_{com}$/$M_{min}$ ratio; \textit{Col5}: number of identified fragments in each clump.}
\label{tab:completeness_limits}
\end{center}
\end{table}

%%%%%%%%%%%%%%%%%%%%%%%%%%%%%%%%%%%%%%%%%%%%%%%%%%%%%%%%%%%%%%%%%%%%%%%%%%%%%%%%%%%%%%%

\section{Fragments properties}\label{sec:fragment_properties}
\begin{figure}
	\includegraphics[width=8.0cm]{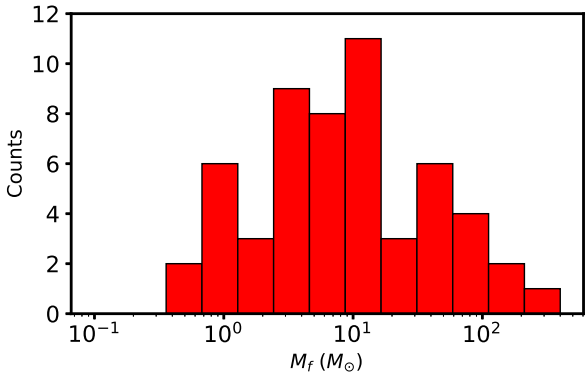}
    \caption{Mass distribution of the 55 fragments identified in our 13 SQUALO sources.} % The grey vertical lines are the mass completeness limits of each clump, as explained in the text.}
    \label{fig:mass_cores_histogram}
\end{figure}

In Figure \ref{fig:mass_cores_histogram} we show the distribution of the mass of the fragments in our clumps. The grey-dotted vertical lines represent the mass completeness of each clump. The fragment mass distribution covers more than two orders of magnitude, with fragments masses between 0.4 M\sun\ and 309 M\sun\ (see Appendix \ref{app:frag_properties}). 

In terms of evolution young, massive clumps are already fragmenting themselves, with the three 70\mum-quiet clumps, HIGALBM24.0116+0.4897, HIGALBM28.1957-0.0724 and HIGALBM31.9462+0.0759 breaking into 2, 4 and 4 fragments, respectively, with masses that ranges between $\simeq2.6$ and $\simeq32$ M\sun. These numbers are roughly in agreement with the results of ALMA surveys specifically designed to investigate the properties of 70\mum-quiet clumps \citep{Svoboda19,Sanhueza19}, which also have a spatial resolution comparable with the SQUALO survey ($\simeq0.015$ pc in \citet{Svoboda19} and $\simeq0.02$ pc on average in \citet{Sanhueza19}, compared to $\simeq 0.028$ pc on average in this work) and found the vast majority of their clumps already fragmented. Our results confirm that a certain degree of fragmentation is already present at the very early stages of formation, although some isolated examples of massive, dense and young objects have been observed in the literature, such as Core \#6 in W43-MM1, with 60 \Msun\ in $\simeq1300$ AU \citep{Nony18}.

In the more evolved regions we found the most massive objects, with 4 fragments with masses above 100 M\sun. These values are higher than those measured in recent ALMA surveys of evolved star-forming regions such as CORE \citep[][]{Beuther18} and ALMA-IMF \citep{Motte18, Pouteau22}, which reach masses up to $\simeq40$ M\sun\ \citep[in the CORE survey, ][]{Beuther18} and $\simeq70$ M\sun\ \citep[in W43 MM2\&MM3 in ALMA-IMF, ][]{Pouteau22}. However, a direct comparison with these projects is not straightforward: there are differences in the source extraction algorithms, the specifics of the ALMA observations and the estimation of the dust mass (due to different assumptions of the dust temperatures and of the parameters used in Equation \ref{eq:mass_estim}). In particular, the spatial resolution of those surveys is higher than that of SQUALO ($\simeq1000$ AU in CORE, \citet[][]{Beuther18}, and $\simeq2000$ AU in ALMA-IMF, \citet[][]{Motte22}), compared to $\simeq 4700$ AU on average of this work, see Table \ref{tab:alma_setup}). Both these works have identified objects smaller than the ones found in our survey.

The impact of the spatial resolution on the identification of the smallest fragments is also evident from the results of the ALMA survey of massive progenitors in ATLASGAL \citep[][]{Csengeri17}. These authors have looked at the fragmentation properties in a sample of 35 objects observed with a resolution $\geq0.06$ pc, $\simeq$a factor of 2 lower than the resolution of SQUALO. At these scales \citet{Csengeri17} found well resolved fragments with masses up to 400 \Msun, suggesting that there are several intermediate steps in the hierarchical process. In several clumps these authors also found evidence of a single, monolithic fragment, which is likely to separate in smaller entities if observed at higher resolution, as observed in this work and in other surveys \citep[e.g.][]{Beuther18,Svoboda19,Sanhueza19}. In fact, even some of our objects may further fragment into smaller objects, and this could be particularly true for those fragments that have been observed in the more evolved regions. This possibility will be further discussed in Section \ref{sec:fragment_clump_props}.

% As already mentioned in Section \ref{sec:clumps_cores_analysis}, 

%%%%%%%%%%%%%%%%%%%%%%%%%%%%%%%%%%%%%%%%%%%%%%%%%%%%%%%%%%%%%%%%%%%%%%%%%%%%%%%%%%
\subsection{A peculiar source: HIGALBM343.7560-0.1629}\label{sec:peculiar_G343}
Among the evolved clumps we also find our single example of an object that does not fragment, HIGALBM343.7560-0.1629 ($L_{cl}/M_{cl}=14.7$ \LM). It contains one of the most massive object we identified, with $M\simeq73$ M\sun. This source is one of the massive protoclusters observed with ALMA at 1.3mm in the work of \citet{Csengeri17}, but with a spatial resolution of $\simeq6840$ AU (re-scaled at the distance adopted in this work for HIGALBM343.7560-0.1629, 2.0 kpc, see Table \ref{tab:completeness_limits}) compared to the $\simeq2130$ AU of our data in the work. It is a good example of the hierarchical process that connects clumps to actual cores, with "intermediate" objects such as the ones observed in the \citet{Csengeri17} sample and, at three times higher resolution, at the SQUALO resolution. In the work of \citet{Csengeri17} they identified two sources in the field, MM1 (corresponding to our fragment) and MM2, with masses of 115.5\Msun\ and 20.2 \Msun\ respectively. Our new data however suggest that our single fragment in this clump is likely truly single at the scales of our observations, and source MM2 is actually part of the diffuse emission associated with the main MM1 source (see their Figure A1 in comparison with our Figure \ref{fig:alma_cores}). If a second fragment would be present, it should be less massive than the mass completeness limit discussed in Section \ref{sec:mass_completeness}, which for this source is $\simeq1.0 $\Msun\ (Table \ref{tab:completeness_limits}). Therefore, if there are other fragments at the SQUALO scales those have to be at least 20 times less massive than the value estimated for MM2, and 70 times less massive than our identified fragment. However HIGALBM327.3918+0.1996, the most distant protostellar clump in our sample, is located $\simeq5.2$ kpc away from us and it shows several fragments (with masses in the range $9\leq M\leq 45$ \Msun). With $L_{cl}/M_{cl}\simeq12.6$ \LM, it is in terms of evolution comparable to HIGALBM343.7560-0.1629. At the same time all our 70\mum-quiet objects are located further away than HIGALBM343.7560-0.1629 and they all already show some degree of fragmentation. 
The behaviour of this source is therefore peculiar with respect to the rest of the sample. It is possible that some particular mechanism is in place in HIGALBM343.7560-0.1629 that allows the condensation of the different fragments observed in 70\mum-quiet objects into a single object, for example a significant role of magnetic fields, as suggested by e.g. \citet[ in][]{Fontani16} or \citet{Palau21}. A more detailed investigation of this peculiar source and its properties is beyond the scope of this work.

\subsection{Jeans analysis}\label{sec:Jeans_analysis}
We start to investigate the driver of the fragmentation at the scales probed by our observations by looking at the Jeans values of our objects. If the fragmentation is thermally driven, we would expect the separation between fragments and the mass of the fragments to be compatible with the Jeans length $\lambda_{J,cl}$ and Jeans mass $M_{J,cl}$ of the clumps respectively, defined as:

\begin{equation}
    \lambda_{J,cl}=\sigma_{th}\bigg(\frac{\pi}{G\rho_{cl}}\bigg)^{1/2} \quad ,
\end{equation}

\noindent
and

\begin{equation}
    M_{J,cl}=\frac{4\pi\rho_{cl}}{3}\bigg(\frac{\lambda_{J,cl}}{2}\bigg)^{3} \quad ,
\end{equation}

\noindent
where $\rho_{cl}$ is the average density of the host clumps (assumed as perfect spheres and derived from mass $M_{cl}$  and radius $R_{cl}$ of the clumps, Table \ref{tab:tab_prop}), and $\sigma_{th}$ is their thermal velocity dispersion. The value of $\sigma_{th}$ is evaluated as $\sigma_{th}=[k_{B}*T_{cl}/(\mu m_{H}]^{1/2}$ with $T_{cl}$ the clump temperature and $\mu=2.33$, assuming that the thermal velocity dispersion is dominated by a gas mixture of H$_{2}$ and He.

To define a value for the separation between the fragments in each clump we measured $d_{min,2D}$ (where the 2D underlines that this value is not corrected for any projection effect), the minimum distance between each fragment and its closest neighbourhoods, using a minimum spanning tree algorithm \citep[MST,][]{Barrow85}. We then introduced the a-dimensional quantity $\lambda_{J_{r},2D}$, defined as the ratio of $d_{min,2D}$ to the thermal Jeans length of the clump:

\begin{equation}
    \lambda_{J_{r},2D}=d_{min,2D}/\lambda_{J,cl}
\end{equation}

In order to account for the distribution of the positions of the fragments along the line of sight (LOS) we perform a set of Monte Carlo simulations, similar to the procedure described in \citet{Svoboda19}. In practice, for a given pair of fragments we extracted a random distance along the LOS from a Gaussian distribution centered on zero and with a variance dispersion equal to one third of the clump radius at 160\mum\ (therefore, there is a 3$\sigma$ possibility that the maximum distance of the two fragments is equal to the 160\mum\ clump size). For each pair of fragments we made 10000 realizations of their relative distances along the LOS and we have taken the mean value of the derived 3D-distances as the representative one. We considered the minimum fragment separation for each clump ($d_{min,3D}$) and we refer to the ratio of this value to the thermal Jeans length as $\lambda_{J_{r},3D}$. Taking into account the projection effects, $\lambda_{J_{r},3D}$ varies in the range $1.06\leq\lambda_{J_{r},3D}\leq7.04$, not consistent on average with pure thermal fragmentation, suggesting that some support against the gravitational collapse at these scales is also likely in these objects.

If we introduce the non-thermal support against the fragmentation in the calculations, and substitute $\sigma_{th}$ with the clumps non-thermal velocity dispersion $\sigma_{nth}$ to estimate the Jeans length, we obtain values in the range $0.28\leq\lambda_{J_{r,nth},3D}\leq1.47$. In particular, except for the source HIGALBM332.6045-0.1674, for which this value is $\lambda_{J_{r,nth},3D}=1.47$, for all the other sources $\lambda_{J_{r,nth},3D}<1$, smaller than "pure" non-thermal fragmentation values, which would give $\lambda_{J_{r,nth},3D}=1$. The Jeans parameters related to the analysis of the Jeans length for each clump are given in Table \ref{tab:Jeans_params}.

Similarly to what we have done for the Jeans length analysis, the comparison of the Jeans mass $M_{J,cl}$ with the value of the most massive fragment $M_{f,max}$ in each clump, shows that for all these clumps the value $M_{J_{r}}=M_{max}/M_{J,cl}>1$, and in some cases $M_{J_{r}}>>1$. If we substitute $\sigma_{th}$ with the clumps non-thermal velocity dispersion $\sigma_{nth}$ to estimate the Jeans mass, we obtain values $0.01\leq M_{J_{r,nth}}\leq0.93$, with the exception of HIGALBM343.7560-0.1629 and HIGALBM331.1314-0.2438 that have $M_{J_{r,nth}}\simeq1.7$, supporting the hypothesis that some mechanism is in place in these sources to support the fragmentation of the clump. All the parameters related to the analysis of the Jeans mass properties of our objects are given in Table \ref{tab:Jeans_params_mass}.

These results suggest that the fragmentation properties observed at the resolution probed by our observations could partially be supported by the turbulent cascade, but part of the observed non-thermal motions could also originate from a different mechanism (such as the gravitational collapse). We will further discuss these findings in the next section.

% This result indicates that a first fragmentation step of the hierarchical process is driven by non-thermal mechanisms, such as turbulence and gravity. 

\begin{table*}
\begin{tabular}{ccccccccccc}
 \hline
DESIGNATION & d$_{min,2D}$ & d$_{min,3D}$ & $\sigma_{th}$ & $\lambda_{J,cl}$ & $\lambda_{J_{r},2D}$ & $\lambda_{J_{r},3D}$ & $\sigma_{nth}$ & $\lambda_{J,nth}$ & $\lambda_{J_{r},nth,2D}$ & $\lambda_{J_{r},nth,3D}$ \\
 & ($\mathrm{pc}$) & ($\mathrm{pc}$) & ($\mathrm{km\,s^{-1}}$) & ($\mathrm{pc}$) &  &  & ($\mathrm{km\,s^{-1}}$) & ($\mathrm{pc}$) &  &  \\
 \hline
 \hline
HIGALBM327.3918+0.19 & 0.030 & 0.077 & 0.28 & 0.025 & 1.23 & 3.13 & 1.44 & 0.124 & 0.24 & 0.62 \\
HIGALBM327.4022+0.44 & 0.030 & 0.075 & 0.32 & 0.025 & 1.22 & 3.05 & 1.66 & 0.127 & 0.24 & 0.59 \\
HIGALBM331.1314-0.24 & 0.037 & 0.081 & 0.33 & 0.018 & 2.02 & 4.43 & 2.04 & 0.113 & 0.33 & 0.72 \\
HIGALBM332.6045-0.16 & 0.136 & 0.142 & 0.26 & 0.020 & 6.73 & 7.04 & 1.26 & 0.097 & 1.41 & 1.47 \\
HIGALBM338.9260+0.63 & 0.073 & 0.096 & 0.26 & 0.023 & 3.17 & 4.15 & 1.66 & 0.147 & 0.50 & 0.65 \\
HIGALBM341.2149-0.23 & 0.021 & 0.065 & 0.27 & 0.045 & 0.46 & 1.43 & 1.09 & 0.181 & 0.12 & 0.36 \\
HIGALBM343.5212-0.51 & 0.025 & 0.051 & 0.28 & 0.025 & 0.99 & 1.98 & 1.25 & 0.116 & 0.22 & 0.44 \\
HIGALBM343.7560-0.16 & 0.000 & 0.000 & 0.29 & 0.018 & 0.00 & 0.00 & 1.22 & 0.076 & 0.00 & 0.00 \\
HIGALBM344.1032-0.66 & 0.069 & 0.084 & 0.28 & 0.028 & 2.51 & 3.04 & 1.64 & 0.164 & 0.42 & 0.51 \\
HIGALBM344.2210-0.59 & 0.014 & 0.035 & 0.35 & 0.033 & 0.41 & 1.06 & 1.33 & 0.128 & 0.11 & 0.28 \\
HIGALBM24.0116+0.489 & 0.083 & 0.115 & 0.20 & 0.029 & 2.84 & 3.94 & 0.91 & 0.137 & 0.61 & 0.84 \\
HIGALBM28.1957-0.072 & 0.127 & 0.182 & 0.21 & 0.056 & 2.27 & 3.25 & 1.07 & 0.281 & 0.45 & 0.65 \\
HIGALBM31.9462+0.075 & 0.071 & 0.105 & 0.19 & 0.024 & 2.97 & 4.39 & 1.19 & 0.149 & 0.48 & 0.70 \\
\hline
\end{tabular}
\caption{Parameters derived from the length Jeans analysis described in Section \ref{sec:fragment_properties}. \textit{Col.1}: Clump designation; \textit{Cols.2-3}: minimum distance between fragments derived with the MST algorithm, uncorrected and corrected for the 3D projection effects along the LOS, respectively; \textit{Col.4}: sound speed derived for each clump; \textit{Col.5}: clump thermal Jeans length; \textit{Cols.6-7}: ratio between the measured Jeans length in each clump and the thermal Jeans length, uncorrected and corrected for the 3D projection effects along the LOS, respectively; \textit{Col. 8}: clump non-thermal sound speed; \textit{Cols. 9-11}: same as Cols. 5-7 but assuming non-thermal velocity dispersion (i.e. substituting $\sigma_{th}$ with $\sigma_{nth}$).}
\label{tab:Jeans_params}
\end{table*}

\begin{table*}
\begin{tabular}{cccccc}
\hline
DESIGNATION & $M_{f,Max}$ & $M_{J,cl}$ & $M_{J_{r}}$ & $M_{J,nth}$ & $M_{J_{r},nth}$ \\
 & ($\mathrm{M_{\odot}}$) & ($\mathrm{M_{\odot}}$) &  & ($\mathrm{M_{\odot}}$) &  \\
 \hline
 \hline
HIGALBM327.3918+0.19 & 44.9 & 0.76 & 58.83 & 98.60 & 0.455 \\
HIGALBM327.4022+0.44 & 124.7 & 0.99 & 125.61 & 133.72 & 0.932 \\
HIGALBM331.1314-0.24 & 309.3 & 0.78 & 396.28 & 179.44 & 1.724 \\
HIGALBM332.6045-0.16 & 18.0 & 0.54 & 33.60 & 58.68 & 0.307 \\
HIGALBM338.9260+0.63 & 68.9 & 0.60 & 113.76 & 154.89 & 0.445 \\
HIGALBM341.2149-0.23 & 10.3 & 1.30 & 7.92 & 82.08 & 0.126 \\
HIGALBM343.5212-0.51 & 2.1 & 0.73 & 2.90 & 69.20 & 0.031 \\
HIGALBM343.7560-0.16 & 73.3 & 0.59 & 125.11 & 43.48 & 1.687 \\
HIGALBM344.1032-0.66 & 2.4 & 0.80 & 3.04 & 168.27 & 0.014 \\
HIGALBM344.2210-0.59 & 9.2 & 1.54 & 5.96 & 86.29 & 0.106 \\
HIGALBM24.0116+0.489 & 31.9 & 0.42 & 75.00 & 43.38 & 0.734 \\
HIGALBM28.1957-0.072 & 7.7 & 0.97 & 7.94 & 123.25 & 0.063 \\
HIGALBM31.9462+0.075 & 25.9 & 0.33 & 77.53 & 80.76 & 0.321 \\
\hline
\end{tabular}
\caption{Parameters derived from the mass Jeans analysis described in Section \ref{sec:fragment_properties}. \textit{Col.1}: Clump designation; \textit{Col.2}: mass of the most massive fragment in each clump; \textit{Col.3}: clump Jeans mass; \textit{Col.4}: ratio between the Jeans mass and the mass of the most massive fragment in each clump. \textit{Cols. 5-6}: same as Cols. 3-4 but assuming non-thermal velocity dispersion (i.e. substituting $\sigma_{th}$ with $\sigma_{nth}$).}
\label{tab:Jeans_params_mass}
\end{table*}

% as expected in more dynamic star formation mechanisms \citep[e.g.][]{Bonnell06}.

%, not simply considering the clump dust temperature similar to what has been done in other ALMA works \citep[][]{Csengeri17,Svoboda19,Sanhueza19,Zhang20}. 

%%%%%%%%%%%%%%%%%%%%%%%%%%%%%%%%%%%%%%%%%%%%%%%%%%%%%%%%%%%%%%%%%%%%%%%%%%%%%%%%%%%%%%%
\section{Fragments vs. clump properties}\label{sec:fragment_clump_props}
In this section we investigate the possible correlations between the properties of the clumps and of their fragments as a function of the clump evolution. Also, whenever possible, we combine our datasets with ancillary data taken from the literature to put our results in a broader context. 

Our sample of clumps is not large enough to allow the use of statistical tools such as the principal component analysis. Instead, we calculated the Pearson's correlation coefficient $\rho$ for each pair of quantities, which estimates the level of linear correlation between our data (from $\rho=-1$ and $\rho=1$, which express total anti-correlation and total correlation, respectively), using the Python package \texttt{Pymccorrelation} \citep{Privon20}. We considered the following parameters for the clumps: mass $M_{cl}$, $L/M$ ratio $L_{cl}/M_{cl}$, surface density $\Sigma_{cl}$, mass accretion rate $\dot{M}_{cl}$ and virial parameter $\alpha_{vir,cl}$; the following ones for the fragments: the number of fragments in each clump $\#_{f}$, the minimum 2D (3D) distances between fragments, $d_{min,2D}$ ($d_{min,3D}$), the total mass of fragments in each clump $M_{f}$, the mass and surface density of the most massive and of the densest fragment in each clump, M$_{f,max}$ and $\Sigma_{f,max}$, respectively. We also considered the following parameters that combine the clump and fragment properties: the instantaneous clump formation efficiency (CFE), defined as the ratio between the total mass of the fragments in each clump and the mass of the clump \citep[e.g. ][]{Anderson21}, and the ratio between the clump 2D Jeans length (3D corrected) and the minimum distance between fragments in each clump, $\lambda_{J_{r},2D}$ ($\lambda_{J_{r},3D}$).

In Figure in Appendix ~\ref{app:pearson_heatmap} we show the correlation matrix (or heatmap) of all these quantities, colour-coded with the value of the Pearson's correlation parameter (reported in each cell). In Appendix \ref{app:pearson_params} we report the values of $\rho$ for all the pairs we considered with the probability that the null hypothesis is true (the p-value) and the 95\% confidence interval (CI 95\%) for each pair of parameters.

Before going into the analysis of the results presented in the heatmap in Appendix \ref{app:pearson_heatmap}, we want to caution that "correlation is not causation", in particular for our relatively small dataset. A high (or low) value of the Pearson coefficient in a given pair of parameters may not be due to some physical process and should be taken "cum grano salis". In particular, the estimation of the uncertainties plays a role to determine the reliability of each Pearson's coefficient (and the corresponding estimation of the p-value). For all the parameters analysed in this work we considered an uncertainties equal to 20\% of the measured value. In most cases this is a conservative estimation, for example the clump properties such as the mass have uncertainties quoted in the \citet{Elia21} catalogue below $10\%$ for 9 out of 13 of our sources. For three sources, HIGALBM341.2149-0.2359, HIGALBM24.0116+0.4897 and HIGALBM28.1957-0.0724 the quoted uncertainties are $\simeq86\%$, $\simeq40\%$ and $\simeq40\%$ respectively. The quoted mass uncertainty for source HIGALBM31.9462+0.0759 in the Hi-GAL catalogue is unrealistically higher because no distance were assigned to the source. Note however that the source mass estimated for this source in \citet{Traficante17} differs by only $3\%$ from the mass estimation in \citet{Elia21}. 

Similarly, the uncertainties on the source distances, which will affect both clumps and fragments properties, are $\leq15\%$ for 10 sources and equal to $20\%$ for the other three \citep[HIGALBM343.7560-0.1629, HIGALBM344.1032-0.6609 and HIGALBM344.2210-0.5932][]{Mege21}.

Even accounting for these uncertainties and limitations, there are some general considerations that can be drawn from the correlation matrix shown in the heatmap in Appendix~\ref{app:pearson_heatmap}, in particular when drawn by comparing the correlation coefficients of multiple pairs of parameters, or when including results from other surveys from the literature. These conclusions will eventually be corroborated by further investigation on larger samples of massive clumps undergoing global collapse (as the one that could be provided by the ALMAGAL survey). %In particular, the main goal of this work is to investigate the correlation between the clump and the fragment properties as function of the clump evolution.

%%%%%%%%%%%%%%%%%%%%%%%%%%%%%%%%%%%%%%%%%%%%%%%%%%%%%%%%%%%%%%%%%%%%%%%%%%%%%%%%%%%%%%%
\subsection{Fragments number and relative distance vs. clump properties}\label{sec:fragments_numbers_location_clump_props}

In this section we will describe the results derived from the properties of the fragments that are independent from their dust temperature. % inferred from Equation \ref{eq:mass_estim}.

\begin{figure}
	\includegraphics[width=8.0cm]{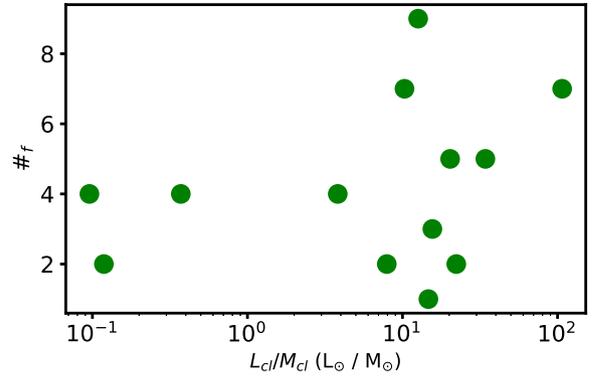}
    \caption{Clumps $L/M$, indicator of their evolutionary phases, as a function of the number of fragments identified in each clump.}
    \label{fig:L_M_n_cores}
\end{figure}

The first pair of parameters that we consider is the degree of fragmentation as a function of the evolutionary phase of these clumps (defined by $L_{cl}/M_{cl}$), shown in Figure~\ref{fig:L_M_n_cores}. Overall, these two quantities are mildly correlated ($\rho=0.37$ with a \textit{p-value} of 0.22). This diagram suggests that the degree of fragmentation at these scales is not simply regulated by the evolutionary phase of each object: we found sources with $L_{cl}/M_{cl}>10$ that embed a single fragment (HIGALBM343.7560-0.1629, Section \ref{sec:peculiar_G343}), sources that embed 2 or more fragments and up to 9 fragments in HIGALBM327.3918+0.1996, the most fragmented clump of our sample. However, one interesting result to note is that, as already discussed in Section \ref{sec:fragment_properties}, at very early stages of formation the systems are already fragmented.

%As already noticed in Section \ref{sec:fragment_properties}, the clump G343.756-00.163, with $L_{cl}/M_{cl}\simeq20$ \LM, shows a single monolithic fragment at our resolution and sensitivity. %During the evolution some mechanism, which can be related to the local environment of each clump and the interplay between the various forces in place, could lead them to different paths.  

\begin{figure}
	\includegraphics[width=8.0cm]{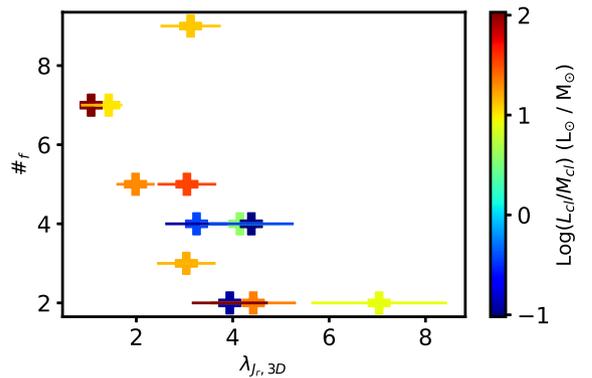}
    \caption{Ratio of the minimum fragment separation and the thermal Jeans length, corrected for the 3D projection effects, $\lambda_{J_{r},3D}$, as a function of the number of fragments in each clump. The distribution is colour-coded for the clumps $L/M$.}
    \label{fig:Jeans_n_cores}
\end{figure}

Figure \ref{fig:Jeans_n_cores} shows the correlation between the number of fragments and the Jeans ratio $\lambda_{J_{r},3D}$, colour-coded for the clump evolution. These parameters present the highest degree of (anti)correlation with the number of fragments, $\#_{f}$, with $\rho=-0.67$, and this (anti-)correlation is more solid then the previous one, with \textit{p-value}=0.02 (assuming an uncertainties of $20\%$ on the estimation of the Jeans ratio, as discussed in the previous Section).

%%% HERE !!! DISCUSSION ABOUT UNCERTAINTIES (AND AS WELL IN THE PRESENTATION OF THE HEATMAP) %%%

This plot suggests that the number of fragments is lower at high values of $\lambda_{J_{r},3D}$ and it increases as the system approaches the thermal Jeans scale, $\lambda_{J_{r},3D}=1$.

\begin{figure}
	\includegraphics[width=8.0cm]{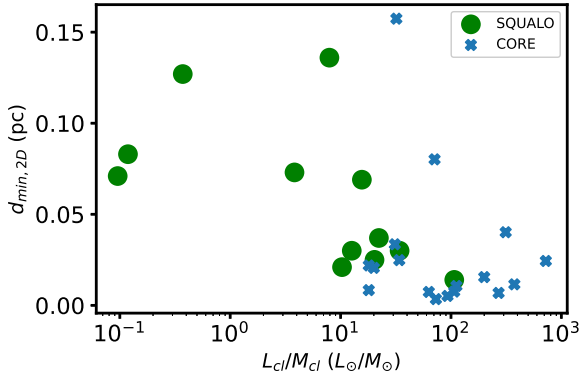}
    \caption{Clump $L/M$, indicator of their evolutionary phases, as a function of the minimum 2D distance between fragments identified in each clump. The green circles is the SQUALO sample, and the blue crosses is the CORE sample taken from \citet{Beuther18}.}
    \label{fig:min_2D_Distance}
\end{figure}

This last result could be a consequence of the number of fragments observed in each clump: it is expected that the higher the number of fragments in each object, the shorter the minimum distance between them (and therefore the smaller $\lambda_{J_{r},3D}$ for clumps of comparable thermal Jeans length, see Table \ref{tab:Jeans_params}). However, this result is more instructive if we put it in an evolutionary context: as showed in Figure \ref{fig:Jeans_n_cores}, the clumps with the lower values of $\lambda_{J_{r},3D}$ are among the most evolved of our sample (with the exception of HIGALBM331.1314-0.2438, which embeds the most massive fragment we have identified). This trend is further confirmed by the plot in Figure \ref{fig:min_2D_Distance}, where we report $d_{min,2D}$ as a function of the $L/M$ of the parent clumps. It is remarkable the lack of points in the bottom-left corner of the plot, i.e. the lack of points with low $L_{cl}/M_{cl}$ and low $d_{min,2D}$. The value of $d_{min,2D}$ for the points with $L_{cl}/M_{cl}<10$ is always above the resolution limit of our survey ($\simeq0.01-0.04$ pc, see Table \ref{tab:alma_setup}). These results suggest that 1) each of the fragments in the youngest clumps may have already produced multiple cores very close to each other, below our resolution limits, or 2) these fragments will get closer with time (and eventually merge or keep fragmenting). Here we have considered the relation between $L_{cl}/M_{cl}$ and $d_{min,2D}$ since we report for comparison also the objects studied in the CORE survey \citep{Beuther18}. The CORE survey targeted evolved regions in the northern sky with NOEMA to resolve down to $\simeq1000$ AU resolution objects up to 6 kpc away from us \citep{Beuther18}, $\simeq4-5$ times better than our resolution. Since a few of their fragments have $d_{min,2D, core}$ smaller than our resolution, it is possible that some of our fragments will further separate down to $\simeq1000$ AU resolution. In particular, this could be true for the more evolved objects: all CORE clumps in fact have $L/M>10$, but note that in their survey all but one of the 8 fragments with $d_{min,2D, core}<2500$ AU have $L/M>70$ \LM, which corroborates the hypothesis that the cores in the most evolved objects are on average closer to each other. %These results altogether are compatible with a scenario where clumps first fragment in (many) different objects.

%  (TO BE ADDED???) in agreement with the lower degree of fragmentation seen in \citet{Csengeri17} at the lower level of the hierarchical process. 

% the clumps with the higher values of $\lambda_{J_{r},3D}$ are either the 70\mum-quiet clumps or the less evolved clumps among the 70\mum\ bright ones (with the exepction of G331.132-00.245, which embeds the most massive fragment in our sample). 

All these results can be interpreted to favour a model where sources evolve in a highly dynamical \textit{clump-fed}, gravo-turbulent scenario, in which the high values of $\lambda_{J_{r},3D}$ (and low values of $d_{min,2D}$)  observed in young clumps may just be a step in the hierarchical evolution. The process starts with the formation of a few non-thermally supported fragments in which there is likely an interplay between gravity and turbulence. With time, the gravitational contraction takes over and the fragments get closer, merge and/or further fragment into more thermally supported objects. In sources such as HIGALBM343.7560-0.1629 and others with few ($\leq3)$ fragments other factors such as magnetic fields may play a more relevant role than in other clumps \citep[][]{Palau21}, leading to a magnetically-regulated collapse that supports, or suppresses the fragmentation \citep{Fontani16}. These hypotheses could be tested against simulations of massive star-forming clumps such as the ones produced by \citet{Lee16a}. % A dedicated set of simulations aimed to explore the fragmentation properties of massive clumps are part of a subsequent paper (Lebrouilly et al., $in prep.$).

%, the latter that partially inhibits or slows down the collapse (as e.g. in the inertial inflow model \citep{Padoan20}), while the former that contribute to the global, parsec-scale collapse, as in the \citet{Ballesteros-Paredes11} and \citet{Vazquez-Semadeni19} models

% ($\rho=-0.52)$, showed in Figure \ref{fig:minimum_distance_c}.  

% These are our first results that we interpret in favour of a model where sources evolve in a highly dynamical \textit{clump-fed}, gravo-turbulent scenario, in which the high values of $\lambda_{J_{r},3D}$ observed in young clumps may just be a step in the hierarchical evolution. The process seems to start with the formation of few fragments in which there is an interplay between gravity and turbulence, the latter that partially inhibits or slows down the collapse, as in the inertial inflow model \citep{Padoan20}, while the former that contribute to the global, parsec-scale collapse, as predicted in \citet{Ballesteros-Paredes11} and \citet{Vazquez-Semadeni19} models. With time, in most of them the gravitational contraction takes over and the clumps continue to fragment and rearrange themselves in more concentrated, more thermally supported fragments. In sources such as G343.756-00.163 and others with few ($\leq3)$ fragments other factors such as magnetic fields may play a role more relevant than in other clumps \citep[][]{Palau21}, leading to a magnetically-regulated collapse which support, or suppress the fragmentation \citep{Fontani16}.

%%%%%%%%%%%%%%%%%%%%%%%%%%%%%%%%%%%%%%%%%%%%%%%%%%%%%%%%%%%%%%%%%%%%%%%%%%%%%%%%%%%%%%%
\subsection{Fragments physical properties vs. clump properties}\label{sec:fragments_temperature_dependent_clump_props}
In this section we will discuss the results derived from the properties of the fragments that are dependent on their dust temperature, such as their mass and surface density. 

In a \textit{clump-fed} scenario the accretion rate from clump to core is $\dot{m}_{cl}\neq 0$ \citep[][]{Peretto20}. The opposite, or at least no direct correlation between core accretion and clump accretion is expected in a \textit{core-fed} scenario. A consequence of the \textit{clump-fed} model is therefore a correlation between the core/fragment mass and the clump mass, which are expected to be dynamically correlated \citep{Anderson21}. Our sample has been chosen to show evidence of clump-scale accretion rate from single-dish surveys, therefore we expected to observe such correlation. As shown in the heatmap in Appendix \ref{app:pearson_heatmap}, we indeed measure a tight correlation between the mass and density of the clumps and the mass and density of the fragments: clumps with the highest surface density also form the most massive fragments and with the highest surface density. 

To put these results more in context, we have compared our sample with three others taken from the literature: the 70\mum\ quiet objects observed in \citet{Sanhueza19} and \citet{Svoboda19} ALMA surveys, plus the 6 hub-filament systems in infrared-dark clouds examined in \citet{Anderson21}. The physical properties of the \citet{Sanhueza19} and \citet{Anderson21} surveys have been extracted from their works. For the \citet{Svoboda19} sample the physical properties of the fragments have been derived from their fluxes, which have been converted into mass using Equation \ref{eq:mass_estim} and the same parameters considered for our 70\mum-quiet objects. We initially tried to re-evaluate the clump properties of these samples from the \citet[][]{Elia21} catalogue, for a consistent comparison between all samples. However, not all the clumps where identified in the \citet[][]{Elia21} catalogue and eventually with a non well-defined distances. Therefore, we decided to consider the clump parameters discussed in the literature, with the note of caution that the comparisons between different catalogues may suffer from biases introduced by the various extraction algorithms and approaches used to derive the physical parameters in the various works \citep[see e.g. the results in][]{Kauffmann10}.

%, and the evolutionary trend is evident only is the single surveys in which it is possible to observe such trend (namely, our SQUALO data and the \citet{Anderson21} sample).

\begin{figure}
	\includegraphics[width=8.0cm]{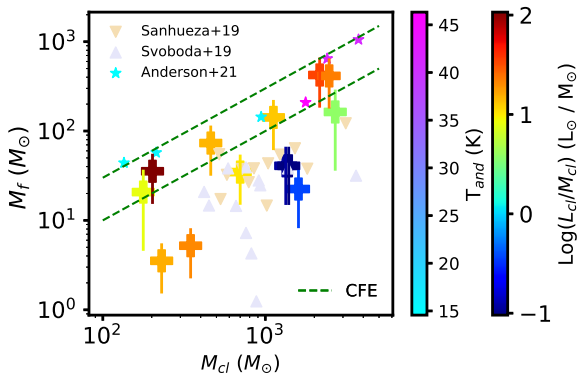}
    \caption{Total mass of the fragments as a function of the mass of the parent clump. The distribution of points from this work is colour-coded for the clumps $L_{cl}/M_{cl}$ and the mass uncertainties derive from the temperature uncertainties as described in Section \ref{sec:temperature_uncertainties}. We also plot the properties of the samples of 70\mum-quiet objects from \citet[][light brown triangles]{Sanhueza19} and \citet[][light blue triangles]{Svoboda19}, and of massive clumps in hub-filament systems taken from \citet[][colour coded for the temperature of the cores]{Anderson21}, as discussed in the main text. The green dashed lines are the CFE with efficiencies of 10\% (bottom line) and 30\% (top line).}
    \label{fig:Mass_clump_mass_cores}
\end{figure}

The relation between mass of the clumps and mass of inner fragments for all the surveys is showed in Figure \ref{fig:Mass_clump_mass_cores}. The error bars in the SQUALO sample derive from the temperature uncertainties as described in Section \ref{sec:temperature_uncertainties}. The filament-hub system in \citet{Anderson21} is colour-coded according to the temperature of the cores as estimated in \citet{Anderson21}, that we assume as evolutionary indicator of their sample. 

Also combining different data sets there is a general indication that these quantities are correlated. As consequence of this plot, we observe a good degree of correlation between the CFE and mass of the fragments ($\rho=0.65$, p-value=0.016, Appendix \ref{app:pearson_heatmap}): the clumps with highest instantaneous formation efficiency embed the most massive fragments in our survey (see Figure \ref{fig:Mass_clump_mass_cores}), in agreement with the results (at slightly larger scales) of \citet[][]{Csengeri17}.

\begin{figure}
	\includegraphics[width=8.0cm]{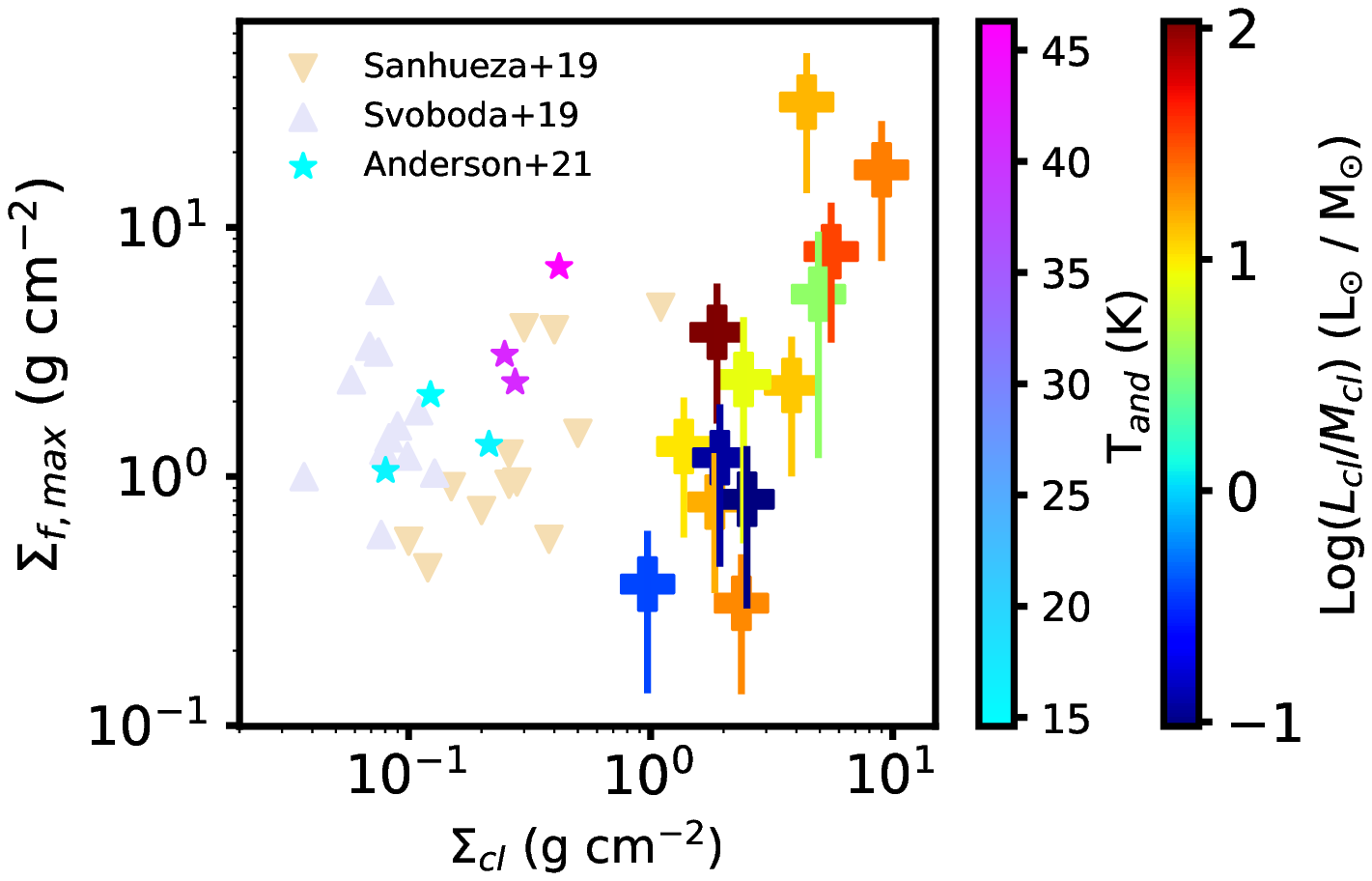}
    \caption{Clump surface density as a function of the surface density of the densest fragment in each clump. The distribution of points from this work is colour-coded for the clumps $L_{cl}/M_{cl}$. We also plot the properties of the samples of \citet[][light brown triangles]{Sanhueza19}, \citet[][light blue triangles]{Svoboda19}, and \citet[][colour coded for the temperature of the cores]{Anderson21}, as discussed in the main text.} 
    \label{fig:Sigma_clump_Sigma_core_max}
\end{figure}

Similarly, in Figure \ref{fig:Sigma_clump_Sigma_core_max} we show the relation between the clump surface density and the surface density of the densest fragment for all surveys. In particular, our new data and the survey of \citet{Anderson21}, the only two samples in this plot with objects spanning different evolutionary phases, show an intrinsic good correlation of these two quantities (Pearson coefficient $\rho=0.60$ for our survey, Appendix \ref{app:pearson_heatmap}). In both these surveys the youngest objects have among the lowest values of $\Sigma_{cl}$ and $\Sigma_{f, max}$. It is worth noting that the densest clumps may form the more massive fragments in a shortest amount of time, so, strictly speaking, the 70\mum-quiet objects and those with high $L_{cl}/M_{cl}$ and higher $\Sigma_{cl}$ may be at a similar age. However, statistically speaking young clumps have, on average, lower column density than more evolved clumps, a result interpreted as a result of evolution at clump scales \citep{Svoboda16, Elia21}. Our results and the \citet{Anderson21} survey suggest a dynamic scenario where both the 70\mum-quiet clumps and their fragments increase their density as a function of evolution. In this view, the 3 SQUALO protostellar clumps with $L_{cl}/M_{cl}>1$ that overlaps in this plot with our 70 \mum-quiet objects could originate from clumps less dense than the ones observed in the SQUALO, \citet{Svoboda19} and \citet{Sanhueza19} surveys.

% \textbf{This result could suggest a dynamic scenario where both the 70\mum-quiet clumps and their fragments increase their density with the evolution. In this scenario, the 3 SQUALO protostellar clumps with $L_{cl}/M_{cl}>1$ that overlaps in this plot our 70 \mum-quiet objects could originate from clumps less dense than the ones observed in the SQUALO, \citet{Svoboda19} and \citet{Sanhueza19} surveys. At the same time, the densest clumps form the more massive fragments in a shortest amount of time so, strictly speaking, the 70\mum-quiet objects and those with high $L_{cl}/M_{cl}>1$ may be at a similar age. But the SQUALO and the \citet{Anderson21} surveys, the only ones for which we have an indication of different evolutionary phases in the samples, are also the only ones that show some degree of correlation in this plot. We interpret this diagram as an indication that young, 70\mum-quiet objects evolve by increasing the clump density and, correspondingly, the most massive fragments.

% that show some degree of correlation, are also the only surveys in this plot for which we have an indication of evolution and in this, the correlation in this plot is also observed as an evolutionary trend. and this suggests a dynamic scenario where both the 70\mum-quiet clumps and their fragments increase their density with the evolution. In this scenario, the 3 SQUALO protostellar clumps with $L_{cl}/M_{cl}>1$ that overlaps in this plot our 70 \mum-quiet objects could originate from clumps less dense than the ones observed in the SQUALO, \citet{Svoboda19} and \citet{Sanhueza19} surveys. 

\begin{figure}
	\includegraphics[width=8.0cm]{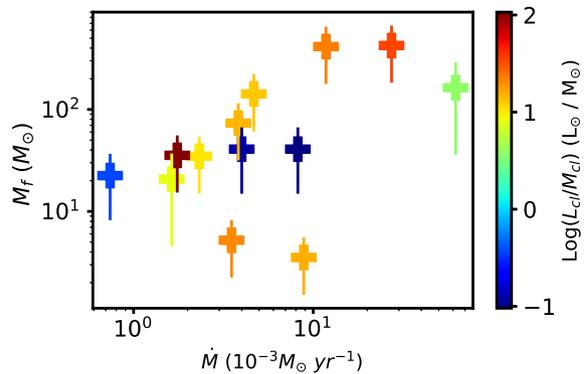}
    \caption{Total mass of the fragments as a function of the clump accretion rate $\dot{M}$ in each clump. The distribution is colour-coded for the clumps $L_{cl}/M_{cl}$.}
    \label{fig:Accr_rate_fragment_mass}
\end{figure}

Interestingly, the accretion rate observed at the clump scales shows only a mild correlation with the properties of the fragments. We observe an indication that higher accretion rates at clump scales feed more mass into the inner fragments ($\rho=0.44$ and p-value=0.14, see also Figure \ref{fig:Accr_rate_fragment_mass}). It is also independent on the number of fragments and on the evolutionary stage of the clump (Appendix \ref{app:pearson_heatmap}). The large-scale accretion seems to feed either one or more objects independently from the rate of accretion (and the clump evolution). It is also possible that the parsec-scale accretion is feeding the intra-clump sub-filaments that we can see in our images (Figures \ref{fig:alma_cores} and \ref{fig:alma_cores_continuum}) and that could be the channels towards which the accretion onto the fragments occurs. In this case, a direct correlation should be found between the accretion at the clump scales and the accretion onto the sub-filaments system. A detailed study of the accretion properties in the intra-clump gas using ALMA Band 3 data in correlation with the accretion properties at the clump scales is the subject of a forthcoming SQUALO paper (Traficante et al. $in\ prep.$).

\begin{figure}
	\includegraphics[width=8.0cm]{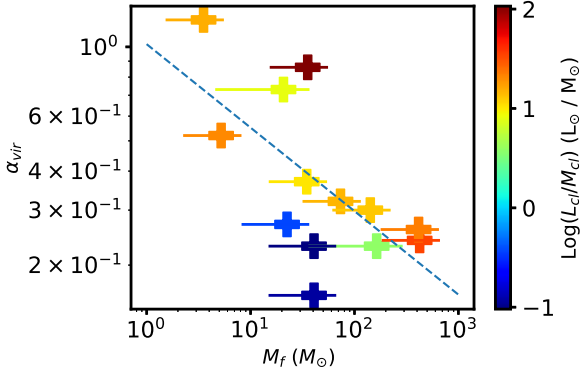}
    \caption{Virial parameter of the clump as a function of the mass of the fragments in each clump. The distribution is colour-coded for the clumps $L_{cl}/M_{cl}$. The blue line is the best power-law fit of the distribution, which gives an exponent $\delta\simeq-0.27$.}
    \label{fig:alpha_vir_core_masses}
\end{figure}

The last correlation we will discuss is between the virial parameter of the clumps ($\alpha_{vir,cl}$) and the mass of our fragments. An anti-correlation between \avir\ and $M$ in massive objects has been discussed several times in the literature, with a typical behaviour $\alpha_{vir}\propto M^{-\delta}$ with $\delta\simeq0.5$ \citep{Kauffmann13,Urquhart18,Traficante18_PII}. The origin of this anti-correlation is not yet understood. It may be a confirmation that the most massive clumps are more gravitationally bound and the gravitational potential overwhelms the local turbulence \citep{Ballesteros-Paredes11}, or it may be an observational bias, as these clumps should collapse maintaining a near-virial equilibrium state, either due to a bias in the observations of the gas kinematics \citep{Traficante18_PIII} or a combination of several systematic \citep{Singh21}. The thirteen clumps of our survey also show a good anti-correlation ($\rho=-0.68$, Appendix \ref{app:pearson_heatmap}), with $\delta\simeq0.54$, in line with the previous observations given the large intrinsic scatter in the relatively small sample. In Figure \ref{fig:alpha_vir_core_masses} instead we show the relation between the virial parameter at the clump scales and the mass of the inner fragments. Therefore, these two parameters are now derived from independent dataset. This diagram shows an anti-correlation, with a Pearson's coefficient of $\rho=-0.41$ and $\delta\simeq0.27$. This result support our previous interpretation, i.e. that fragments are fed by their native clumps. If there is a bias in the estimation of the virial parameter of these clumps, this relation simply confirms that the mass of the fragments is tightly connected with the mass of the parental clumps, since they independently produce a similar $\alpha_{vir}\propto M$ relation. Alternatively, this relation implies that the stronger the gravitational potential of the clumps, the bigger the mass of the inner fragments fed by their reservoir. It is worth noting that the mass of the fragments for the three 70\mum-quiet clumps is below the values expected from the average fit in this diagram. This may be a further indication that fragments form at early stages and they will keep accreting with time.

% This anti-correlation, combined with the tight correlation found between mass of the clumps and mass of the fragments, can have two possible origins: 1) a bias in the estimation of the virial parameter of the clumps, in which case all clumps should have in reality \avir$\simeq 1$ and the mass of the fragments derive directly from the mass of the clumps, as already discussed; 2) a stronger gravitational potential in the more massive clumps, which justifies the lower values of their virial parameter but also explains the formation of more massive fragments in these objects only if they are fed from the clump reservoir and the larger is this reservoir, the bigger is the mass of the inner fragments.

%. Either way, the fact that a very similar correlation is found between the virial parameter of the clumps and both the mass of the clumps and of the cores strongly suggest that fragments are fed from the clump reservoir and the larger is this reservoir, the bigger is the mass of the inner fragments.

%\begin{figure}
%	\includegraphics[width=8.0cm]{Minimum_distance_n_cores.eps}
%    \caption{Minimum distance between cores vs. number of fragments}
%    \label{fig:minimum_distance_cores}
%\end{figure}

%%%%%%%%%%%%%%%%%%%%%%%%%%%%%%%%%%%%%%%%%%%%%%%%%%%%%%%%%%%%%%%%%%%%%%%%%%%%%%%%%%%%%%%
\section{Conclusions}\label{sec:conclusions}
In this work we have presented the first results from the SQUALO (Star formation in QUiescent And Luminous Objects) survey. SQUALO is an ALMA 1 mm and 3 mm survey of 13 massive star-forming clumps at various evolutionary phases, from young 70\mum-quiet up to evolved HII regions, all selected to exhibit signs of large, parsec scale infall motions. In this first paper we have presented the catalogue of objects in the survey and we have analyzed the properties of the fragments identified in the ALMA 1 mm dust continuum data in correlation with the properties of the parent clumps. The fragments have been extracted in the primary beam corrected maps using the \Hyp\ algorithm. A total of 55 robust fragments have been identified in the ALMA images of the 13 SQUALO clumps, 51 of them within the 250\mum\ Hi-GAL footprint and 44 within the 160\mum\ footprint. %We inferred the physical properties of the fragments assuming a dust temperature derived from the evolutionary phase of their parent clumps, following the evolutionary path described in \citet[][]{Molinari16_l_m}: $T=20$ K for clumps with $L_{cl}/M_{cl}<1$ \LM\ (i.e. 70\mum-quiet clumps), $T=30$ K for clumps with $1\leq L_{cl}/M_{cl}\leq 10$ \LM\ and $T=40$ K for clumps with $L_{cl}/M_{cl}>10$ \LM. 

The main results of this work are the following:

\begin{itemize}
    \item[-] 12 out of 13 sources show some degree of fragmentation, with up to 9 fragments in source HIGALBM327.3918+0.1996. Given the resolution of our survey (from $\simeq0.01$ pc up to $\simeq0.04$ pc at the distance of the farthest source of our sample, $d=5.5$ kpc), we expect that some of these fragments will be resolved into smaller objects, as observed in higher resolution images \citep[e.g.][]{Avison16}. HIGALBM343.7560-0.1629 embeds one of the most massive fragment of our sample, with a mass $M\simeq73$ \Msun, and does not show any other fragment down to the minimum detectable mass that for this clump is equal to$\simeq1.0$ \Msun. 
    
    \item[-] All 70\mum-quiet sources are already fragmented into 2 or 4 pieces, which excludes the existence of common isolated, massive pre-stellar cores, in agreement with the findings of \citet{Svoboda19} and \citet{Sanhueza19}. 
    
    \item[-] The ratio between the Jeans length of the fragments and the thermal Jeans length, corrected for the 3D-projection effects, $\lambda_{J_{r},3D}$, decreases as the number of fragments in each clump increases. Also, the youngest clumps in our sample have on average the highest values of $\lambda_{J_{r},3D}$, and only the more evolved ones approaches values of $\lambda_{J_{r},3D}$ close to unity (i.e. the expected value in case of pure thermal Jeans fragmentation). Similarly, the minimum 2D distance between fragments appears to decrease with the clumps evolution, a result also supported by the fragmentation properties observed in the clumps of the CORE survey \citep{Beuther18}. We interpret these numbers as the effect of a gravo-turbulent, hierarchical process: massive clumps initially produce non-thermally supported fragments which get closer with time and eventually continue to internally fragment until they reach the thermal fragmentation level ($\lambda_{J_{r},3D}\simeq 1$). Clumps with a low degree of fragmentation at various scales may represent those cases where the local turbulence play a major role in the process and they are likely supported by magnetic fields, which can limit or completely suppress the fragmentation process \citep{Palau21}. 
    
 %   as the effect of the gravity that overcomes the local turbulence during the evolutionary process: massive clumps initially produce non-thermally supported fragments, which are dynamically active, get closer with time, and eventually continue to fragment until they reach the thermal fragmentation level ($\lambda_{J_{r},3D}\simeq 1$). Clumps with a low degree of fragmentation may represent those cases where the local turbulence play a major role in the process and they are likely supported by magnetic fields, which can limit or completely suppress the fragmentation process \citep{Palau21}. 
    
    %     \item[-]  But their evolution towards higher degree of fragmentation or towards an assembly in single, more massive objects is yet unclear. The magnetic fields may play a predominant role in some sources and not in others, as observed in \citet{Palau21}.

    \item[-] The mass and surface density of the fragments are tightly correlated with the mass and surface density of the parent clumps: the most massive fragments are embedded in the most massive clumps, and the instantaneous clump formation efficiency CFE is higher in clumps with higher fragment mass. The data shows a possible evolutionary path where clump surface density and the density of the fragments increase from 70 \mum-quiet objects to more evolved regions. These trends are in agreement with Galactic Plane surveys of clumps which demonstrated that their surface density increases with evolution \citep{Svoboda16,Elia21}, and they are also observed at the fragment scales in the ALMA survey of filament-hub systems at various evolutionary phases \citep{Anderson21}.
  
    \item[-] We found an indication that the high accretion rates measured at the clump scales produces more massive fragments (correlation coefficient $\rho=0.44$). The accretion rate is also independent from the number of fragments identified at the various evolutionary phases. A detailed analysis of the infall properties at clump and fragment scales will be the topic of a dedicated SQUALO work.
 
    \item[-] The anti-correlation observed in several surveys between the virial parameter and the mass of the clumps, \avir\ $\propto M^{-\delta}$, with $\delta\simeq0.5$, is also observed in our sample. We have also combined the virial parameter of the clumps with the mass of the fragments, building an \avir\ vs. $M$ relation from independent observations. We found a similar anti-correlation, with a value of $\delta\simeq0.27$. We interpret this result as a further suggestion that the dynamics of the clumps and of the inner fragments are tightly connected: if there is no bias in the estimation of the virial parameter of our clumps, the more gravitationally bound clumps are the most massive ones and the ones that have build-up the most massive fragments.
    
\end{itemize}

We interpret our results as combined evidence in favour of the \textit{clump-fed} scenario, as opposed to the \textit{core-fed} scenario for the formation of massive stars. 

Our data suggest a hierarchical, gravo-turbulent process: young, massive clumps initially fragment under the influence of non-thermal motions in interplay with the gravitational fields. With time the fragments condense, gather mass from their surroundings and eventually keep fragmenting until some of them, where the gravitational potential overcome the turbulence, evolve to reach pure thermal Jeans fragmentation. Future surveys such as ALMAGAL, which include a more statistically significant sample of objects, can confirm the scenario that we propose in this work.

% The physical properties of the inner fragments are intimately connected with those of the parent clumps, and we found evidences of dynamical evolution of the properties of the fragments moving from 70\mum\ quiet clumps to more evolved ones.    

\section*{Acknowledgements}
This paper makes use of the following ALMA data: ADS/JAO.ALMA\#2018.1.00443.S. ALMA is a partnership of ESO (representing its member states), NSF (USA) and NINS (Japan), together with NRC (Canada), MOST and ASIAA (Taiwan), and KASI (Republic of Korea), in cooperation with the Republic of Chile. The Joint ALMA Observatory is operated by ESO, AUI/NRAO and NAOJ. A.A. acknowledges support from STFC grants ST/T001488/1 and ST/P000827/1. G.A.F acknowledges support from the Collaborative Research Centre 956, funded by the Deutsche Forschungsgemeinschaft (DFG) project ID 184018867. G.A.F also acknowledges financial support from the State Agency for Research of the Spanish MCIU through the AYA 2017-84390-C2-1-R grant (co-funded by FEDER) and through the "Center of Excellence Severo Ochoa" award for the Instituto de Astrof\'isica de Andalucia (SEV-2017-0709). R.J.S acknowledges funding from an STFC ERF (grant ST/N00485X/1) and HPC from the DiRAC facility (ST/P002293/1).

%%%%%%%%%%%%%%%%%%%% DATA AVAILABILITY %%%%%%%%%%%%%%%%%%
\section*{Data Availability}
The data underlying this article are available in the article and in its online supplementary material.

% The inclusion of a Data Availability Statement is a requirement for articles published in MNRAS. Data Availability Statements provide a standardised format for readers to understand the availability of data underlying the research results described in the article. The statement may refer to original data generated in the course of the study or to third-party data analysed in the article. The statement should describe and provide means of access, where possible, by linking to the data or providing the required accession numbers for the relevant databases or DOIs.

%%%%%%%%%%%%%%%%%%%% REFERENCES %%%%%%%%%%%%%%%%%%

% The best way to enter references is to use BibTeX:

\bibliographystyle{mnras}
\bibliography{bibliography} % if your bibtex file is called example.bib

% Alternatively you could enter them by hand, like this:
% This method is tedious and prone to error if you have lots of references
%\begin{thebibliography}{99}
%\bibitem[\protect\citeauthoryear{Author}{2012}]{Author2012}
%Author A.~N., 2013, Journal of Improbable Astronomy, 1, 1
%\bibitem[\protect\citeauthoryear{Others}{2013}]{Others2013}
%Others S., 2012, Journal of Interesting Stuff, 17, 198
%\end{thebibliography}

%%%%%%%%%%%%%%%%%%%%%%%%%%%%%%%%%%%%%%%%%%%%%%%%%%

%%%%%%%%%%%%%%%%% APPENDICES %%%%%%%%%%%%%%%%%%%%%

\appendix

\section{Comparison between photometry codes}\label{app:codes}
In this Appendix we explore the differences between \Hyp\ and astrodendro applied to our ALMA data.

These codes were designed for different purposes: \Hyp\ \citep{Traficante15b} was initially written to identify and perform the photometry of dense star-forming regions in the crowded, background-dominated \textit{Herschel} images. Astrodendro \citep{Rosolowsky08} was designed to compute dendograms and decompose the images into various hierarchical layers. The main differences of these two approaches can be separated in the differences in the source identification and in the source photometry.

%%%%%%%%%%%%%%%%%%%%%%%%%%%%%%%%%%%%%%%%%%%%%%%%%%%%%%%%%%%%%%%%%%%%%%%%%%%%%%%%%%%%%%%%%%%%
\subsection{Source identification}
The \Hyp\ algorithm uses a finding algorithm which identifies peaks in a high-pass filtered image, and for each peak it performs a 2d-Gaussian fit on the real image. Therefore a "source" is defined as a 2d-Gaussian structure. On the other hand, astrodendro performs a dendogram analysis which identifies leaves and branches in a hierarchical structure based on the emission across the field, with no impositions to the source shape. The smallest leaves are identified as "source" and are not necessarily 2d-Gaussians. 

The different extraction algorithms can produce a different number of sources identified in each field, depending on the intensity of the main peaks and on the distribution of the source fluxes, which influence the separation between leaves and trunks (for astrodendro).

Astrodendro has three parameters that can be tuned to optimize the source extraction and photometry: \textit{min\_value}, the lowest value that can be considered in the map as a potential source; \textit{min\_delta}, i.e. the difference between the peak flux and each pixel flux, needed to identify up to which level a leaf independent from another, and when it should be merged in the tree; \textit{min\_npix}, the minimum number of pixels needed to consider each identified entity as a single and independent leaf.

After several tests with all our sources, we fixed these parameters to min\_value=5 times the map \textit{rms} estimated with a sigma-clipping algorithm, min\_delta=0.2$\sigma$ and min\_npix=2 for the entire sample.

\begin{figure}
	\includegraphics[width=8.0cm]{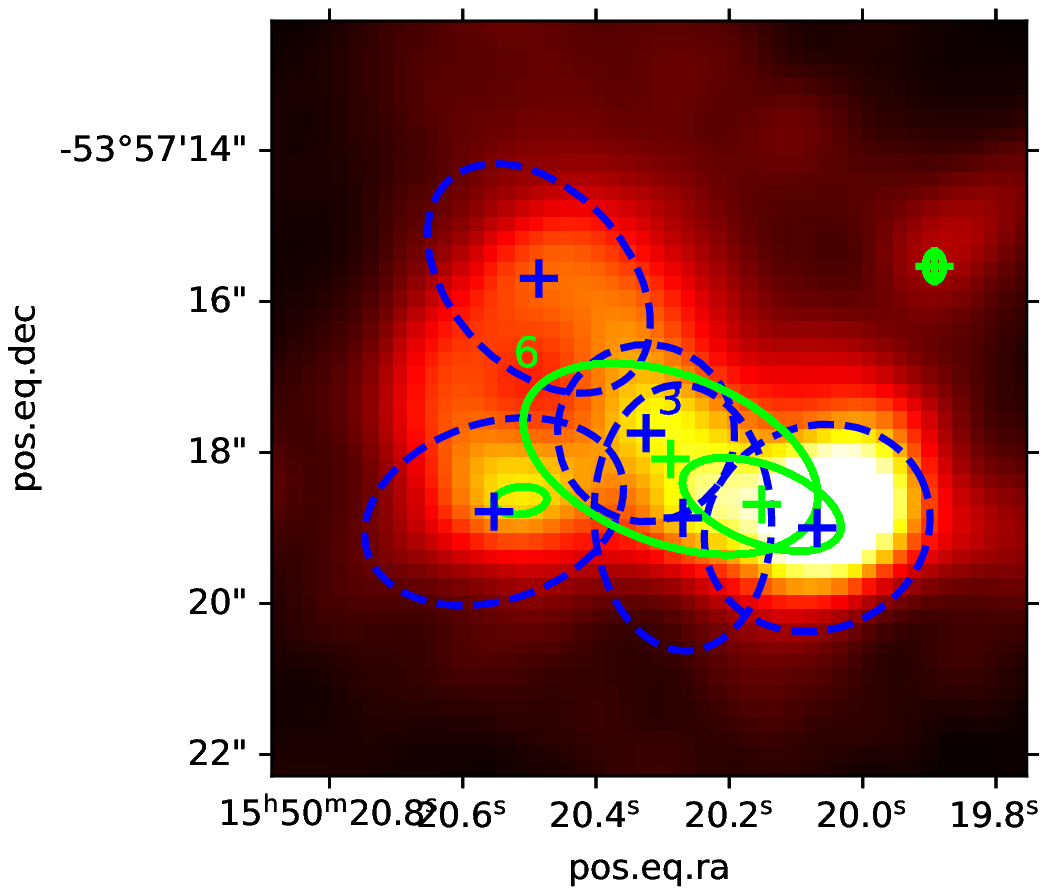}
    \caption{Central region of source HIGALBM327.3918+0.1996 as seen by ALMA at 1.3mm. The blue ellipses are the sources identified by \Hyp. The green ellipses are those corresponding to the astrodendro source identification.}
    \label{fig:G327_zoom_photometry}
\end{figure}

With these numbers we identify 47 sources to be compared with the 55 of the \Hyp\ extraction. Astrodendro works reasonably well in relatively crowded regions, but with these parameters it fails to identify single fragments in highly confused regions. In Figure \ref{fig:G327_zoom_photometry} we show the sources identified with \Hyp\ and Astrodendro (also represented with a 2d-Gaussian derived from the astrodendro parameters) in the central region of source HIGALBM327.3918+0.1996, where the different fragments that \Hyp\ identifies are merged in a single leaf. Although the astrodendro parameters can be fine-tuned for each specific source, we did not find a combination of values that allowed us to recover the 5 sources that \Hyp\ identified in the central region of HIGALBM327.3918+0.1996.

%%%%%%%%%%%%%%%%%%%%%%%%%%%%%%%%%%%%%%%%%%%%%%%%%%%%%%%%%%%%%%%%%%%%%%%%%%%%%%%%%%%%%%%%%%%%
\subsection{Source photometry}
The main differences in the flux estimation between the two tested algorithms are that astrodendro does not account for any background emission, and it is not able to deblend sources that are overlapping along the line of sight. These two conditions, in particular a strong background contribution to the source fluxes, can be critical in our ALMA data. The example of source HIGALBM327.3918+0.1996 in Figure \ref{fig:G327_zoom_photometry} is a good example.

To compare the astrodendro and \Hyp\ fluxes in a coherent way, we have followed two approaches: first, we have run \Hyp\ without subtracting any background, therefore making the \Hyp\ fluxes coherent with the astrodendro results; then, we have used the astrodendro background estimation done with the sigma-clipping approach as a background value to subtract at the whole map, in order to make these fluxes as much coherent as possible with the \Hyp\ background-subtracted fluxes. 

\begin{figure}
	\includegraphics[width=8.0cm]{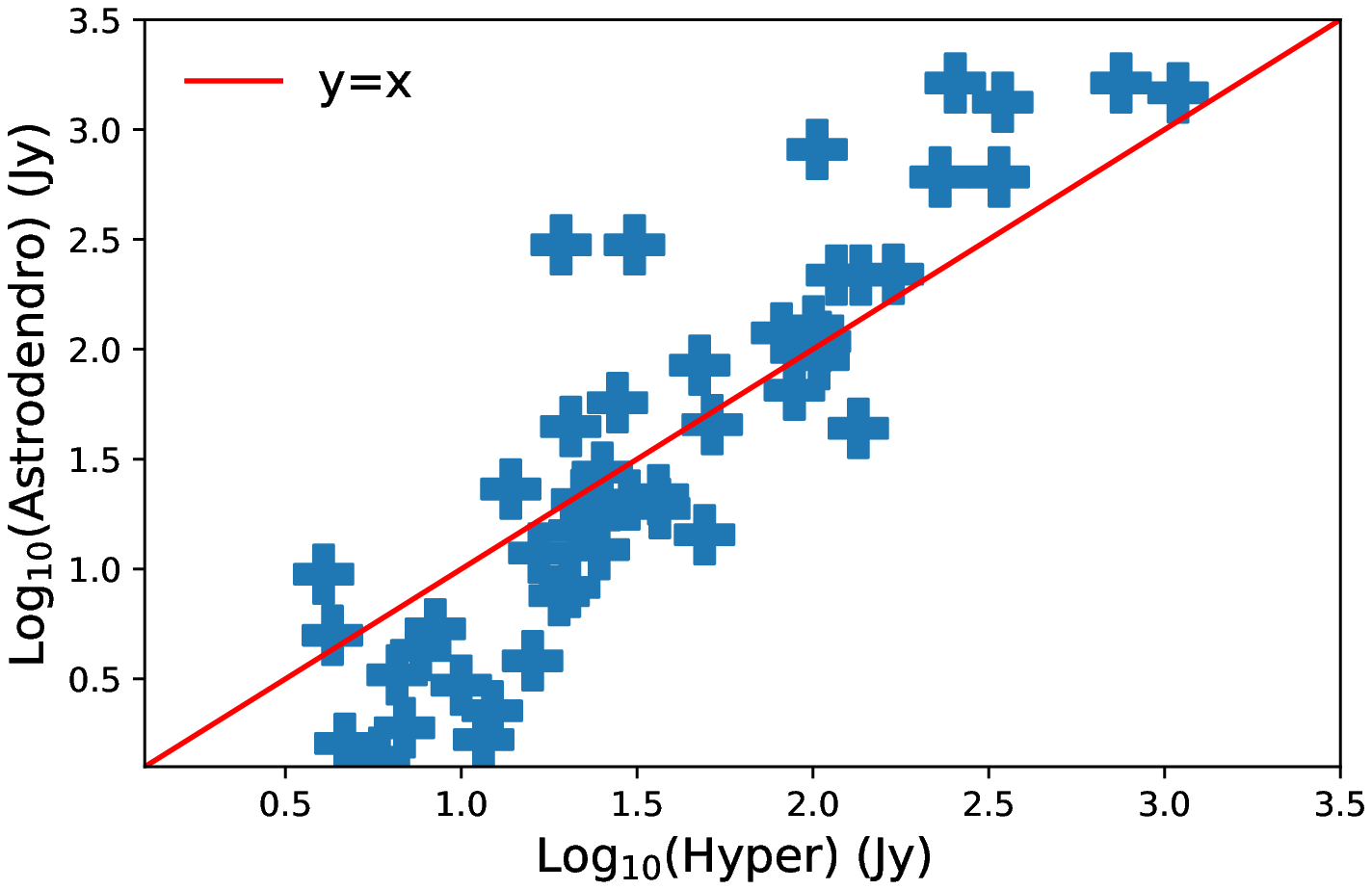}
	\includegraphics[width=8.0cm]{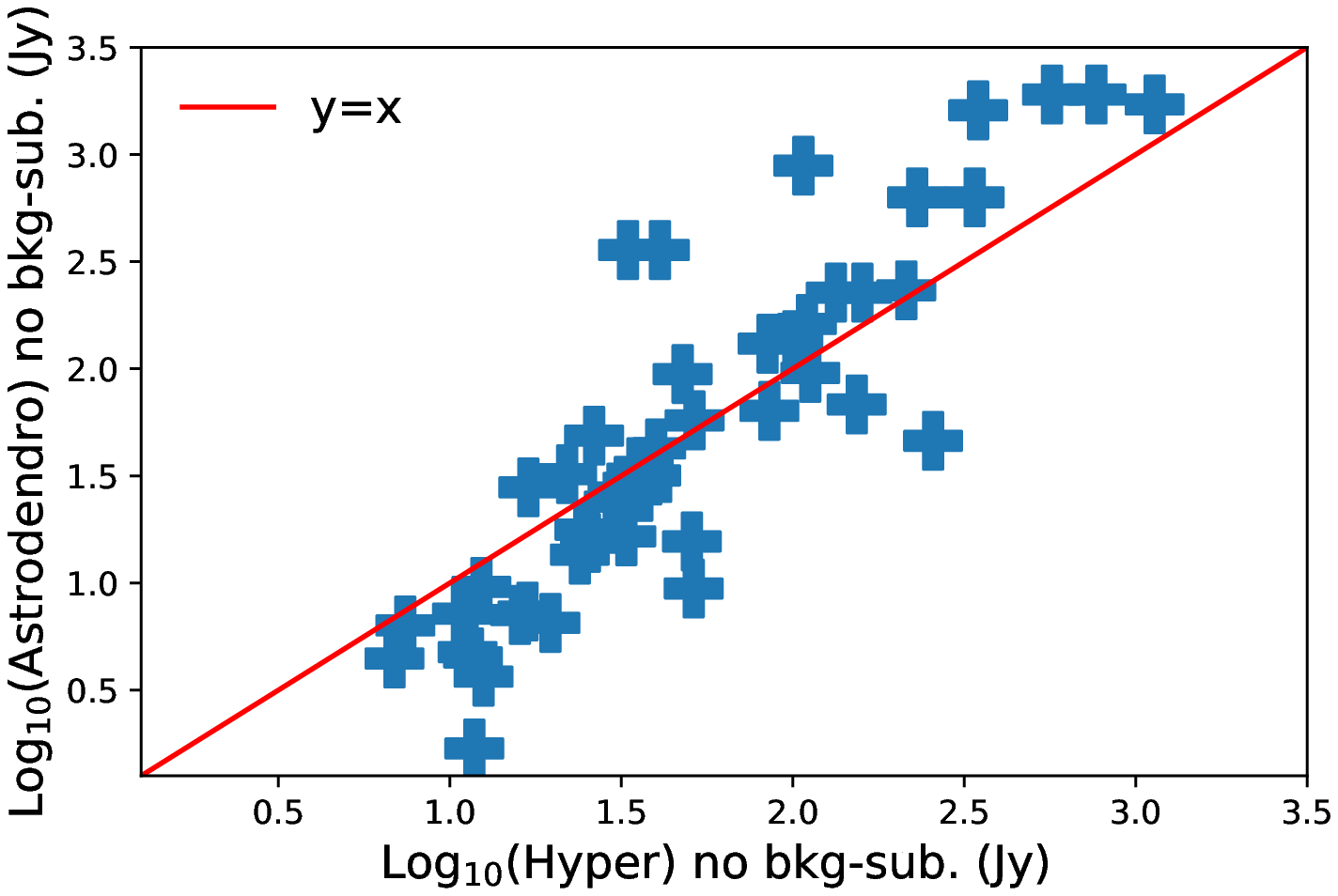}
    \caption{\textit{Upper panel}: \Hyp\ fluxes versus astrodendro fluxes for the 47 fragments in common. Both fluxes have been background subtracted as explained in the text. \textit{Lower panel}: same plot, but with fluxes integrated without any background subtraction for both \Hyp\ and astrodendro sources.}
    \label{fig:photometry_differences}
\end{figure}

\begin{figure}
	\includegraphics[width=8.0cm]{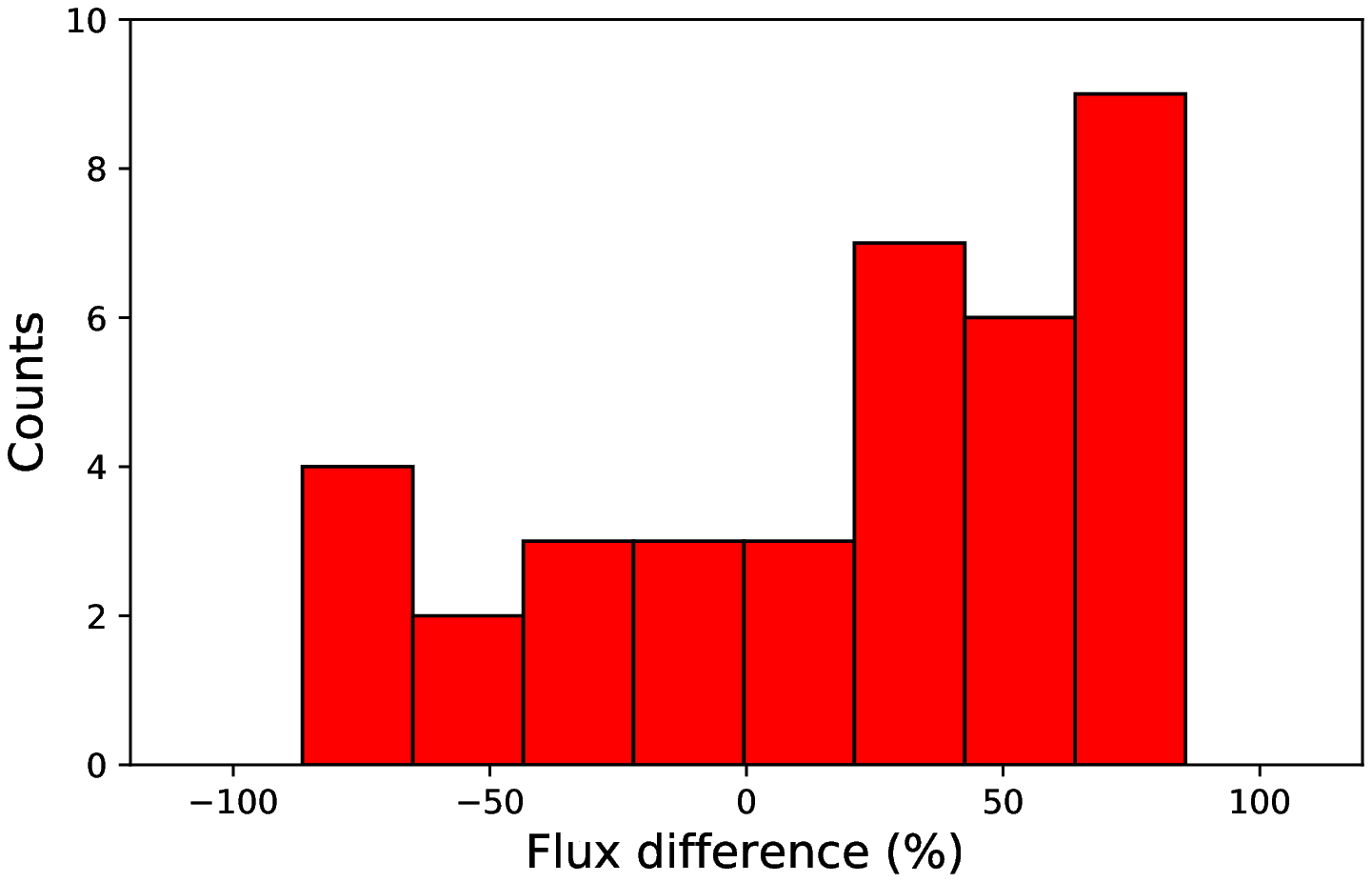}
    \caption{Distribution of fluxes difference (in \%) between \Hyp\ and astrodendro fluxes for the 37 cases where the fragments are relatively isolated and well identified with both algorithms.}
    \label{fig:flux_difference_histogram}
\end{figure}

In Figure \ref{fig:photometry_differences} we show the comparison of the fluxes of the 47 sources in common with both algorithms in the two cases described. The red-line in both plots is the bisector line. The fluxes are statistically in good agreement with correlation coefficients of $\rho=0.88$ and $\rho=0.87$ in case of no background subtraction and with background subtraction, respectively. In 10 cases the astrodendro flux is significantly larger than the \Hyp\ flux (more than 100\% of difference), but these are cases where astrodendro cannot disentangle \Hyp\ sources like in Figure \ref{fig:G327_zoom_photometry}. In Figure \ref{fig:flux_difference_histogram} we show the histogram of the flux difference (in \%) between the two methods for the remaining 37 sources, excluding the 10 cases just discussed. The absolute average flux differences in this (relatively small) sample is $\simeq48\%$, and relatively well distributed around the center (the mean value of the distribution is $\simeq18\%$). 

These simple tests show that the main differences between the algorithms may lead to significantly different results on single objects, but these differences may be averaged-out on large samples. At the same time, the fine-tuning for each single source of the astrodendro parameters seems necessary to properly identify ALMA fragments in complex and crowded regions. 

We want to stress that there is no reason to prefer one algorithm or another \textit{a-priori}, as long as the results are discussed within the framework of the chosen algorithm which will produce a definition of the ALMA fragments and of their properties (e.g. derived radius, integrated flux) which can be substantially different from one approach to another.  

We decided to choose the \Hyp\ algorithm to work with a definition of the fragment that is consistent with the definition of the parent clump that we extracted from the original Hi-GAL images.

%%%%%%%%%%%%%%%%%%%%%%%%%%%%%%%%%%%%%%%%%%%%%%%%%%%%%%%%%%%%%%%%%%%%%%%%%%%%%%%%%%%%%%%
\section{Fragments properties}\label{app:frag_properties}
In this Appendix we report the Table with all the properties of the 55 fragments obtained as described in the main text. 

\begin{landscape}
\begin{table}\label{tab:frag_properties}
\begin{tabular}{cccccccccccc}
\hline
DESIGNATION & ALMA\_ID & $\alpha$ & $\delta$ & FWHM 1 & FWHM 2 & PA & Peak Flux & Flux & $R_{f}$ & $M_{f}$ & $\Sigma_{f}$ \\
 &  &  &  & (") & (") & ($\deg$) & ($\mathrm{mJy\,beam^{-1}}$) & ($\mathrm{mJy}$) & ($\mathrm{pc}$) & ($\mathrm{M_{\odot}}$) & ($\mathrm{g\,cm^{-2}}$) \\
\hline
\hline
HIGALBM327.3918+0.1996 & 1 & 15:50:18.2328 & -53:57:07.2 & 1.35 & 1.52 & 114.17 & 56.62 & 100.414 & 0.036 & 44.89 & 2.32 \\
HIGALBM327.3918+0.1996 & 2 & 15:50:18.4344 & -53:57:07.056 & 1.17 & 1.76 & 182.4 & 16.32 & 31.091 & 0.036 & 13.9 & 0.72 \\
HIGALBM327.3918+0.1996 & 3 & 15:50:18.4896 & -53:57:05.94 & 1.17 & 1.17 & 254.49 & 13.09 & 19.236 & 0.029 & 8.6 & 0.67 \\
HIGALBM327.3918+0.1996 & 4 & 15:50:18.7176 & -53:57:06.984 & 1.17 & 1.76 & 108.08 & 20.87 & 49.191 & 0.036 & 21.99 & 1.13 \\
HIGALBM327.3918+0.1996 & 5 & 15:50:20.0832 & -53:57:07.668 & 1.35 & 1.76 & 127.21 & 14.33 & 25.177 & 0.039 & 11.25 & 0.5 \\
HIGALBM327.3918+0.1996 & 6 & 15:50:18.6504 & -53:57:03.888 & 1.17 & 1.76 & 222.8 & 12.31 & 31.422 & 0.036 & 14.04 & 0.72 \\
HIGALBM327.3918+0.1996 & 7 & 15:50:19.2864 & -53:57:07.524 & 1.17 & 1.62 & 102.54 & 15.27 & 24.684 & 0.034 & 11.04 & 0.62 \\
HIGALBM327.3918+0.1996 & 8 & 15:50:19.6296 & -53:57:07.488 & 1.29 & 1.76 & 263.43 & 9.50 & 20.338 & 0.038 & 9.09 & 0.43 \\
HIGALBM327.3918+0.1996 & 9 & 15:50:16.776 & -53:57:02.232 & 1.41 & 1.76 & 126.06 & 10.01 & 15.952 & 0.039 & 7.13 & 0.31 \\
HIGALBM327.4022+0.4449 & 1 & 15:49:19.3248 & -53:45:14.256 & 1.29 & 1.76 & 101.74 & 225.63 & 338.342 & 0.034 & 121.76 & 7.08 \\
HIGALBM327.4022+0.4449 & 2 & 15:49:19.4736 & -53:45:14.364 & 1.17 & 1.76 & 97.95 & 177.66 & 346.478 & 0.032 & 124.69 & 7.99 \\
HIGALBM327.4022+0.4449 & 3 & 15:49:19.2168 & -53:45:13.212 & 1.17 & 1.76 & 133.53 & 127.53 & 229.981 & 0.032 & 82.77 & 5.3 \\
HIGALBM327.4022+0.4449 & 4 & 15:49:19.7112 & -53:45:14.58 & 1.17 & 1.76 & 101.18 & 84.93 & 134.643 & 0.032 & 48.46 & 3.1 \\
HIGALBM327.4022+0.4449 & 5 & 15:49:19.356 & -53:45:11.448 & 1.65 & 1.76 & 113.6 & 40.81 & 127.368 & 0.038 & 45.84 & 2.08 \\
HIGALBM331.1314-0.2438 & 1 & 16:10:59.7408 & -51:50:22.812 & 1.2 & 1.75 & 168.77 & 526.54 & 751.879 & 0.035 & 309.28 & 17.0 \\
HIGALBM331.1314-0.2438 & 2 & 16:10:59.76 & -51:50:24.36 & 1.16 & 1.56 & 177.97 & 178.71 & 254.008 & 0.032 & 104.49 & 6.66 \\
HIGALBM332.6045-0.1674 & 1 & 16:17:29.3256 & -50:46:12.612 & 1.26 & 1.75 & 266.6 & 39.39 & 81.416 & 0.022 & 18.03 & 2.45 \\
HIGALBM332.6045-0.1674 & 2 & 16:17:30.2832 & -50:46:11.82 & 1.17 & 1.75 & 251.81 & 4.85 & 12.259 & 0.021 & 2.71 & 0.4 \\
HIGALBM338.9260+0.6340 & 1 & 16:40:13.9296 & -45:38:29.58 & 1.29 & 1.62 & 102.51 & 82.15 & 169.367 & 0.029 & 68.87 & 5.39 \\
HIGALBM338.9260+0.6340 & 2 & 16:40:14.256 & -45:38:30.696 & 1.2 & 1.37 & 115.84 & 65.69 & 104.051 & 0.026 & 42.31 & 4.21 \\
HIGALBM338.9260+0.6340 & 3 & 16:40:13.5648 & -45:38:33.216 & 1.32 & 1.62 & 140.9 & 47.40 & 105.084 & 0.029 & 42.73 & 3.27 \\
HIGALBM338.9260+0.6340 & 4 & 16:40:12.9384 & -45:38:26.592 & 1.08 & 1.62 & 143.47 & 15.80 & 23.299 & 0.027 & 9.48 & 0.89 \\
HIGALBM341.2149-0.2359 & 1 & 16:52:23.736 & -44:27:55.008 & 1.07 & 1.6 & 146.02 & 22.89 & 47.639 & 0.022 & 9.52 & 1.32 \\
HIGALBM341.2149-0.2359 & 2 & 16:52:22.3608 & -44:28:02.532 & 1.28 & 1.6 & 166.54 & 21.92 & 51.711 & 0.024 & 10.33 & 1.2 \\
HIGALBM341.2149-0.2359 & 3 & 16:52:23.6784 & -44:27:53.964 & 1.07 & 1.6 & 146.8 & 14.33 & 20.481 & 0.022 & 4.09 & 0.57 \\
HIGALBM341.2149-0.2359 & 4 & 16:52:23.9232 & -44:27:56.232 & 1.07 & 1.07 & 137.16 & 15.27 & 21.614 & 0.018 & 4.32 & 0.9 \\
HIGALBM341.2149-0.2359 & 5 & 16:52:24.0672 & -44:27:54.54 & 1.07 & 1.07 & 235.1 & 11.91 & 16.619 & 0.018 & 3.32 & 0.69 \\
HIGALBM341.2149-0.2359 & 6 & 16:52:22.4136 & -44:27:43.776 & 1.07 & 1.07 & 180.0 & 2.91 & 4.062 & 0.018 & 0.81 & 0.17 \\
HIGALBM341.2149-0.2359 & 7 & 16:52:22.0656 & -44:27:55.08 & 1.07 & 1.6 & 117.85 & 6.81 & 11.578 & 0.022 & 2.32 & 0.32 \\
HIGALBM343.5212-0.5172 & 1 & 17:01:34.0104 & -42:50:20.04 & 1.32 & 1.6 & 257.92 & 8.11 & 13.828 & 0.021 & 2.13 & 0.31 \\
HIGALBM343.5212-0.5172 & 2 & 17:01:33.7872 & -42:50:19.428 & 1.07 & 1.11 & 129.75 & 3.08 & 5.852 & 0.016 & 0.9 & 0.23 \\
HIGALBM343.5212-0.5172 & 3 & 17:01:33.5736 & -42:50:10.104 & 1.28 & 1.54 & 230.77 & 3.08 & 4.666 & 0.021 & 0.71 & 0.11 \\
HIGALBM343.5212-0.5172 & 4 & 17:01:33.8304 & -42:50:21.3 & 1.39 & 1.6 & 131.13 & 1.47 & 4.928 & 0.022 & 0.75 & 0.1 \\
HIGALBM343.5212-0.5172 & 5 & 17:01:33.7848 & -42:50:17.736 & 1.07 & 1.07 & 229.02 & 3.69 & 4.695 & 0.016 & 0.72 & 0.19 \\
HIGALBM343.7560-0.1629 & 1 & 17:00:49.8792 & -42:26:09.096 & 1.19 & 1.37 & 137.15 & 708.58 & 1092.309 & 0.012 & 73.35 & 31.81 \\
HIGALBM344.1032-0.6609 & 1 & 17:04:06.6864 & -42:27:57.852 & 1.37 & 1.6 & 134.08 & 17.18 & 36.354 & 0.014 & 2.44 & 0.79 \\
HIGALBM344.1032-0.6609 & 2 & 17:04:07.5072 & -42:28:02.82 & 1.11 & 1.59 & 101.01 & 6.48 & 9.997 & 0.013 & 0.67 & 0.27 \\
HIGALBM344.1032-0.6609 & 3 & 17:04:06.972 & -42:28:06.708 & 1.22 & 1.6 & 143.01 & 4.04 & 6.317 & 0.014 & 0.42 & 0.15 \\
\hline
\end{tabular}

\caption{Properties of the 55 fragments identified in our 13 massive clumps. \textit{col.1}: Clump designation; \textit{col.2}: ALMA ID following the \Hyp\ identification number in each clump; \textit{col.3-4}: coordinates of the fragments peak position; \textit{cols. 5-7}: FWHMs and position angles of the 2d-Gaussian fit used to estimate the source integrated flux; \textit{col.8}: peak flux (background subtracted); \textit{col.9}: integrated flux within the ellipse defined by the 2 FWHMs and the PA; \textit{col.10-12}: radius, mass and surface density of each fragment derived as described in the text.}

\end{table}

\end{landscape}

\begin{landscape}
\begin{table}
\begin{tabular}{cccccccccccc}
\hline
DESIGNATION & ALMA\_ID & $\alpha$ & $\delta$ & FWHM 1 & FWHM 2 & PA & Peak Flux & Flux & $R_{f}$ & $M_{f}$ & $\Sigma_{f}$ \\
 &  &  &  & (") & (") & ($\deg$) & ($\mathrm{mJy\,beam^{-1}}$) & ($\mathrm{mJy}$) & ($\mathrm{pc}$) & ($\mathrm{M_{\odot}}$) & ($\mathrm{g\,cm^{-2}}$) \\
\hline
\hline
HIGALBM344.2210-0.5932 & 1 & 17:04:12.936 & -42:19:52.14 & 1.07 & 1.6 & 112.52 & 66.03 & 136.753 & 0.013 & 9.18 & 3.79 \\
HIGALBM344.2210-0.5932 & 2 & 17:04:12.6048 & -42:19:54.192 & 1.18 & 1.6 & 253.73 & 39.46 & 88.574 & 0.013 & 5.95 & 2.23 \\
HIGALBM344.2210-0.5932 & 3 & 17:04:12.8184 & -42:19:52.716 & 1.07 & 1.6 & 222.54 & 44.18 & 116.642 & 0.013 & 7.83 & 3.24 \\
HIGALBM344.2210-0.5932 & 4 & 17:04:12.7728 & -42:19:54.624 & 1.07 & 1.6 & 171.74 & 38.88 & 102.784 & 0.013 & 6.9 & 2.85 \\
HIGALBM344.2210-0.5932 & 5 & 17:04:12.9672 & -42:19:57.36 & 1.33 & 1.59 & 163.19 & 20.95 & 27.813 & 0.014 & 1.87 & 0.63 \\
HIGALBM344.2210-0.5932 & 6 & 17:04:12.3504 & -42:19:46.884 & 1.07 & 1.6 & 145.7 & 16.23 & 36.709 & 0.013 & 2.47 & 1.02 \\
HIGALBM344.2210-0.5932 & 7 & 17:04:12.636 & -42:19:49.08 & 1.6 & 1.6 & 172.08 & 6.62 & 18.98 & 0.016 & 1.27 & 0.35 \\
HIGALBM24.0116+0.4897 & 1 & 18:33:18.2952 & -7:42:26.136 & 1.38 & 2.01 & 237.29 & 14.87 & 30.044 & 0.042 & 31.86 & 1.19 \\
HIGALBM24.0116+0.4897 & 2 & 18:33:18.516 & -7:42:25.74 & 1.34 & 2.01 & 103.04 & 6.04 & 8.438 & 0.042 & 8.95 & 0.34 \\
HIGALBM28.1957-0.0724 & 1 & 18:43:02.8344 & -4:14:50.784 & 1.58 & 2.03 & 241.43 & 3.13 & 4.302 & 0.047 & 4.81 & 0.15 \\
HIGALBM28.1957-0.0724 & 2 & 18:43:03.2592 & -4:15:08.892 & 1.39 & 1.39 & 180.0 & 4.17 & 6.571 & 0.036 & 7.35 & 0.37 \\
HIGALBM28.1957-0.0724 & 3 & 18:43:01.9992 & -4:14:48.084 & 1.39 & 2.09 & 92.84 & 3.51 & 6.897 & 0.044 & 7.71 & 0.26 \\
HIGALBM28.1957-0.0724 & 4 & 18:43:02.5104 & -4:14:51.072 & 1.41 & 2.09 & 121.0 & 1.37 & 2.292 & 0.045 & 2.57 & 0.09 \\
HIGALBM31.9462+0.0759 & 1 & 18:49:22.104 & -0:50:32.352 & 1.55 & 1.94 & 95.71 & 12.32 & 21.993 & 0.046 & 25.9 & 0.81 \\
HIGALBM31.9462+0.0759 & 2 & 18:49:22.1448 & -0:50:45.708 & 1.61 & 2.11 & 227.25 & 4.35 & 7.667 & 0.049 & 9.03 & 0.25 \\
HIGALBM31.9462+0.0759 & 3 & 18:49:22.08 & -0:50:34.98 & 1.41 & 2.11 & 182.88 & 1.95 & 2.758 & 0.046 & 3.25 & 0.1 \\
HIGALBM31.9462+0.0759 & 4 & 18:49:22.3104 & -0:50:31.56 & 1.41 & 1.59 & 231.19 & 1.62 & 2.217 & 0.04 & 2.61 & 0.11 \\
\\\hline
\end{tabular}

%\caption{Properties of the 55 fragments identified in our 13 massive clumps. \textit{col.1}: Clump ID; \textit{col.2}: ALMA ID following the \Hyp\ identification number in each clump; \textit{col.3-4}: coordinates of the fragments peak position; \textit{cols. 5-7}: FWHMs and position angles of the 2d-Gaussian fit used to estimate the source integrated flux; \textit{col.8}: peak flux (background subtracted); \textit{col.9}: integrated flux within the ellipse defined by the 2 FWHMs and the PA; \textit{col.10-12}: radius, mass and surface density of each fragment derived as described in the text.}

\end{table}

\end{landscape}

%%%%%%%%%%%%%%%%%%%%%%%%%%%%%%%%%%%%%%%%%%%%%%%%%%%%%%%%%%%%%%%%%%%%%%%%%%%%%%%%%%%%%%%
\newpage
\section{Heatmap of Pearson's coefficients}\label{app:pearson_heatmap}
In this Appendix we show the correlation matrix, or heatmap for all the pairs of parameters discussed in the main text. It includes parameters of the clumps: mass $M_{cl}$, $L/M$ ratio $L_{cl}/M_{cl}$, surface density $\Sigma_{cl}$, mass accretion rate $\dot{M}_{cl}$ and virial parameter $\alpha_{vir,cl}$; parameters of the fragments: the number of fragments in each clump $\#_{f}$, the minimum 2D (3D) distances between fragments, $d_{min,2D}$ ($d_{min,3D}$), the total mass of fragments in each clump $M_{f}$, the mass and surface density of the most massive and of the densest fragment in each clump, M$_{f,max}$ and $\Sigma_{f,max}$, respectively; parameters that combine clumps and fragments properties: the instantaneous clump formation efficiency (CFE), i.e. the ratio between the total mass of the fragments in each clump and the mass of the clump \citep[e.g. ][]{Anderson21}, and the ratio between the clump 2D Jeans length (3D corrected) and the minimum distance between fragments in each clump, $\lambda_{J_{r},2D}$ ($\lambda_{J_{r},3D}$).

\begin{figure*}
	\includegraphics[width=16.0cm]{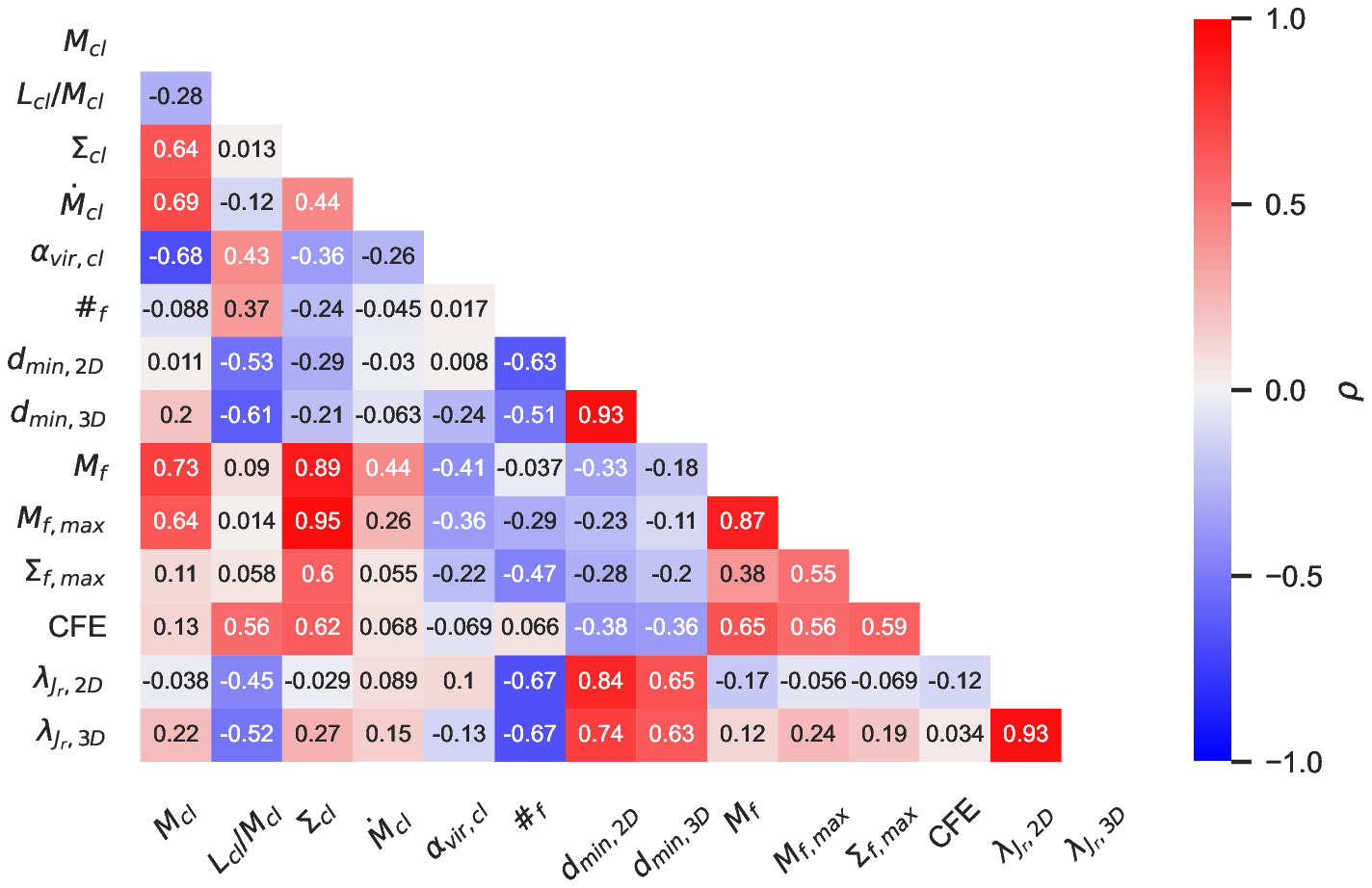}
    \caption{Heatmap of the Pearson's correlation coefficients $\rho$ between several properties derived for clumps and fragments, as described in the text. The heatmap is colour-coded such that the bluest cells represents the highest degree of anti-correlation (approaching $\rho=-1$), and the reddest cells the highest degree of correlation (approaching $\rho=1$). The value of $\rho$ for each given pair of parameters is reported within each cell.}
    \label{fig:Pearson_heatmap}
\end{figure*}

%%%%%%%%%%%%%%%%%%%%%%%%%%%%%%%%%%%%%%%%%%%%%%%%%%%%%%%%%%%%%%%%%%%%%%%%%%%%%%%%%%%%%%%
\newpage
\section{Temperature uncertainties}\label{app:Pearson_MC_statistics}
\begin{figure*}
	\includegraphics[width=7.0cm]{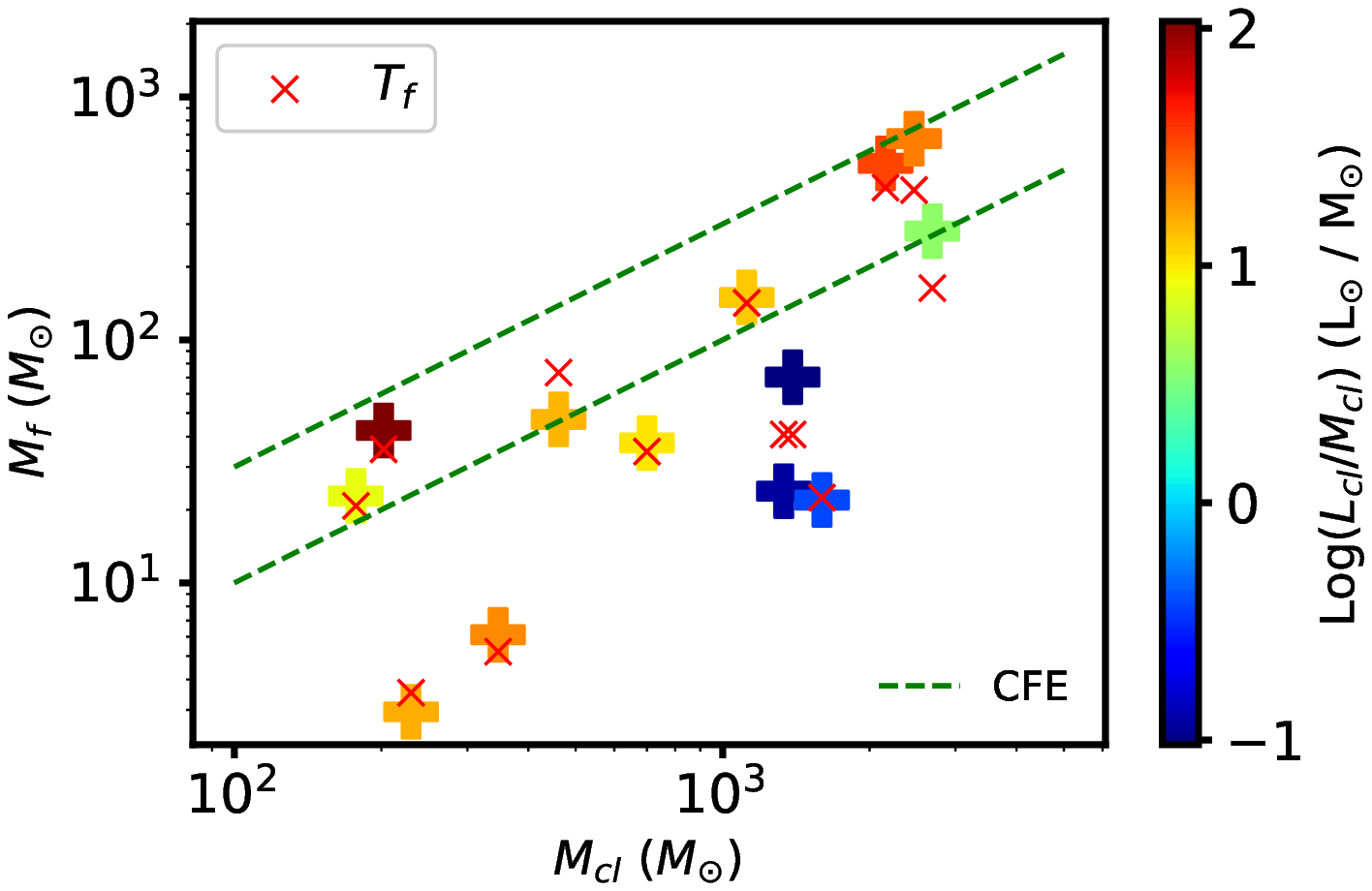}
	\includegraphics[width=7.0cm]{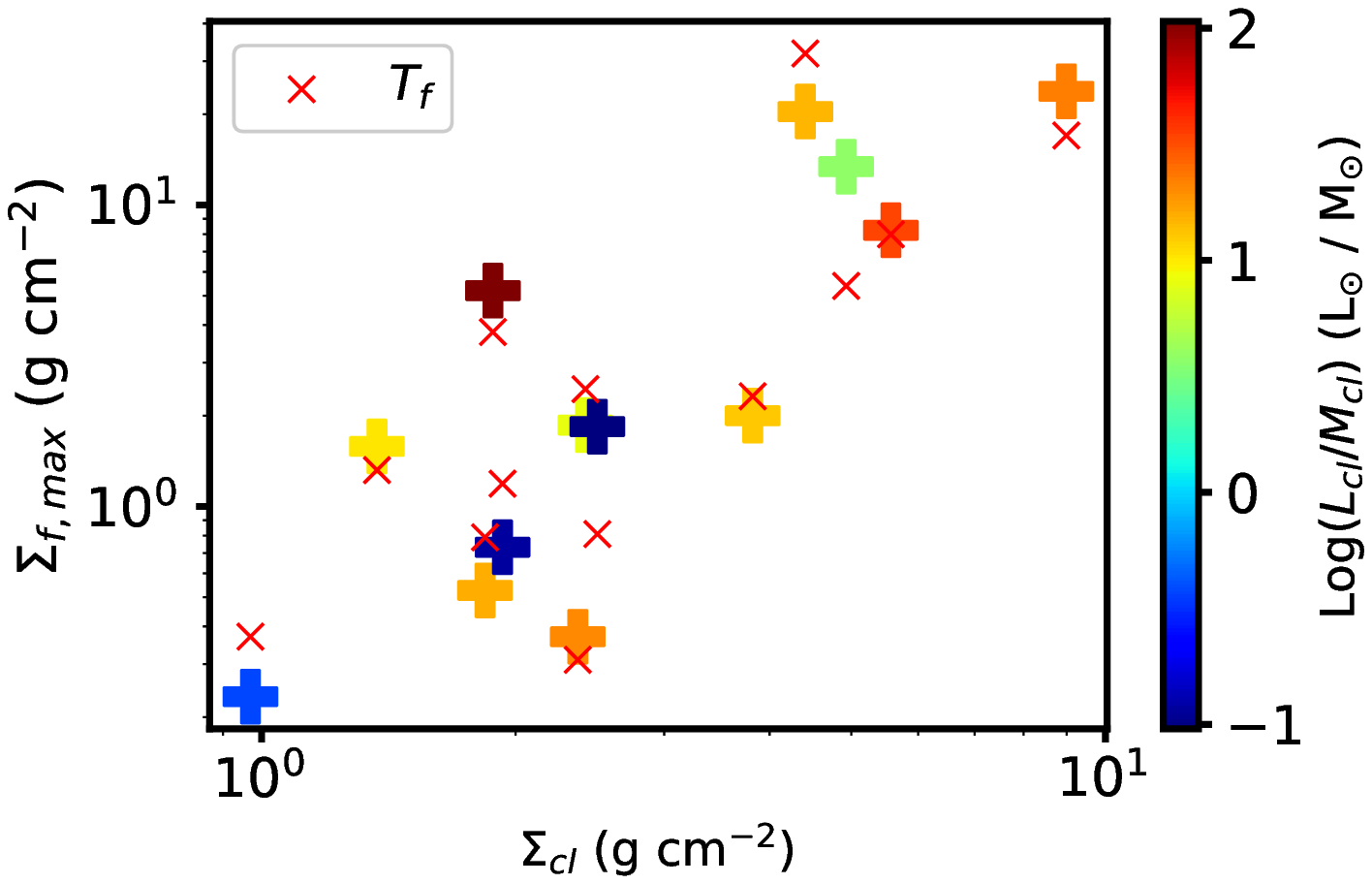} \\
	\includegraphics[width=7.0cm]{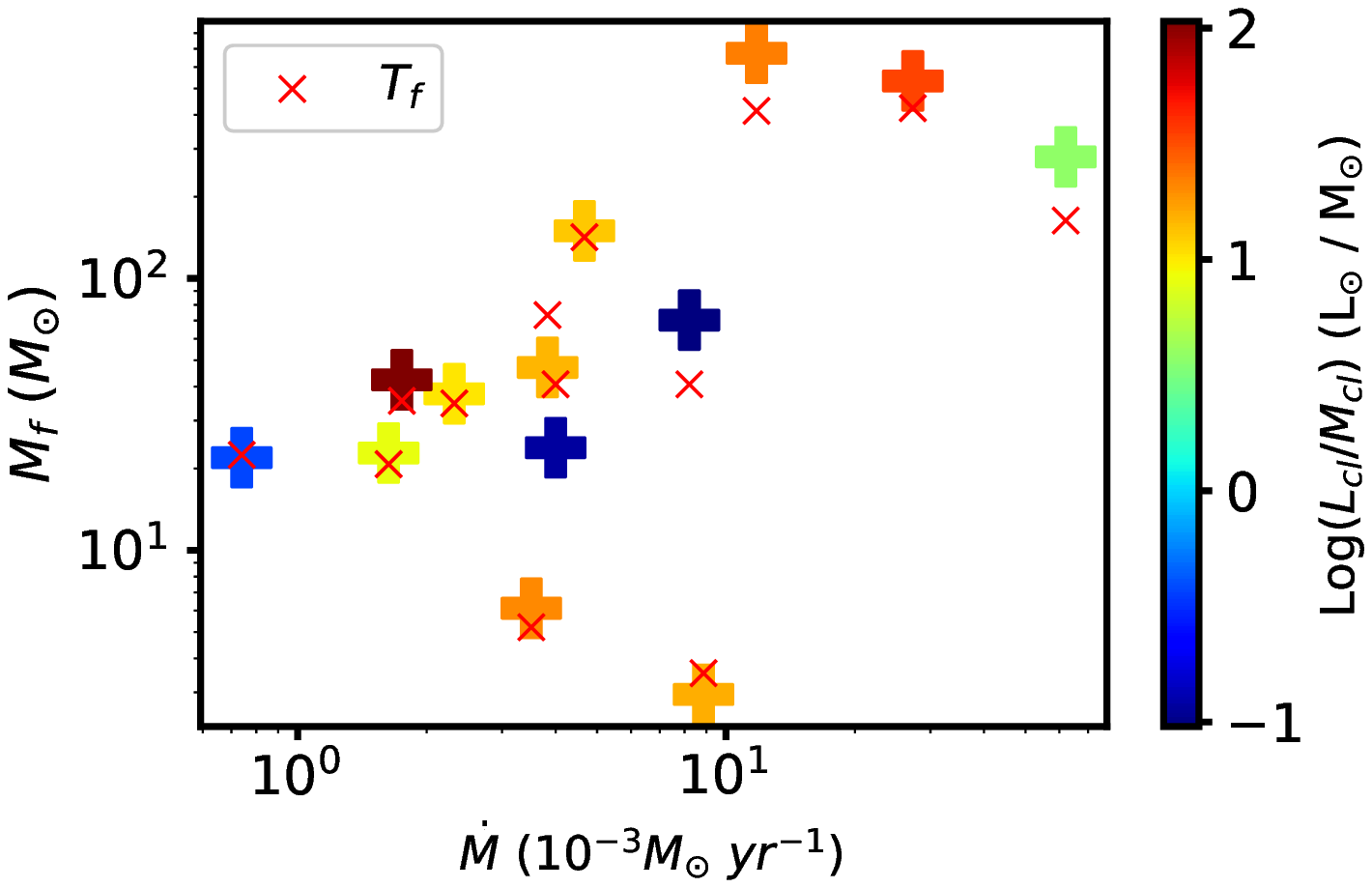}
	\includegraphics[width=7.0cm]{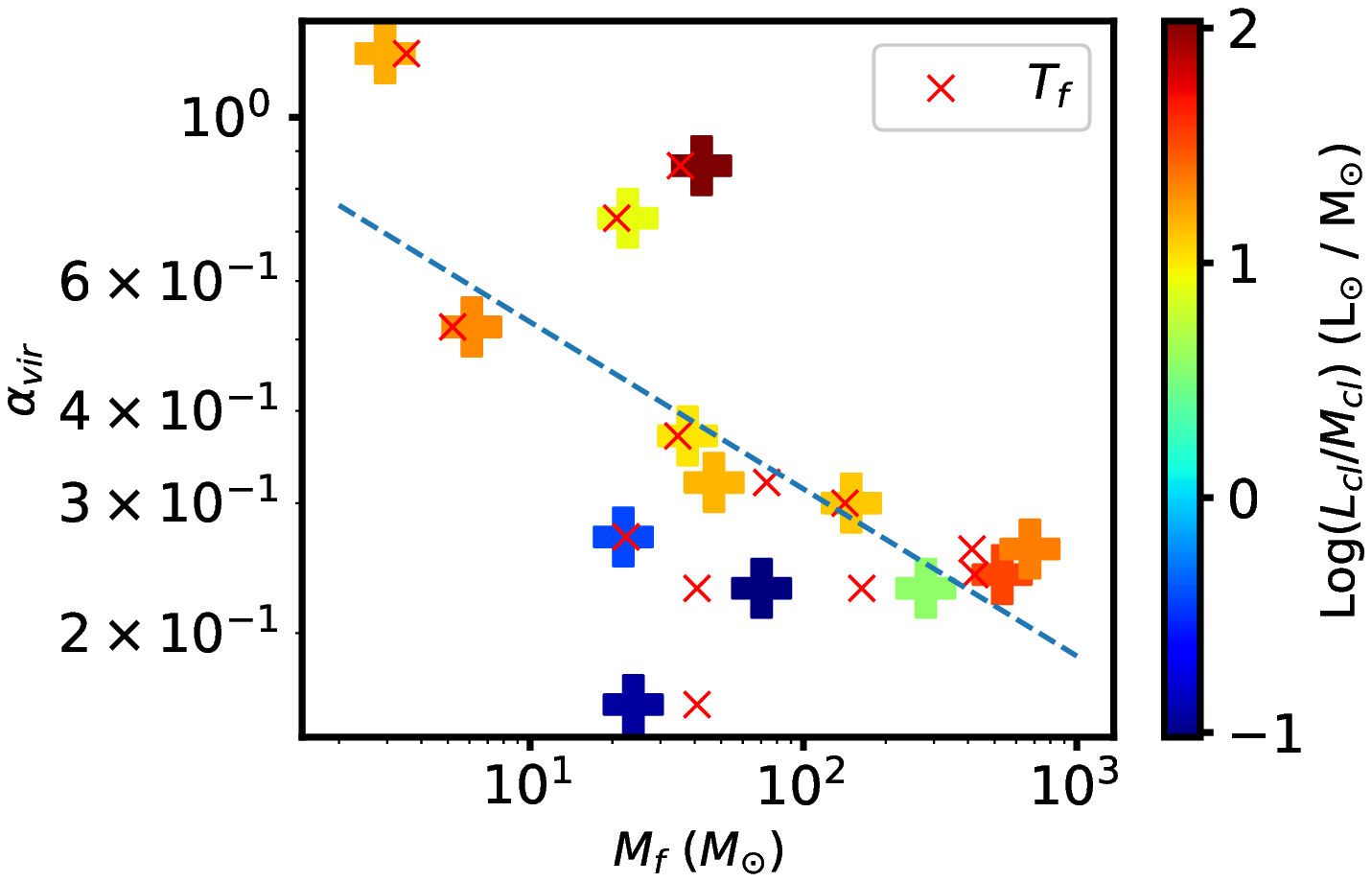}
    \caption{Same distributions of Figures \ref{fig:Mass_clump_mass_cores}, \ref{fig:Sigma_clump_Sigma_core_max}, \ref{fig:Accr_rate_fragment_mass} and \ref{fig:alpha_vir_core_masses} but for a different realization of the dust temperature of each fragment (and therefore a different estimation of the temperature-dependent parameters). The red crosses are the reference values assumed in the main text assuming T = $T_{f}$.}
    \label{fig:pearsons_MC_realizations}
\end{figure*}

In this Appendix we describe how the uncertainties in the dust temperatures of our fragments may impact our results. To investigate the impact on the temperature uncertainties, we have re-evaluated the temperature-dependent properties of our fragments described in Section \ref{sec:fragments_temperature_dependent_clump_props} using a different set of temperatures for each fragment. Namely, we have 
run a set of 100 Monte Carlo simulations and in each realization we have assigned to the fragments a random dust temperature $T_{lim}$ within the boundaries described in Table \ref{tab:temperature_limits}. For each run we have re-evaluated the scatter-plots in Figures \ref{fig:Mass_clump_mass_cores}, \ref{fig:Sigma_clump_Sigma_core_max}, \ref{fig:Accr_rate_fragment_mass} and \ref{fig:alpha_vir_core_masses} and the Pearson's correlation coefficients. In Figure \ref{fig:pearsons_MC_realizations} there is an example of one of this realization for each of the 4 scatter-plots, compared with the reference values assumed in the main text (using T = $T_{f}$), and in Table \ref{tab:pearson_mc_realizations} there are the mean, minimum and maximum values of the Pearson's coefficients for the 100 runs and for the reference values assumed in the main text using T = $T_{f}$. We also show the different slopes of the linear fit (in log-log space) to the $\alpha_{vir,cl}$ versus $M_{f}$ relation for the 100 realizations. Although the mass estimation can significantly vary for each single fragment (see Section \ref{sec:temperature_uncertainties}), the statistical (anti)correlations are preserved in all runs (as well as the negative slope in the fit of the $\alpha_{vir,cl}$ versus $M_{f}$ distribution), which assures that our conclusions are robust against the temperature uncertainties.

\begin{table*}\label{tab:pearson_mc_realizations}
\begin{tabular}{cccccc}
\hline
$\rho_{val}$ & $M_{f}$ vs. $M_{cl}$ & $\Sigma_{f,max}$ vs. $\Sigma_{cl}$ & $M_{f}$ vs. $\dot{M}_{cl}$ & $\alpha_{vir,cl}$ vs. $M_{f}$ & Slope ($\alpha_{vir,cl}$ vs. $M_{f}$)  \\
\hline
\hline

MEAN & 0.72 & 0.61 & 0.48 & -0.40 & -0.24 \\
MIN & 0.63 & 0.32 & 0.21 & -0.48 & -0.33 \\
MAX & 0.86 & 0.90 & 0.81 & -0.33 & -0.17 \\
REF & 0.73 & 0.60 & 0.44 & -0.41 & -0.29 \\

\hline
\end{tabular}
\caption{Distribution of the Pearson's correlation coefficients for the temperature dependent parameters described in Section \ref{sec:fragments_temperature_dependent_clump_props} for the 100 simulations. \textit{Col. 1}: mean ,min, max and reference value of the Pearson's coefficients ($\rho_{val}$); \textit{Cols. 2-5}: ($\rho_{val}$) for $M_{f}$ vs. $M_{cl}$, $\Sigma_{f,max}$ vs. $\Sigma_{cl}$, $M_{f}$ vs. $\dot{M}_{cl}$ and $\alpha_{vir,cl}$ vs. $M_{f}$ relations respectively; \textit{Col. 6}: slopes of the linear fit in the log-log space at the $\alpha_{vir,cl}$ vs. $M_{f}$ distribution.}
\end{table*}

% for fragments with assigned $T_{d,f}= 20K$: $10\leq T_{rand}\leq40$ K; for fragments with assigned $T_{d,f}= 30K$: $10\leq T_{rand}\leq50$ K; for fragments with assigned $T_{d,f}= 40K$: $20\leq T_{rand}\leq60$ K.

%%%%%%%%%%%%%%%%%%%%%%%%%%%%%%%%%%%%%%%%%%%%%%%%%%%%%%%%%%%%%%%%%%%%%%%%%%%%%%%%%%%%%%%
\newpage
\section{Pearson's parameters}\label{app:pearson_params}
In this Appendix we report the values of the Pearson's correlation coefficients for each pair of parameter considered in the heatmap in Appendix \ref{app:pearson_heatmap}. The Table includes the p-val percentiles and confidence intervals at 95\% level of confidence assuming uncertainties of each parameter equal to 20\% of its value.

\begin{table}\label{tab:pearson}
\begin{tabular}{ccccc}
\hline
Param\_1 & Param\_2 & $\rho$ & p-val & CI 95\% \\
\hline
\hline
$M_{cl}$ & $M_{cl}$ & 1.0 & 0.000 & [1.0, 1.0] \\
$M_{cl}$ & $L_{cl}/M_{cl}$ & -0.279 & 0.356 & [-0.72, 0.32] \\
$M_{cl}$ & $\Sigma_{cl}$ & 0.643 & 0.018 & [0.14, 0.88] \\
$M_{cl}$ & $\dot{M}_{cl}$ & 0.691 & 0.009 & [0.23, 0.9] \\
$M_{cl}$ & $\alpha_{vir,cl}$ & -0.684 & 0.010 & [-0.9, -0.21] \\
$M_{cl}$ & $\#_{f}$ & -0.088 & 0.776 & [-0.61, 0.49] \\
$M_{cl}$ & $d_{min,2D}$ & 0.011 & 0.972 & [-0.57, 0.58] \\
$M_{cl}$ & $d_{min,3D}$ & 0.202 & 0.528 & [-0.42, 0.7] \\
$M_{cl}$ & $M_{f}$ & 0.728 & 0.005 & [0.29, 0.91] \\
$M_{cl}$ & $M_{f,max}$ & 0.64 & 0.018 & [0.14, 0.88] \\
$M_{cl}$ & $\Sigma_{f,max}$ & 0.106 & 0.730 & [-0.47, 0.62] \\
$M_{cl}$ & CFE & 0.13 & 0.673 & [-0.45, 0.64] \\
$M_{cl}$ & $\lambda_{J_{r},2D}$ & -0.038 & 0.906 & [-0.6, 0.55] \\
$M_{cl}$ & $\lambda_{J_{r},3D}$ & 0.22 & 0.492 & [-0.41, 0.7] \\
$L_{cl}/M_{cl}$ & $M_{cl}$ & -0.279 & 0.356 & [-0.72, 0.32] \\
$L_{cl}/M_{cl}$ & $L_{cl}/M_{cl}$ & 1.0 & 0.000 & [1.0, 1.0] \\
$L_{cl}/M_{cl}$ & $\Sigma_{cl}$ & 0.013 & 0.966 & [-0.54, 0.56] \\
$L_{cl}/M_{cl}$ & $\dot{M}_{cl}$ & -0.125 & 0.685 & [-0.63, 0.46] \\
$L_{cl}/M_{cl}$ & $\alpha_{vir,cl}$ & 0.425 & 0.147 & [-0.16, 0.79] \\
$L_{cl}/M_{cl}$ & $\#_{f}$ & 0.367 & 0.218 & [-0.23, 0.76] \\
$L_{cl}/M_{cl}$ & $d_{min,2D}$ & -0.533 & 0.074 & [-0.85, 0.06] \\
$L_{cl}/M_{cl}$ & $d_{min,3D}$ & -0.615 & 0.033 & [-0.88, -0.06] \\
$L_{cl}/M_{cl}$ & $M_{f}$ & 0.09 & 0.769 & [-0.48, 0.61] \\
$L_{cl}/M_{cl}$ & $M_{f,max}$ & 0.014 & 0.964 & [-0.54, 0.56] \\
$L_{cl}/M_{cl}$ & $\Sigma_{f,max}$ & 0.058 & 0.851 & [-0.51, 0.59] \\
$L_{cl}/M_{cl}$ & CFE & 0.56 & 0.046 & [0.01, 0.85] \\
$L_{cl}/M_{cl}$ & $\lambda_{J_{r},2D}$ & -0.446 & 0.146 & [-0.81, 0.17] \\
$L_{cl}/M_{cl}$ & $\lambda_{J_{r},3D}$ & -0.519 & 0.084 & [-0.84, 0.08] \\
$\Sigma_{cl}$ & $M_{cl}$ & 0.643 & 0.018 & [0.14, 0.88] \\
$\Sigma_{cl}$ & $L_{cl}/M_{cl}$ & 0.013 & 0.966 & [-0.54, 0.56] \\
$\Sigma_{cl}$ & $\Sigma_{cl}$ & 1.0 & 0.000 & [1.0, 1.0] \\
$\Sigma_{cl}$ & $\dot{M}_{cl}$ & 0.438 & 0.135 & [-0.15, 0.8] \\
$\Sigma_{cl}$ & $\alpha_{vir,cl}$ & -0.36 & 0.227 & [-0.76, 0.24] \\
$\Sigma_{cl}$ & $\#_{f}$ & -0.238 & 0.434 & [-0.7, 0.36] \\
$\Sigma_{cl}$ & $d_{min,2D}$ & -0.288 & 0.363 & [-0.74, 0.34] \\
$\Sigma_{cl}$ & $d_{min,3D}$ & -0.213 & 0.507 & [-0.7, 0.41] \\
$\Sigma_{cl}$ & $M_{f}$ & 0.887 & 0.000 & [0.66, 0.97] \\
$\Sigma_{cl}$ & $M_{f,max}$ & 0.946 & 0.000 & [0.82, 0.98] \\
$\Sigma_{cl}$ & $\Sigma_{f,max}$ & 0.6 & 0.030 & [0.07, 0.87] \\
$\Sigma_{cl}$ & CFE & 0.62 & 0.024 & [0.11, 0.87] \\
$\Sigma_{cl}$ & $\lambda_{J_{r},2D}$ & -0.029 & 0.929 & [-0.59, 0.55] \\
$\Sigma_{cl}$ & $\lambda_{J_{r},3D}$ & 0.267 & 0.401 & [-0.36, 0.73] \\
$\dot{M}_{cl}$ & $M_{cl}$ & 0.691 & 0.009 & [0.23, 0.9] \\
$\dot{M}_{cl}$ & $L_{cl}/M_{cl}$ & -0.125 & 0.685 & [-0.63, 0.46] \\
$\dot{M}_{cl}$ & $\Sigma_{cl}$ & 0.438 & 0.135 & [-0.15, 0.8] \\
$\dot{M}_{cl}$ & $\dot{M}_{cl}$ & 1.0 & 0.000 & [1.0, 1.0] \\
$\dot{M}_{cl}$ & $\alpha_{vir,cl}$ & -0.257 & 0.398 & [-0.71, 0.34] \\
$\dot{M}_{cl}$ & $\#_{f}$ & -0.045 & 0.885 & [-0.58, 0.52] \\
$\dot{M}_{cl}$ & $d_{min,2D}$ & -0.03 & 0.927 & [-0.59, 0.55] \\
$\dot{M}_{cl}$ & $d_{min,3D}$ & -0.063 & 0.845 & [-0.61, 0.53] \\
$\dot{M}_{cl}$ & $M_{f}$ & 0.436 & 0.136 & [-0.15, 0.8] \\
$\dot{M}_{cl}$ & $M_{f,max}$ & 0.256 & 0.398 & [-0.34, 0.71] \\
$\dot{M}_{cl}$ & $\Sigma_{f,max}$ & 0.055 & 0.858 & [-0.51, 0.59] \\
$\dot{M}_{cl}$ & CFE & 0.068 & 0.825 & [-0.5, 0.6] \\
$\dot{M}_{cl}$ & $\lambda_{J_{r},2D}$ & 0.089 & 0.784 & [-0.51, 0.63] \\
$\dot{M}_{cl}$ & $\lambda_{J_{r},3D}$ & 0.15 & 0.641 & [-0.46, 0.67] \\
\hline
\end{tabular}
\caption{Perason's coefficients and statistical values for each pair of parameters described in the heatmap in Appendix \ref{app:pearson_heatmap}. \textit{Cols 1-2}: pair of parameter considered; \textit{Col3} value of the Pearson's correlation coefficient for the given pair of parameters; \textit{Cols 4-5} p-value percentiles and confidence interval at 95\% for the given pair of parameters.}
\end{table}

\begin{table}
\begin{tabular}{ccccc}
\hline
Param\_1 & Param\_2 & $\rho$ & p-val & CI 95\% \\
\hline
\hline
$\alpha_{vir,cl}$ & $M_{cl}$ & -0.684 & 0.010 & [-0.9, -0.21] \\
$\alpha_{vir,cl}$ & $L_{cl}/M_{cl}$ & 0.425 & 0.147 & [-0.16, 0.79] \\
$\alpha_{vir,cl}$ & $\Sigma_{cl}$ & -0.36 & 0.227 & [-0.76, 0.24] \\
$\alpha_{vir,cl}$ & $\dot{M}_{cl}$ & -0.257 & 0.398 & [-0.71, 0.34] \\
$\alpha_{vir,cl}$ & $\alpha_{vir,cl}$ & 1.0 & 0.000 & [1.0, 1.0] \\
$\alpha_{vir,cl}$ & $\#_{f}$ & 0.017 & 0.956 & [-0.54, 0.56] \\
$\alpha_{vir,cl}$ & $d_{min,2D}$ & 0.008 & 0.980 & [-0.57, 0.58] \\
$\alpha_{vir,cl}$ & $d_{min,3D}$ & -0.243 & 0.446 & [-0.72, 0.38] \\
$\alpha_{vir,cl}$ & $M_{f}$ & -0.407 & 0.168 & [-0.78, 0.19] \\
$\alpha_{vir,cl}$ & $M_{f,max}$ & -0.363 & 0.223 & [-0.76, 0.24] \\
$\alpha_{vir,cl}$ & $\Sigma_{f,max}$ & -0.222 & 0.465 & [-0.69, 0.37] \\
$\alpha_{vir,cl}$ & CFE & -0.069 & 0.824 & [-0.6, 0.5] \\
$\alpha_{vir,cl}$ & $\lambda_{J_{r},2D}$ & 0.104 & 0.748 & [-0.5, 0.64] \\
$\alpha_{vir,cl}$ & $\lambda_{J_{r},3D}$ & -0.132 & 0.683 & [-0.66, 0.48] \\
$\#_{f}$ & $M_{cl}$ & -0.088 & 0.776 & [-0.61, 0.49] \\
$\#_{f}$ & $L_{cl}/M_{cl}$ & 0.367 & 0.218 & [-0.23, 0.76] \\
$\#_{f}$ & $\Sigma_{cl}$ & -0.238 & 0.434 & [-0.7, 0.36] \\
$\#_{f}$ & $\dot{M}_{cl}$ & -0.045 & 0.885 & [-0.58, 0.52] \\
$\#_{f}$ & $\alpha_{vir,cl}$ & 0.017 & 0.956 & [-0.54, 0.56] \\
$\#_{f}$ & $\#_{f}$ & 1.0 & 0.000 & [1.0, 1.0] \\
$\#_{f}$ & $d_{min,2D}$ & -0.626 & 0.029 & [-0.88, -0.08] \\
$\#_{f}$ & $d_{min,3D}$ & -0.51 & 0.090 & [-0.84, 0.09] \\
$\#_{f}$ & $M_{f}$ & -0.037 & 0.904 & [-0.58, 0.52] \\
$\#_{f}$ & $M_{f,max}$ & -0.295 & 0.328 & [-0.73, 0.31] \\
$\#_{f}$ & $\Sigma_{f,max}$ & -0.468 & 0.107 & [-0.81, 0.11] \\
$\#_{f}$ & CFE & 0.066 & 0.830 & [-0.5, 0.6] \\
$\#_{f}$ & $\lambda_{J_{r},2D}$ & -0.668 & 0.018 & [-0.9, -0.15] \\
$\#_{f}$ & $\lambda_{J_{r},3D}$ & -0.668 & 0.018 & [-0.9, -0.15] \\
$d_{min,2D}$ & $M_{cl}$ & 0.011 & 0.972 & [-0.57, 0.58] \\
$d_{min,2D}$ & $L_{cl}/M_{cl}$ & -0.533 & 0.074 & [-0.85, 0.06] \\
$d_{min,2D}$ & $\Sigma_{cl}$ & -0.288 & 0.363 & [-0.74, 0.34] \\
$d_{min,2D}$ & $\dot{M}_{cl}$ & -0.03 & 0.927 & [-0.59, 0.55] \\
$d_{min,2D}$ & $\alpha_{vir,cl}$ & 0.008 & 0.980 & [-0.57, 0.58] \\
$d_{min,2D}$ & $\#_{f}$ & -0.626 & 0.029 & [-0.88, -0.08] \\
$d_{min,2D}$ & $d_{min,2D}$ & 1.0 & 0.000 & [1.0, 1.0] \\
$d_{min,2D}$ & $d_{min,3D}$ & 0.933 & 0.000 & [0.77, 0.98] \\
$d_{min,2D}$ & $M_{f}$ & -0.331 & 0.293 & [-0.76, 0.3] \\
$d_{min,2D}$ & $M_{f,max}$ & -0.23 & 0.472 & [-0.71, 0.4] \\
$d_{min,2D}$ & $\Sigma_{f,max}$ & -0.278 & 0.383 & [-0.73, 0.35] \\
$d_{min,2D}$ & CFE & -0.384 & 0.217 & [-0.79, 0.24] \\
$d_{min,2D}$ & $\lambda_{J_{r},2D}$ & 0.837 & 0.001 & [0.51, 0.95] \\
$d_{min,2D}$ & $\lambda_{J_{r},3D}$ & 0.735 & 0.006 & [0.28, 0.92] \\
$d_{min,3D}$ & $M_{cl}$ & 0.202 & 0.528 & [-0.42, 0.7] \\
$d_{min,3D}$ & $L_{cl}/M_{cl}$ & -0.615 & 0.033 & [-0.88, -0.06] \\
$d_{min,3D}$ & $\Sigma_{cl}$ & -0.213 & 0.507 & [-0.7, 0.41] \\
$d_{min,3D}$ & $\dot{M}_{cl}$ & -0.063 & 0.845 & [-0.61, 0.53] \\
$d_{min,3D}$ & $\alpha_{vir,cl}$ & -0.243 & 0.446 & [-0.72, 0.38] \\
$d_{min,3D}$ & $\#_{f}$ & -0.51 & 0.090 & [-0.84, 0.09] \\
$d_{min,3D}$ & $d_{min,2D}$ & 0.933 & 0.000 & [0.77, 0.98] \\
$d_{min,3D}$ & $d_{min,3D}$ & 1.0 & 0.000 & [1.0, 1.0] \\
$d_{min,3D}$ & $M_{f}$ & -0.185 & 0.566 & [-0.69, 0.44] \\
$d_{min,3D}$ & $M_{f,max}$ & -0.113 & 0.726 & [-0.65, 0.49] \\
$d_{min,3D}$ & $\Sigma_{f,max}$ & -0.202 & 0.528 & [-0.7, 0.42] \\
$d_{min,3D}$ & CFE & -0.362 & 0.248 & [-0.77, 0.27] \\
$d_{min,3D}$ & $\lambda_{J_{r},2D}$ & 0.651 & 0.022 & [0.12, 0.89] \\
$d_{min,3D}$ & $\lambda_{J_{r},3D}$ & 0.631 & 0.028 & [0.09, 0.88] \\
\hline
\end{tabular}
\end{table}

\begin{table}
\begin{tabular}{ccccc}
\hline
Param\_1 & Param\_2 & $\rho$ & p-val & CI 95\% \\
\hline
\hline
$M_{f}$ & $M_{cl}$ & 0.728 & 0.005 & [0.29, 0.91] \\
$M_{f}$ & $L_{cl}/M_{cl}$ & 0.09 & 0.769 & [-0.48, 0.61] \\
$M_{f}$ & $\Sigma_{cl}$ & 0.887 & 0.000 & [0.66, 0.97] \\
$M_{f}$ & $\dot{M}_{cl}$ & 0.436 & 0.136 & [-0.15, 0.8] \\
$M_{f}$ & $\alpha_{vir,cl}$ & -0.407 & 0.168 & [-0.78, 0.19] \\
$M_{f}$ & $\#_{f}$ & -0.037 & 0.904 & [-0.58, 0.52] \\
$M_{f}$ & $d_{min,2D}$ & -0.331 & 0.293 & [-0.76, 0.3] \\
$M_{f}$ & $d_{min,3D}$ & -0.185 & 0.566 & [-0.69, 0.44] \\
$M_{f}$ & $M_{f}$ & 1.0 & 0.000 & [1.0, 1.0] \\
$M_{f}$ & $M_{f,max}$ & 0.87 & 0.000 & [0.61, 0.96] \\
$M_{f}$ & $\Sigma_{f,max}$ & 0.381 & 0.199 & [-0.21, 0.77] \\
$M_{f}$ & CFE & 0.65 & 0.016 & [0.15, 0.88] \\
$M_{f}$ & $\lambda_{J_{r},2D}$ & -0.173 & 0.590 & [-0.68, 0.44] \\
$M_{f}$ & $\lambda_{J_{r},3D}$ & 0.123 & 0.704 & [-0.49, 0.65] \\
$M_{f,max}$ & $M_{cl}$ & 0.64 & 0.018 & [0.14, 0.88] \\
$M_{f,max}$ & $L_{cl}/M_{cl}$ & 0.014 & 0.964 & [-0.54, 0.56] \\
$M_{f,max}$ & $\Sigma_{cl}$ & 0.946 & 0.000 & [0.82, 0.98] \\
$M_{f,max}$ & $\dot{M}_{cl}$ & 0.256 & 0.398 & [-0.34, 0.71] \\
$M_{f,max}$ & $\alpha_{vir,cl}$ & -0.363 & 0.223 & [-0.76, 0.24] \\
$M_{f,max}$ & $\#_{f}$ & -0.295 & 0.328 & [-0.73, 0.31] \\
$M_{f,max}$ & $d_{min,2D}$ & -0.23 & 0.472 & [-0.71, 0.4] \\
$M_{f,max}$ & $d_{min,3D}$ & -0.113 & 0.726 & [-0.65, 0.49] \\
$M_{f,max}$ & $M_{f}$ & 0.87 & 0.000 & [0.61, 0.96] \\
$M_{f,max}$ & $M_{f,max}$ & 1.0 & 0.000 & [1.0, 1.0] \\
$M_{f,max}$ & $\Sigma_{f,max}$ & 0.546 & 0.054 & [-0.01, 0.84] \\
$M_{f,max}$ & CFE & 0.558 & 0.047 & [0.01, 0.85] \\
$M_{f,max}$ & $\lambda_{J_{r},2D}$ & -0.056 & 0.862 & [-0.61, 0.53] \\
$M_{f,max}$ & $\lambda_{J_{r},3D}$ & 0.242 & 0.449 & [-0.39, 0.72] \\
$\Sigma_{f,max}$ & $M_{cl}$ & 0.106 & 0.730 & [-0.47, 0.62] \\
$\Sigma_{f,max}$ & $L_{cl}/M_{cl}$ & 0.058 & 0.851 & [-0.51, 0.59] \\
$\Sigma_{f,max}$ & $\Sigma_{cl}$ & 0.6 & 0.030 & [0.07, 0.87] \\
$\Sigma_{f,max}$ & $\dot{M}_{cl}$ & 0.055 & 0.858 & [-0.51, 0.59] \\
$\Sigma_{f,max}$ & $\alpha_{vir,cl}$ & -0.222 & 0.465 & [-0.69, 0.37] \\
$\Sigma_{f,max}$ & $\#_{f}$ & -0.468 & 0.107 & [-0.81, 0.11] \\
$\Sigma_{f,max}$ & $d_{min,2D}$ & -0.278 & 0.383 & [-0.73, 0.35] \\
$\Sigma_{f,max}$ & $d_{min,3D}$ & -0.202 & 0.528 & [-0.7, 0.42] \\
$\Sigma_{f,max}$ & $M_{f}$ & 0.381 & 0.199 & [-0.21, 0.77] \\
$\Sigma_{f,max}$ & $M_{f,max}$ & 0.546 & 0.054 & [-0.01, 0.84] \\
$\Sigma_{f,max}$ & $\Sigma_{f,max}$ & 1.0 & 0.000 & [1.0, 1.0] \\
$\Sigma_{f,max}$ & CFE & 0.592 & 0.033 & [0.06, 0.86] \\
$\Sigma_{f,max}$ & $\lambda_{J_{r},2D}$ & -0.069 & 0.831 & [-0.62, 0.53] \\
$\Sigma_{f,max}$ & $\lambda_{J_{r},3D}$ & 0.189 & 0.555 & [-0.43, 0.69] \\
CFE & $M_{cl}$ & 0.13 & 0.673 & [-0.45, 0.64] \\
CFE & $L_{cl}/M_{cl}$ & 0.56 & 0.046 & [0.01, 0.85] \\
CFE & $\Sigma_{cl}$ & 0.62 & 0.024 & [0.11, 0.87] \\
CFE & $\dot{M}_{cl}$ & 0.068 & 0.825 & [-0.5, 0.6] \\
CFE & $\alpha_{vir,cl}$ & -0.069 & 0.824 & [-0.6, 0.5] \\
CFE & $\#_{f}$ & 0.066 & 0.830 & [-0.5, 0.6] \\
CFE & $d_{min,2D}$ & -0.384 & 0.217 & [-0.79, 0.24] \\
CFE & $d_{min,3D}$ & -0.362 & 0.248 & [-0.77, 0.27] \\
CFE & $M_{f}$ & 0.65 & 0.016 & [0.15, 0.88] \\
CFE & $M_{f,max}$ & 0.558 & 0.047 & [0.01, 0.85] \\
CFE & $\Sigma_{f,max}$ & 0.592 & 0.033 & [0.06, 0.86] \\
CFE & CFE & 1.0 & 0.000 & [1.0, 1.0] \\
CFE & $\lambda_{J_{r},2D}$ & -0.124 & 0.700 & [-0.65, 0.48] \\
CFE & $\lambda_{J_{r},3D}$ & 0.034 & 0.915 & [-0.55, 0.6] \\
\hline
\end{tabular}
\end{table}

\begin{table}
\begin{tabular}{ccccc}
\hline
Param\_1 & Param\_2 & $\rho$ & p-val & CI 95\% \\
\hline
\hline
$\lambda_{J_{r},2D}$ & $M_{cl}$ & -0.038 & 0.906 & [-0.6, 0.55] \\
$\lambda_{J_{r},2D}$ & $L_{cl}/M_{cl}$ & -0.446 & 0.146 & [-0.81, 0.17] \\
$\lambda_{J_{r},2D}$ & $\Sigma_{cl}$ & -0.029 & 0.929 & [-0.59, 0.55] \\
$\lambda_{J_{r},2D}$ & $\dot{M}_{cl}$ & 0.089 & 0.784 & [-0.51, 0.63] \\
$\lambda_{J_{r},2D}$ & $\alpha_{vir,cl}$ & 0.104 & 0.748 & [-0.5, 0.64] \\
$\lambda_{J_{r},2D}$ & $\#_{f}$ & -0.668 & 0.018 & [-0.9, -0.15] \\
$\lambda_{J_{r},2D}$ & $d_{min,2D}$ & 0.837 & 0.001 & [0.51, 0.95] \\
$\lambda_{J_{r},2D}$ & $d_{min,3D}$ & 0.651 & 0.022 & [0.12, 0.89] \\
$\lambda_{J_{r},2D}$ & $M_{f}$ & -0.173 & 0.590 & [-0.68, 0.44] \\
$\lambda_{J_{r},2D}$ & $M_{f,max}$ & -0.056 & 0.862 & [-0.61, 0.53] \\
$\lambda_{J_{r},2D}$ & $\Sigma_{f,max}$ & -0.069 & 0.831 & [-0.62, 0.53] \\
$\lambda_{J_{r},2D}$ & CFE & -0.124 & 0.700 & [-0.65, 0.48] \\
$\lambda_{J_{r},2D}$ & $\lambda_{J_{r},2D}$ & 1.0 & 0.000 & [1.0, 1.0] \\
$\lambda_{J_{r},2D}$ & $\lambda_{J_{r},3D}$ & 0.933 & 0.000 & [0.77, 0.98] \\
$\lambda_{J_{r},3D}$ & $M_{cl}$ & 0.22 & 0.492 & [-0.41, 0.7] \\
$\lambda_{J_{r},3D}$ & $L_{cl}/M_{cl}$ & -0.519 & 0.084 & [-0.84, 0.08] \\
$\lambda_{J_{r},3D}$ & $\Sigma_{cl}$ & 0.267 & 0.401 & [-0.36, 0.73] \\
$\lambda_{J_{r},3D}$ & $\dot{M}_{cl}$ & 0.15 & 0.641 & [-0.46, 0.67] \\
$\lambda_{J_{r},3D}$ & $\alpha_{vir,cl}$ & -0.132 & 0.683 & [-0.66, 0.48] \\
$\lambda_{J_{r},3D}$ & $\#_{f}$ & -0.668 & 0.018 & [-0.9, -0.15] \\
$\lambda_{J_{r},3D}$ & $d_{min,2D}$ & 0.735 & 0.006 & [0.28, 0.92] \\
$\lambda_{J_{r},3D}$ & $d_{min,3D}$ & 0.631 & 0.028 & [0.09, 0.88] \\
$\lambda_{J_{r},3D}$ & $M_{f}$ & 0.123 & 0.704 & [-0.49, 0.65] \\
$\lambda_{J_{r},3D}$ & $M_{f,max}$ & 0.242 & 0.449 & [-0.39, 0.72] \\
$\lambda_{J_{r},3D}$ & $\Sigma_{f,max}$ & 0.189 & 0.555 & [-0.43, 0.69] \\
$\lambda_{J_{r},3D}$ & CFE & 0.034 & 0.915 & [-0.55, 0.6] \\
$\lambda_{J_{r},3D}$ & $\lambda_{J_{r},2D}$ & 0.933 & 0.000 & [0.77, 0.98] \\
$\lambda_{J_{r},3D}$ & $\lambda_{J_{r},3D}$ & 1.0 & 0.000 & [1.0, 1.0] \\
\hline
\end{tabular}
\end{table}

%%%%%%%%%%%%%%%%%%%%%%%%%%%%%%%%%%%%%%%%%%%%%%%%%%

% Don't change these lines
\bsp	% typesetting comment
\label{lastpage}
\end{document}